\pdfoutput=1

\documentclass[onecolumn]{aa}
\usepackage[varg]{txfonts}
\usepackage{amsmath,bm}
\usepackage{graphicx}
\usepackage[colorlinks=true,linkcolor=blue,citecolor=blue,filecolor=blue,urlcolor=blue]{hyperref}

\makeatletter
\renewcommand*\aa@pageof{, page \thepage{} of \pageref*{LastPage}}
\makeatother

\DeclareMathOperator\arctanh{arctanh}

\begin{document}

\title{Gravitational lensing by an ellipsoidal Navarro--Frenk--White dark-matter halo: An analytic solution and its properties}

\author{David Heyrovsk\'y \and Michal Karamazov}

\institute{Institute of Theoretical Physics, Faculty of Mathematics and Physics, Charles University,\\ V Hole\v{s}ovi\v{c}k\'ach 2, 18000~Praha 8, Czech Republic\\ \email{david.heyrovsky@mff.cuni.cz, michal.karamazov@gmail.com}}

\date{Received 29 March 2024 / Accepted 4 June 2024}

\abstract{The analysis of gravitational lensing by galaxies and galaxy clusters typically relies on ellipsoidal lens models to describe the deflection of light by the involved dark-matter halos. These models are most often based on the isothermal density profile---not an optimal description of the halo, but easy to use because it leads to an analytic deflection-angle formula.}{Dark-matter halos are better described by the Navarro--Frenk--White (hereafter NFW) density profile. We set out to study lensing by a general triaxial ellipsoidal NFW halo, with the aim of providing an analytic model that would be more consistent with the current understanding of dark-matter halos.}{We computed the conversion between the properties of a triaxial ellipsoidal lens model and its elliptical surface-density profile. In the case of the NFW lens model, its angular scale is defined by the projected scale semi-major axis of the halo, while its lensing regime depends on two parameters: the projected eccentricity $e$ and the convergence parameter $\kappa_\text{s}$. We employed the Bourassa \& Kantowski formalism to compute the complex scattering function of the model, which yields the deflection-angle components when separated into its real and imaginary parts.}{We present the obtained closed-form expressions for the deflection-angle components, valid for an arbitrary eccentricity of the surface-density profile. We use them to compute and describe the lensing properties of the model, including: the shear, its components, and the phase; the critical curves, caustics, and the parameter-space mapping of their different geometries; the deformations and orientations of images.}{The analytically solved ellipsoidal NFW lens model is available for implementation in gravitational lensing software. The techniques introduced here such as the image-plane analysis can prove to be useful for understanding the properties of other lens models as well.}


\titlerunning{Gravitational Lensing by Ellipsoidal NFW Halo}
\authorrunning{D. Heyrovsk\'y \& M. Karamazov}

\maketitle

\nolinenumbers

\section{Introduction}
\label{sec:Intro}

Gravitational lensing provides a unique tool for studying the distribution of matter in galaxies and galaxy clusters. In the case of clusters, detailed maps of their surface density have been obtained by combined analyses of strong lensing in their inner and weak lensing in their outer regions \citep[e.g.,][]{bradac_etal05,limousin_etal07,finney_etal18,jauzac_etal18}. In the strong-lensing analysis, the mass distribution is typically reconstructed using building-block lens models describing the cluster halo and the subhalos corresponding to cluster-member galaxies \citep{natarajan_etal24}. The analysis of strong lensing by galaxies follows a similar modular approach, albeit with fewer components. The stellar and dark-matter components may be modeled separately or by a combined model; these are usually supplemented by external shear \citep{treu10,shajib_etal22} to account for the lensing influence of other nearby galaxies or larger structures. Subhalos due to satellite galaxies or other substructure may be added to explain asymmetries or flux-ratio anomalies.

The individual halos, subhalos, or different matter components are generally modeled by simple density or surface-density profiles \citep[for a theoretical overview, see][]{wagner20}. In practice, the range of possible profiles is narrowed down by the requirement of an analytic expression for the light-deflection angle. Without it the angle has to be integrated numerically over the surface-density distribution for each light ray, which is computationally prohibitive. As a result, density profiles that cannot be integrated analytically are shunned, even though they may provide a better description of the mass distribution of the lens. Analytic expressions for the deflection angle are available for a number of spherically symmetric lens models \citep{keeton01}. However, in the case of ellipsoidal models, one geometric step closer to reality, analytic formulae for the deflection-angle components have been obtained only for a few.

All widely used ellipsoidal models are based on the isothermal density profile. The basic versions described by \cite{kassiola_kovner93} and \cite{kormann_etal94} are the singular isothermal ellipsoid (SIE, SIEMD), with a centrally divergent density, and the non-singular isothermal (pseudoisothermal) ellipsoid (NIE, PIEMD), which is softened by introducing a central core. Subtracting a second NIE with a larger core leads to an asymptotically attenuated density profile with a finite total mass. This model appears in the literature under different names: PIEMD with smooth cutoff \citep{kassiola_kovner93}, superposition of PIEMD components \citep{natarajan_kneib97}, pseudo-Jaffe ellipsoid \citep{keeton01}, or dual pseudoisothermal elliptical mass distribution \citep[dPIE,][]{eliasdottir07}. At present this is the model of choice in cluster-lensing analysis, while galaxy-lensing analysis often uses a simple SIE with external shear \citep{shajib_etal22}. As shown by \cite{etherington_etal24}, rather than describing the influence of nearby matter distributions, the included ``external shear'' tends to play the role of a prop compensating for an inadequate lens model.

Another class of analytic models is based on the power-law density profile: the singular power-law ellipsoid \citep{grogin_narayan96,tessore_metcalf15,oriordan_etal20}, the broken-power-law-density ellipsoid \citep{du_etal20}, the broken power-law-surface-density elliptical model \citep{oriordan_etal21}, and the truncated power-law-surface-density elliptical model \citep{oriordan_etal21}. In all these variants the deflection angle is expressed in terms of the hypergeometric function of a complex argument, which requires a suitable numerical technique for efficient evaluation \citep{tessore_metcalf15}. The character of some of these models is approaching our current understanding of the structure of dark-matter-halo density profiles. In particular, \cite{du_etal20} showed that the broken power-law-density profile can be used to approximate the Navarro--Frenk--White \citep[hereafter NFW,][]{navarro_etal96} and the Einasto \citep{einasto65} density profiles.

The NFW profile was originally introduced as an empirical law describing the density profile of dark-matter halos arising from cosmological structure-formation simulations. Observational support has built up gradually over the years, showing that the profile described adequately the continuous matter (i.e., dark matter + baryonic gas) in galaxy clusters \citep[e.g.,][]{newman_etal13}, and agreed with the analyses of cluster lensing \citep{okabe_etal13,umetsu_diemer17} and cluster X-ray emission \citep{ettori_etal13}. On the scale of individual galaxies, it is suitable for describing their halos in cosmological simulations \citep{ludlow_etal13}, and it has been observationally supported as an approximation for elliptical-galaxy \citep{shajib_etal21} or massive-spiral-galaxy \citep{rodrigues_etal17} halos.

The spherical NFW lens model was described by \cite{bartelmann96} and by \cite{wright_brainerd00} shortly after the discovery of the profile. The ellipsoidal NFW model proved to be a harder challenge; its sought analytic solution has been elusive despite its simple homoeoidal symmetry. In the studies that implemented the model the deflection angle was either integrated numerically \citep[e.g.,][]{suyu_etal12}, or it was constructed from analytic solutions of various approximate models. One common approximation is based on elliptical distortions of the spherical NFW lens potential \citep[e.g.,][]{golse_kneib02,meneghetti_etal03,meena_bagla23}, an approach that leads to unphysical dumbbell-shaped surface-density contours for all but the lowest ellipticities \citep{kassiola_kovner93}. The approximation can be improved by adjusting the ellipticity of the potential to yield the desired ellipticity of the innermost surface-density contours \citep{golse_kneib02}. However, the ellipticity of the surface-density contours changes outward right from the halo center even in the case of low ellipticities, already breaking the homoeoidal symmetry before deviations from elliptical symmetry become prominent \citep{gomer_etal23}. Another method of approximation is based on emulating the ellipsoidal NFW by a sum or series of simpler analytic models. Recently \cite{oguri21} proposed to decompose the NFW surface-density profile into a series of cored-steep-ellipsoid components and presented sums of 13 and 44 components approximating the lensing characteristics of the ellipsoidal NFW model to relative accuracy $10^{-2}$ and $10^{-4}$, respectively.

In this work we derive analytic expressions for the deflection angle of the ellipsoidal NFW model and describe its lensing properties. We start in Sect.~\ref{sec:triaxial} from a triaxial homoeoidally symmetric mass distribution with an arbitrary spatial orientation and compute the properties of its corresponding elliptically symmetric surface-density distribution. In Sect.~\ref{sec:Bourassa} we review the complex formalism of \cite{bourassa_kantowski75} for computing the deflection angle of a lens with an elliptically symmetric surface-density distribution. In Sect.~\ref{sec:eNFW} we apply the methodology introduced in the previous sections to the ellipsoidal NFW lens, describing: the convergence profile (Sect.~\ref{sec:kappa}); the complex scattering function and the deflection angle (Sect.~\ref{sec:alpha}); the shear and phase (Sect.~\ref{sec:shear}) with particular attention to the halo center (Sect.~\ref{sec:shear-halo-center}) and the zero-shear points (Sect.~\ref{sec:zero-shear-points}); the critical curves and caustics and their variation in the lens-parameter space (Sect.~\ref{sec:curves}); the properties of images (Sect.~\ref{sec:images}). We discuss the observational relevance and possible further modifications of the model in Sect.~\ref{sec:discussion}, and summarize the main results in Sect.~\ref{sec:summary}. Of the four appendices we would like to draw attention to Appendix~\ref{sec:Appendix-image-plane} in which we introduce the technique of image-plane analysis of critical curves and caustics, a powerful visual tool for tracking the variations of these curves in the parameter space of the lens.

\section{Surface density of a triaxial ellipsoidal mass distribution}
\label{sec:triaxial}

We consider a model of an astrophysical object with a triaxial ellipsoidal mass distribution and homoeoidal symmetry: its density is constant on concentric nestled ellipsoids of a fixed shape. The density
\begin{equation}
\label{eq:density}
\rho=\rho(\hat{a})
\end{equation}
can thus be written as a function of the semi-major axis of these ellipsoids,
\begin{equation}
\label{eq:semi-major-3D}
\hat{a}(\hat{x}_1, \hat{x}_2, \hat{x}_3)=\sqrt{\hat{x}_1^2+\hat{x}_2^2/(1-e_1^2)+\hat{x}_3^2/(1-e_2^2)}\,,
\end{equation}
where the constant eccentricities $e_1$, $e_2$ satisfy $0\leq e_1 \leq e_2 < 1$. In this notation the Cartesian coordinates $\hat{x}_1$, $\hat{x}_2$, and $\hat{x}_3$ correspond to the principal-axes frame with orientations along the major, median, and minor axis of the ellipsoids, respectively.

\begin{figure}
\centering
\resizebox{\hsize/2}{!}{\includegraphics{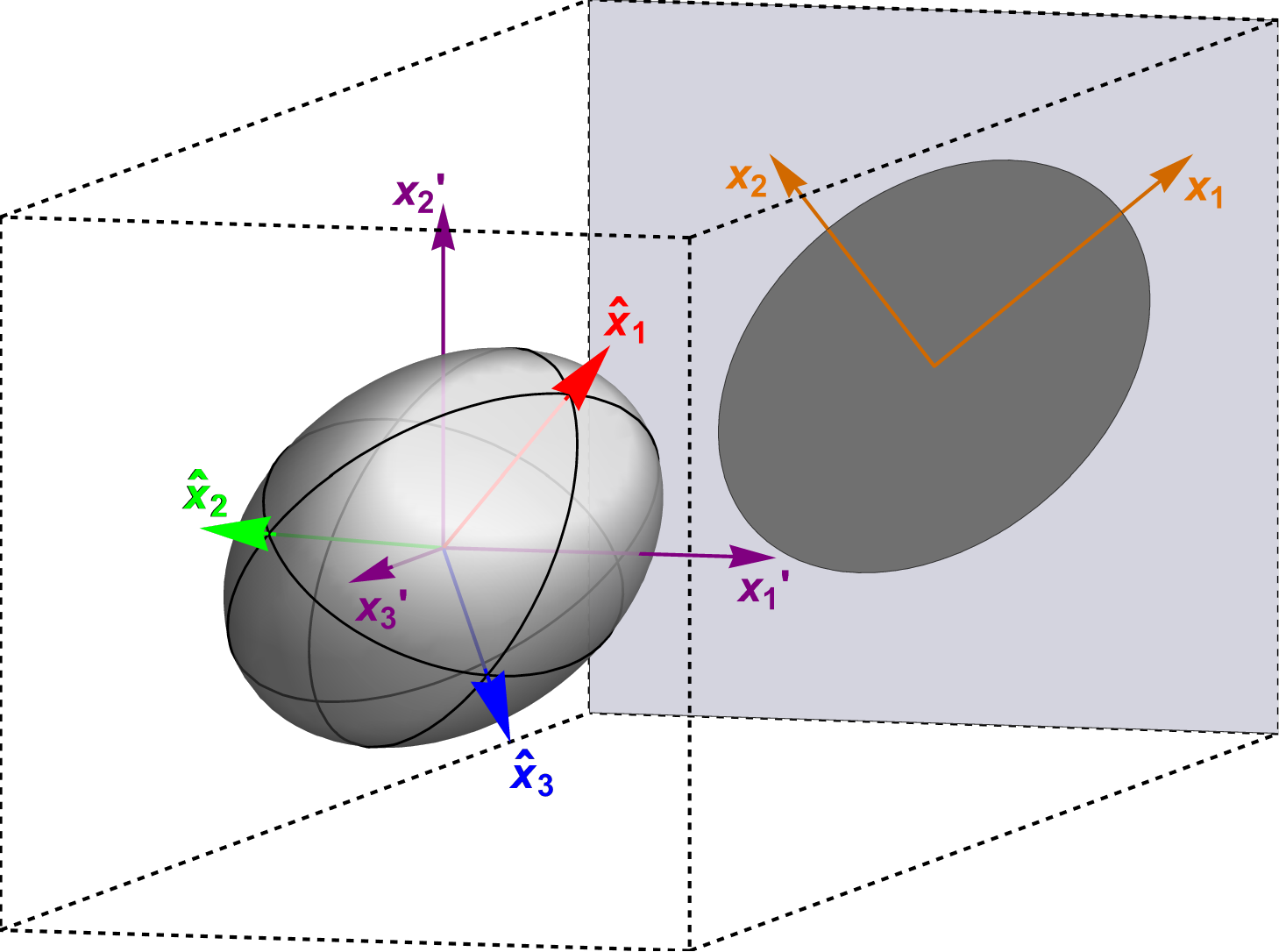}}
\caption{Reference frames of a triaxially ellipsoidally symmetric mass distribution and its projected surface density. The transparent shaded gray ellipsoid represents a surface of constant density, with principal major $\hat{x}_1$, median $\hat{x}_2$, and minor $\hat{x}_3$ axes marked red, green, and blue, respectively. The observer's-frame axes are marked purple, with $x_1'$ and $x_2'$ in the plane of the sky and $x_3'$ oriented along the line of sight to the observer. The perimeter of the dark-gray ellipse in the background represents a corresponding contour of constant surface density, with plane-of-the-sky axes marked brown: $x_1$ along the major, and $x_2$ along the minor axis of the ellipse.\label{fig:ellipsoid}}
\end{figure}

In order to study gravitational lensing by such an object in the thin-lens approximation, we need to integrate the density profile along the line of sight to obtain its surface density. In doing so, we have to account for the general spatial orientation of the object \citep[see also][]{oguri_etal03,van_de_Ven_etal09}. We present the necessary formulae in Appendix~\ref{sec:Appendix-transformation}, where we use Euler angles $\varphi, \vartheta, \psi$ to transform from the principal-axes frame to an observer's frame defined by plane-of-the-sky and line-of-sight coordinates.

Taking these transformations into account, the surface density can be written as
\begin{equation}
\label{eq:surface_density}
\Sigma=\int_{x_{3-}'(\hat{a}_\text{\tiny{T}})}^{x_{3+}'(\hat{a}_\text{\tiny{T}})}\, \rho(\hat{a})\,\text{d} x_{3}'\,,
\end{equation}
where the integration is carried out over the interval defined by the line-of-sight intersections with the outermost ellipsoid enclosing the mass distribution. Its semi-major axis $\hat{a}_\text{\tiny{T}}$ can be replaced by $\infty$ for mass-distribution models ranging to infinity. The intersections are given explicitly by setting $\hat{a}=\hat{a}_\text{\tiny{T}}$ in Eq.~(\ref{eq:intersections}), expressed as a function of plane-of-the-sky position, eccentricities $e_1$, $e_2$, and the Euler angles. Due to symmetry it is sufficient to evaluate the integral over half of the interval, from the point at which the line of sight is tangent to the smallest ellipsoid,
\begin{equation}
\label{eq:surface_density-2}
\Sigma=2\,\int_{[x_{3-}'(\hat{a}_\text{\tiny{T}})+x_{3+}'(\hat{a}_\text{\tiny{T}})]/2}^{x_{3+}'(\hat{a}_\text{\tiny{T}})}\, \rho(\hat{a})\,\text{d} x_{3+}'(\hat{a})\,.
\end{equation}
Next, we use Eq.~(\ref{eq:derivative}) to change the integration variable to the semi-major axis $\hat{a}$, yielding
\begin{equation}
\label{eq:surface_density-3}
\Sigma(\tilde{x}_1,\tilde{x}_2)=2\,\sqrt{\frac{(1-e_1^2)(1-e_2^2)} {\lambda_1\,\lambda_2}}\, \int_{\sqrt{\tilde{x}_1^2/\lambda_1+{\tilde{x}_2^2/\lambda_2}}}^{\hat{a}_\text{\tiny{T}}}\, \frac{\rho(\hat{a})\,\hat{a}\,\text{d}\hat{a}} {\sqrt{\hat{a}^2-\frac{\tilde{x}_1^2}{\lambda_1}-\frac{\tilde{x}_2^2}{\lambda_2}}}\,,
\end{equation}
where the parameters $\lambda_1$ and $\lambda_2$ are defined in Eq.~(\ref{eq:axis_factors}).
A closer inspection of Eq.~(\ref{eq:surface_density-3}) reveals the elliptical symmetry of the surface density which can be written as a function of the semi-major axis $\tilde{a}$ in the plane of the sky,
\begin{equation}
\label{eq:semi-major}
\tilde{a}(\tilde{x}_1,\tilde{x}_2)=\sqrt{\tilde{x}_1^2+\frac{\tilde{x}_2^2}{1-e^2}}\,.
\end{equation}
The major axes of the constant-surface-density ellipses lie along the $\tilde{x}_1$ axis, and the minor axes along the $\tilde{x}_2$ axis. These axes have the same orientation as $x_1$ and $x_2$, respectively, illustrated in Fig.~\ref{fig:ellipsoid}.

The ellipses have a fixed shape, with eccentricity $e$ given by
\begin{equation}
\label{eq:eccentricity}
e=\sqrt{1-\frac{\lambda_2}{\lambda_1}}=\sqrt{2}\,\left\{1+ \frac{2-e_2^2\,\sin^2\vartheta-(1-\cos^2\psi\,\sin^2\vartheta)\,e_1^2} {\sqrt{\left[e_2^2\,\sin^2\vartheta+(1-\cos^2\psi\,\sin^2\vartheta)\,e_1^2\,\right]^2- 4\,e_1^2\,e_2^2\,\sin^2\psi\,\sin^2\vartheta}} \right\}^{-1/2}\,.
\end{equation}
As a function of general spatial orientation, for any fully triaxial object with $e_2>e_1>0$ the eccentricity of the surface-density contours ranges from a minimum value of $0$ for Euler angles $\vartheta=\arcsin(e_1/e_2)$ and $\psi=\pm\pi/2$ (tilted orientation with the median axis in the plane of the sky) to a maximum value of $e_2$ for $\vartheta=\pi/2$ and $\psi=0$ or $\pi$ (orientation with the median axis along the line of sight). For an axially symmetric prolate object with $e_2=e_1>0$ Eq.~(\ref{eq:eccentricity}) simplifies to
\begin{equation}
\label{eq:eccentricity-prolate}
e_\text{prol} = e_1\,\left\{\frac{1-\sin^2\psi\,\sin^2\vartheta}{1-e_1^2\,\sin^2\psi\,\sin^2\vartheta}  \right\}^{1/2}\,,
\end{equation}
while for an axially symmetric oblate object with $e_2>e_1=0$ we get
\begin{equation}
\label{eq:eccentricity-oblate}
e_\text{obl} = e_2\,\sin\vartheta\,.
\end{equation}
In view of its elliptical symmetry the surface density of a general triaxial ellipsoidal mass distribution can be written as
\begin{equation}
\label{eq:surface_density-final}
\Sigma(\tilde{a})=2\,\sqrt{\frac{(1-e_1^2)(1-e_2^2)} {(1-e_1^2)\,\cos^2\vartheta+(1-e_2^2)(1-e_1^2\,\sin^2\psi)\,\sin^2\vartheta}}\, \int_{\tilde{a}/\sqrt{\lambda_1}}^{\hat{a}_\text{\tiny{T}}}\, \frac{\rho(\hat{a})\,\hat{a}\,\text{d}\hat{a}} {\sqrt{\hat{a}^2-\tilde{a}^2/\lambda_1}}\,.
\end{equation}
In the special case of an axially symmetric oblate object (with $e_1=0$ and $\lambda_1=1$) Eq.~(\ref{eq:surface_density-final}) reduces to the simpler form found in equation~(14) of \cite{bourassa_kantowski75}.

\section{Deflection angle in the Bourassa \& Kantowski formalism}
\label{sec:Bourassa}

For a general mass distribution with surface density $\Sigma(\boldsymbol w)$, the gravitational deflection angle at lens-plane position $\boldsymbol \tilde{x}$ is given by
\begin{equation}
\label{eq:deflection-general}
\boldsymbol \alpha(\boldsymbol \tilde{x})=\frac{4\,G}{c^2}\,\int\,\Sigma(\boldsymbol w)\, \frac{\boldsymbol \tilde{x} - \boldsymbol w}{|\boldsymbol \tilde{x} - \boldsymbol w|^2}\,\text{d}^2\boldsymbol w \,,
\end{equation}
where $G$ is the gravitational constant and $c$ is the speed of light \citep{schneider_etal92}. For surface densities with elliptical homoeoidal symmetry the two-dimensional integrals can be analytically reduced to one dimension, as shown for example by \cite{schramm90}.

We follow another method based on complex analysis, described by \cite{bourassa_kantowski75} and fine-tuned by \cite{bray84}. The components of the deflection angle are computed as the real and imaginary parts from the equation
\begin{equation}
\label{eq:complex-deflection}
\alpha_1(\tilde{x}_1, \tilde{x}_2)+\text{i}\,\alpha_2(\tilde{x}_1, \tilde{x}_2)=\frac{4\,G}{c^2}\,I^*(\tilde{x}_1, \tilde{x}_2)\,,
\end{equation}
where $\text{i}$ is the imaginary unit and $I^*$ is the complex conjugate of the scattering function
\begin{equation}
\label{eq:BK-scattering-function}
I(\tilde{x}_1, \tilde{x}_2)=\frac{2\,\pi\,\sqrt{1-e^2}}{\tilde{x}_1+\text{i}\,\tilde{x}_2}\, \int_{0}^{\tilde{a}(\tilde{x}_1,\tilde{x}_2)} \frac{\Sigma(a')\,a'\,\text{d}a'}{\sqrt{1-\frac{e^2}{(\tilde{x}_1+\text{i}\,\tilde{x}_2)^2}\,a'^2}}\,.
\end{equation}
Here $\tilde{x}_1$ and $\tilde{x}_2$ are coordinates along the major and minor axis of the ellipses, $e$ is their eccentricity, and the semi-major axis $\tilde{a}(\tilde{x}_1, \tilde{x}_2)$ is defined in Eq.~(\ref{eq:semi-major}). The upper limit of integration is technically valid for a surface-density distribution extending to infinity. In the case of distributions with a finite extent, it should be replaced by the limiting semi-major axis for positions $\boldsymbol x$ outside the mass distribution. We note that while it may appear tempting to transfer the first denominator into the square-root expression inside the integral, doing so would cancel natural symmetries of the result such as $I(-\boldsymbol \tilde{x})=-I(\boldsymbol \tilde{x})$ and others \citep{bray84}.

Distributions for which the integral in Eq.~(\ref{eq:BK-scattering-function}) can be solved analytically yield the scattering function $I$ in the form of a complex function. In order to obtain real formulae for the deflection-angle components, this function has to be separated into its real and imaginary parts.

\section{Lensing by an ellipsoidal NFW halo}
\label{sec:eNFW}

\subsection{Surface density and convergence}
\label{sec:kappa}

The NFW density profile \citep{navarro_etal96} describes the mass distribution in a dark-matter halo of galaxy clusters as well as galaxies. While the model in its original form describes a spherical halo, we study its generalization to triaxial ellipsoidal symmetry with a density profile
\begin{equation}
\label{eq:NFW-density}
\rho(\hat{a})=\rho_\text{s}\,\left(\frac{\hat{a}}{\hat{a}_\text{s}}\right)^{-1}\, \left(1+\frac{\hat{a}}{\hat{a}_\text{s}}\right)^{-2}\,,
\end{equation}
where $\hat{a}$ is the semi-major axis of a constant-density ellipsoid passing through a given point as defined in Eq.~(\ref{eq:semi-major-3D}), $\hat{a}_\text{s}$ is the scale semi-major axis at which $\text{d}\ln{\rho}/\text{d}\ln{\hat{a}}=-2$, and $\rho_\text{s}$ is a characteristic density such that $\rho(\hat{a}_\text{s})=\rho_\text{s}/4$.

Substituting the density profile $\rho(\hat{a})$ in Eq.~(\ref{eq:surface_density-final}) and setting the upper limit $\hat{a}_\text{\tiny{T}}\rightarrow\infty$ leads to an integral of the same form as for the spherical NFW profile, yielding the surface density
\begin{equation}
\label{eq:NFW-surface-density}
\Sigma(\tilde{a})=2\,\rho_\text{s}\,\hat{a}_\text{s}\,\sqrt{\frac{(1-e_1^2)(1-e_2^2)} {(1-e_1^2)\,\cos^2\vartheta+(1-e_2^2)(1-e_1^2\,\sin^2\psi)\,\sin^2\vartheta}}\; \left[1-\mathcal{F}\left(\frac{\tilde{a}}{\hat{a}_\text{s}\sqrt{\lambda_1}}\right)\right]\, \left(\frac{\tilde{a}^2}{\hat{a}_\text{s}^2\lambda_1}-1\right)^{-1}\, ,
\end{equation}
where the monotonically decreasing function $\mathcal{F}(\xi)$ is defined piecewise as follows:
\begin{equation}
\label{eq:NFW-F}
\mathcal{F}(\xi)=\begin{cases}
\cfrac{\displaystyle\arctanh{\sqrt{1-\xi^2}}}{\displaystyle\sqrt{1-\xi^2}} & \text{for $\xi<1$}\,,\\[9pt]
\hfil 1 & \text{for $\xi=1$}\,,\\[6pt]
\cfrac{\displaystyle\arctan{\sqrt{\xi^2-1}}}{\displaystyle\sqrt{\xi^2-1}} & \text{for $\xi>1$}\,.
\end{cases}
\end{equation}
We note that for a given triaxial mass distribution in a fixed orientation the entire expression before the square brackets is a constant; only the square brackets and the final parentheses are functions of the plane-of-the-sky position via the semi-major axis $\tilde{a}$. In addition, we note that it is advantageous to write the plane-of-the-sky separations and positions in units of the projected scale semi-major axis of the halo,
\begin{equation}
\label{eq:semi-major-scale-projected}
\tilde{a}_\text{s}=\hat{a}_\text{s}\sqrt{\lambda_1}\,,
\end{equation}
as shown in Appendix~\ref{sec:Appendix-transformation}. In these units
\begin{equation}
\label{eq:positions-dimensionless}
\{a,x_1,x_2\}=\{\tilde{a},\tilde{x}_1, \tilde{x}_2\}/\tilde{a}_\text{s}
\end{equation}
denote the semi-major axis and the coordinates along the major and minor axes of the ellipse in the plane of the sky, satisfying
\begin{equation}
\label{eq:semi-major-dimensionless}
a(x_1,x_2)=\sqrt{x_1^2+\frac{x_2^2}{1-e^2}}\,.
\end{equation}

We account for these properties in the expression for the convergence, which is obtained by dividing the surface density by the critical surface density
\begin{equation}
\label{eq:critical-density}
\Sigma_{\text{cr}}=\frac{c^2}{4\pi G}\frac{D_{\text{s}}}{D_{\text{l}}\,D_{\text{ls}}}\,,
\end{equation}
where $D_\text{s}$, $D_\text{l}$, and $D_\text{ls}$ are the angular diameter distances between the observer and the source, between the observer and the lens, and between the lens and the source, respectively. The convergence of an NFW halo with triaxial ellipsoidal symmetry can be written as
\begin{equation}
\label{eq:convergence}
\kappa(a)=\begin{cases}
2\,\kappa_{\text{s}}\;\cfrac{\displaystyle 1-\mathcal{F}(a)}{\displaystyle a^2-1} & \text{for $a\neq 1$}\,,\\[9pt]
\hfil \cfrac{2}{3}\,\kappa_\text{s} & \text{for $a=1$}\,.
\end{cases}
\end{equation}
While the functional form is exactly the same as for a spherical NFW halo \citep[e.g.,][]{bartelmann96,keeton01}, the variable is generalized from the radial position in units of halo scale radius to the semi-major axis in units of projected halo scale semi-major axis. In addition, the halo convergence parameter
\begin{equation}
\label{eq:kappa_s}
\kappa_{\text{s}}=\frac{\rho_\text{s}\,\hat{a}_\text{s}}{\Sigma_{\text{cr}}}\, \sqrt{\frac{(1-e_1^2)(1-e_2^2)} {(1-e_1^2)\,\cos^2\vartheta+(1-e_2^2)(1-e_1^2\,\sin^2\psi)\,\sin^2\vartheta}}
\end{equation}
depends not only on the characteristic density and scale semi-major axis of the halo, but also on its shape and spatial orientation through the square-root term which reduces to unity in the spherical case ($e_1=e_2=0$).

As a function of orientation, for a given triaxially ellipsoidal halo the square root reaches its maximum value of 1 when viewed down the major axis ($\vartheta=\pi/2$ and $\psi=\pi/2$ or $3\pi/2$), its minimum value of $\sqrt{1-e_2^2}$ down the minor axis ($\vartheta=0$ or $\pi$), and the intermediate value of $\sqrt{1-e_1^2}$ down the median axis ($\vartheta=\pi/2$ and $\psi=0$ or $\pi$). For a more intuitive understanding of the term, we note that the volume of an ellipsoid is proportional to the square root of the numerator, while the area of its elliptical projection is proportional to the square root of the denominator, as indicated by Eq.~(\ref{eq:factor_product}). We may thus re-write Eq.~(\ref{eq:kappa_s}) in a form that is easier to interpret,
\begin{equation}
\label{eq:kappa_s-better}
\kappa_{\text{s}}=\frac{3}{4}\,\frac{\rho_\text{s}}{\Sigma_{\text{cr}}}\, \frac{V(\hat{a}_\text{s})}{S_\text{proj}(\hat{a}_\text{s})}\,,
\end{equation}
where $V(\hat{a}_\text{s})=4\pi\hat{a}_\text{s}^3\sqrt{(1-e_1^2)(1-e_2^2)}\,/\,3$ is the volume of an ellipsoid with semi-major axis $\hat{a}_\text{s}$, and \mbox{$S_\text{proj}(\hat{a}_\text{s})=\pi\hat{a}_\text{s}^2\sqrt{\lambda_1\lambda_2}$} is the area of its plane-of-the-sky projection (see Appendix~\ref{sec:Appendix-transformation} and Fig.~\ref{fig:ellipsoid}).

\begin{figure*}
\centering
\includegraphics[width=18 cm]{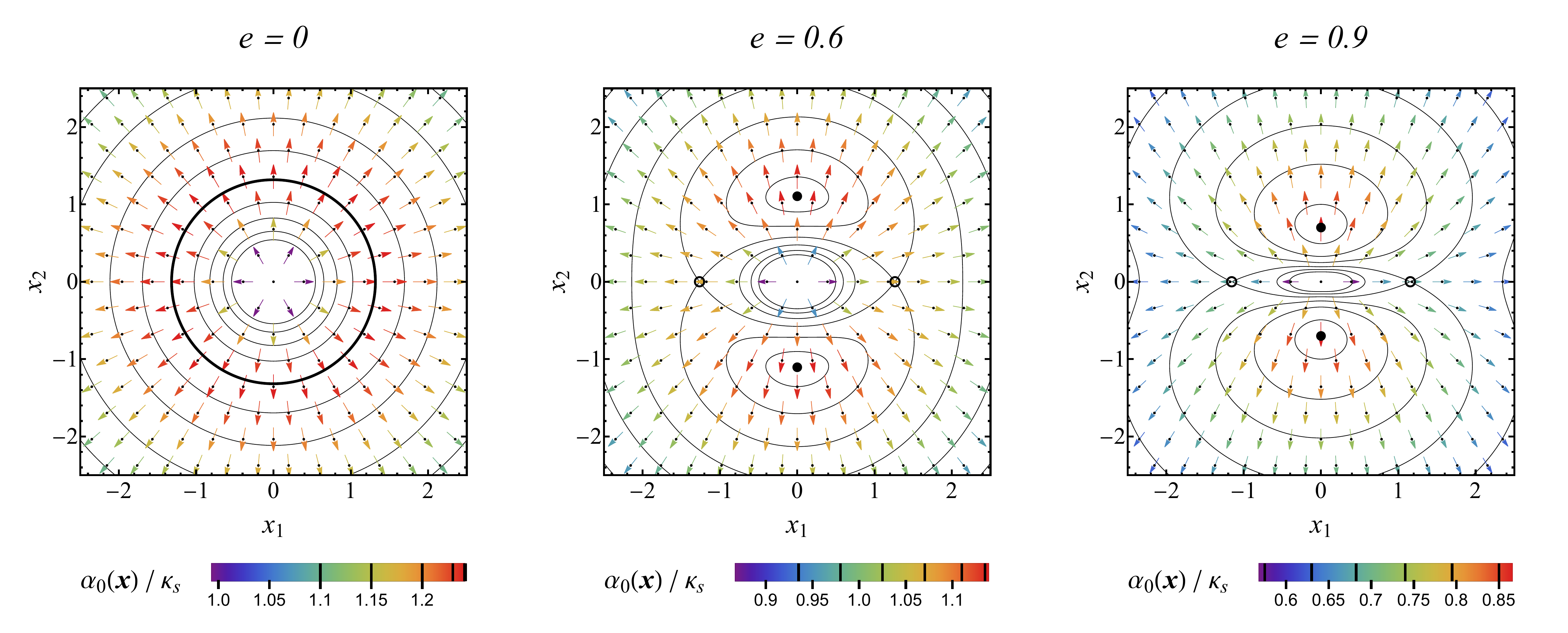}
\caption{Lens-plane plots of the deflection angle $\boldsymbol \alpha_0 (\boldsymbol x)/\kappa_\text{s}$ of the ellipsoidal NFW lens with components given by Eqs.~(\ref{eq:deflection-1}) and (\ref{eq:deflection-2}). The three panels correspond to eccentricities $e\in\{0, 0.6, 0.9\}$, as marked above the plots. The arrows mark the deflection-angle vectors at the position of their central dots. The amplitude of each vector is indicated by its color (with a separate color bar for each panel), and by constant-amplitude contours at values marked vertically across the color bars. Positions with maximum deflection amplitude are marked by the large bold black circle in the left panel, and by the two solid black dots along the minor axis in the second and third panels. The two small black circles in these panels indicate saddle points with maximum deflection along the major axis. We note that in the case of non-zero eccentricity the deflection at positions off the major and minor axes has a non-radial orientation.\label{fig:deflection-vectorplot}}
\end{figure*}

\subsection{Deflection angle}
\label{sec:alpha}

In order to obtain the deflection angle we first re-write Eq.~(\ref{eq:BK-scattering-function}) in terms of a dimensionless integration variable $\bar{a}=a'/\tilde{a}_\text{s}$ and the dimensionless quantities from Eq.~(\ref{eq:positions-dimensionless}),
\begin{equation}
\label{eq:BK-scattering-function-NFW}
I(x_1, x_2)=\frac{2\,\pi\,\tilde{a}_\text{s}\,\sqrt{1-e^2}}{x_1+\text{i}\,x_2}\, \int_{0}^{a(x_1,x_2)} \frac{\Sigma(\bar{a})\,\bar{a}\,\text{d}\bar{a}}{\sqrt{1- \frac{e^2}{(x_1+\text{i}\,x_2)^2}\,\bar{a}^2}}\,.
\end{equation}
In our case, $\Sigma(\bar{a})=\Sigma_{\text{cr}}\,\kappa(\bar{a})$ with the convergence given by Eq.~(\ref{eq:convergence}). The integral can be computed analytically, yielding the complex scattering function
\begin{multline}
\label{eq:NFW-scattering-function}
I(x_1, x_2) = \frac{4\,\pi\,\Sigma_\text{cr}\,\kappa_\text{s}\,\tilde{a}_\text{s}\,\sqrt{1-e^2}} {(x_1+\text{i}\,x_2)^2-e^2}\,\left[\sqrt{(x_1+\text{i}\,x_2)^2 - e^2\, \left(x_1^2+\frac{x_2^2}{1-e^2}\right)}\quad \mathcal{F}\left(\sqrt{x_1^2+\frac{x_2^2}{1-e^2}}\;\right) \right.\\ \left. +(x_1+\text{i}\,x_2)\,\ln{\frac{\sqrt{x_1^2+\frac{x_2^2}{1-e^2}}} {1+\sqrt{1-\frac{e^2}{(x_1+\text{i}\,x_2)^2}\left(x_1^2+\frac{x_2^2}{1-e^2}\right)}}}\, \right]\,.
\end{multline}
This result can be substituted on the r.h.s. of Eq.~(\ref{eq:complex-deflection}) to obtain the components of the deflection angle. Checking the result in the limiting $e=0$ case, we get
\begin{equation}
\label{eq:complex-deflection-spherical}
\alpha_1(x_1, x_2)+\text{i}\,\alpha_2(x_1,x_2) =\frac{4\,D_{\text{s}}\,\tilde{a}_\text{s}\,\kappa_\text{s}} {D_{\text{l}}\,D_{\text{ls}}}\, \frac{x_1+\text{i}\,x_2}{x_1^2+x_2^2}\, \left[ \mathcal{F}\left(\sqrt{x_1^2+x_2^2}\;\right)+\ln{\frac{\sqrt{x_1^2+x_2^2}}{2}}\,\right]\,,
\end{equation}
the expression for the deflection angle of a spherical NFW halo \citep[e.g.,][]{karamazov_etal21}. For simplicity, we present the remaining results in terms of the reduced deflection angle expressed in units of the angular projected scale semi-major axis of the halo,
\begin{equation}
\label{eq:reduced-deflection-angle}
\boldsymbol \alpha_0(\boldsymbol x) = \frac{D_{\text{ls}}\,D_{\text{l}}}{D_{\text{s}}\,\tilde{a}_\text{s}}\, \boldsymbol\alpha(\boldsymbol x)\,.
\end{equation}
For an arbitrary projected eccentricity $e$, Eq.~(\ref{eq:NFW-scattering-function}) also directly yields simple formulae for the deflection angle along the major axis
\begin{equation}
\label{eq:deflection-major-axis}
\boldsymbol \alpha_0(x_1, 0)= \left( 4\,\kappa_\text{s}\, \frac{x_1}{x_1^2-e^2}\,\left\{(1-e^2)\, \mathcal{F}(x_1)+\sqrt{1-e^2}\,\ln{\frac{|x_1|}{1+\sqrt{1-e^2}}}\, \right\},\,0\right)\,,
\end{equation}
and along the minor axis
\begin{equation}
\label{eq:deflection-minor-axis}
\boldsymbol \alpha_0(0, x_2)= \left( 0,\,4\,\kappa_\text{s}\, \frac{x_2}{x_2^2+e^2}\,\left\{\mathcal{F}\left(\frac{x_2}{\sqrt{1-e^2}}\right) +\sqrt{1-e^2}\,\ln{\frac{|x_2|}{1+\sqrt{1-e^2}}}\, \right\}\right)\,.
\end{equation}

In order to obtain real expressions for the deflection angle at an arbitrary lens-plane position $\boldsymbol x$, Eq.~(\ref{eq:NFW-scattering-function}) has to be decomposed into its real and imaginary parts. This procedure leads to the following general results for the deflection-angle component parallel to the major axis:
\begin{multline}
\label{eq:deflection-1}
\alpha_{01}(\boldsymbol x)=\frac{4\,\kappa_\text{s}\,\sqrt{1-e^2}} {\left[(x_1-e)^2+x_2^2\right]\left[(x_1+e)^2+x_2^2\right]}\,\left\{ x_1\left[(x_1^2-e^2)(1-e^2)+x_2^2(1+e^2)\right]\,f_1(x_1,x_2)\qquad\qquad\qquad \right. \\ \left. +x_1 (x_1^2+x_2^2-e^2)\,f_2(x_1,x_2) - x_2 (x_1^2+x_2^2+e^2)\,f_3(x_1,x_2)\right\}\,,
\end{multline}
and for the deflection-angle component parallel to the minor axis:
\begin{multline}
\label{eq:deflection-2}
\alpha_{02}(\boldsymbol x)=\frac{4\,\kappa_\text{s}\,\sqrt{1-e^2}} {\left[(x_1-e)^2+x_2^2\right]\left[(x_1+e)^2+x_2^2\right]}\,\left\{ x_2\left[x_1^2(1-2\,e^2)+x_2^2+e^2\right]\,f_1(x_1,x_2)\qquad\qquad\qquad\qquad\qquad \right. \\ \left. + x_2 (x_1^2+x_2^2+e^2)\,f_2(x_1,x_2) + x_1 (x_1^2+x_2^2-e^2)\,f_3(x_1,x_2) \right\}\,.
\end{multline}
In both expressions we introduced the auxiliary functions
\begin{eqnarray}
\label{eq:deflection-functions}
\nonumber f_1(x_1,x_2)&=&(1-e^2)^{-1/2}\,\mathcal{F}\left(\sqrt{x_1^2+x_2^2\,/(1-e^2)}\;\right)\\
f_2(x_1,x_2)&=&\ln{\frac{\sqrt{x_1^2+x_2^2}}{1+\sqrt{1-e^2}}}\\
\nonumber f_3(x_1,x_2)&=&\arctan{\frac{x_1 x_2(1-\sqrt{1-e^2})}{x_1^2\sqrt{1-e^2}+x_2^2}}\,.
\end{eqnarray}

\begin{figure*}
\centering
\includegraphics[width=18 cm]{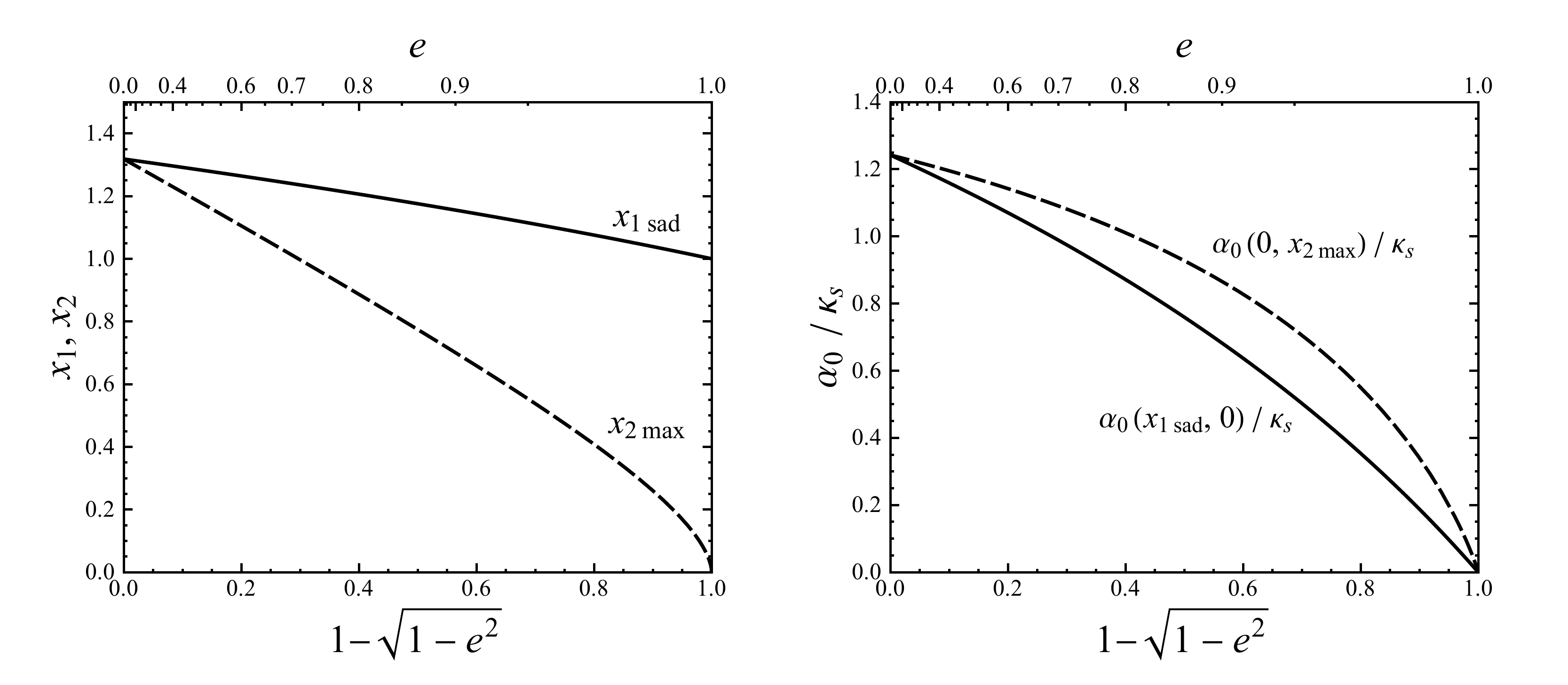}
\caption{Scaling of the deflection-angle maps with eccentricity. Left panel: Position of deflection-angle maximum $x_{2 \text{max}}$ (dashed line) and position of saddle point $x_{1 \text{sad}}$ (solid line) as a function of eccentricity $e$ (non-linear top axis) and ellipticity $1-\sqrt{1-e^2}$ (linear bottom axis). Right panel: Deflection-angle amplitudes $|\boldsymbol \alpha_0|$ at these positions in units of convergence parameter $\kappa_\text{s}$ as a function of eccentricity and ellipticity. \label{fig:max-sad-deflection}}
\end{figure*}

The formulae for both components have the same prefactor before the curly braces. Its denominator is equal to the product of the squared distances from the two foci of the $a=1$ ellipse. Nevertheless, the deflection angle is not divergent at these points; it is continuous and smooth there due to the vanishing expressions in the curly braces. This can be seen by expanding Eq.~(\ref{eq:deflection-major-axis}) near $x_1=\pm e$, as shown also in Appendix~\ref{sec:Appendix-foci}.

Both expressions in the curly braces are formed by combinations of the same three transcendental functions defined in Eq.~(\ref{eq:deflection-functions}), each weighted by a polynomial in terms of $x_i$. The formulae for the two components thus differ only in these polynomials. We note also that $x_1$ cannot be factored from $\alpha_{01}$, just as $x_2$ cannot be factored from $\alpha_{02}$ --- in both cases due to the $\arctan$ term. This implies that the general orientation of $\boldsymbol \alpha$ is not radial. An analysis of Eqs.~(\ref{eq:deflection-1}) and (\ref{eq:deflection-2}) shows that the deflection is oriented radially only along the major and minor axes.

The pattern of the deflection angle $\boldsymbol \alpha_0(\boldsymbol x)$ as a function of lens-plane position is shown in Fig.~\ref{fig:deflection-vectorplot} for eccentricities $e=0, 0.6, 0.9$. The arrows indicate the orientation of the deflection angle at their centers (marked by dots); their color and the contours indicate its amplitude. The color bars with values in units of the convergence parameter $\kappa_\text{s}$ are set for each panel separately. The values corresponding to the contours are marked by black lines in the bars.

For all values of the eccentricity $e$, the deflection amplitude is zero at the center, from where it increases outward and drops back to zero asymptotically. In the first panel of Fig.~\ref{fig:deflection-vectorplot} illustrating the $e=0$ case, the ring corresponding to maximum deflection is marked by the bold contour. For non-zero eccentricity the symmetry is broken and the global maximum of deflection occurs along the minor (here vertical) axis, at points marked by black dots in the two right panels. The maximum value along the major axis
occurs at saddle points which are marked in the two right panels by small black circles.

The change of the deflection plots with eccentricity is further illustrated by Fig.~\ref{fig:max-sad-deflection}, with the curves in both panels plotted as a function of the ellipticity (flattening) $1-\sqrt{1-e^2}$ with the eccentricity $e$ marked along the non-linear top axis. The solid curve in the left panel shows the position $x_{2 \text{max}}$ of the global maximum along the minor axis: it starts at $x_{2 \text{max}}\approx1.32$ for $e=0$ and decreases with eccentricity, reaching $x_{2 \text{max}}=0$ at $e=1$. The dashed curve in the same panel shows the position $x_{1 \text{sad}}$ of the saddle point along the major axis: it starts at the same distance of $x_{1 \text{sad}}\approx1.32$ for $e=0$ and decreases with eccentricity. However, unlike in the case of the maximum, at $e=1$ it reaches $x_{1 \text{sad}}=1$, the point along the major axis at a distance of the projected scale semi-major axis from the halo center.

The right panel of Fig.~\ref{fig:max-sad-deflection} shows the deflection-angle amplitude at the global maximum (solid curve) and at the saddle point (dashed curve). Both curves start at $\alpha_0\approx 1.24\,\kappa_\text{s}$ for $e=0$ and decrease to $\alpha_0 = 0$ for $e=1$, with the decline particularly pronounced at higher projected eccentricities.

From the perspective of the three-dimensional halo, it can be shown from Eq.~(\ref{eq:eccentricity}) that the extreme value $e=1$ can be achieved only for a halo with $e_2=1$ and either $\vartheta=\pi/2$ (flat halo viewed edge-on) or $e_1=1$ (extremely prolate or filament-like halo). Except in the case of an exactly edge-on orientation, for a halo with $e_2\rightarrow1$ the convergence parameter given by Eq.~(\ref{eq:kappa_s}) drops to zero. The decline of the deflection-angle amplitude at high projected eccentricities is thus even more pronounced in absolute terms.

\subsection{Shear and phase}
\label{sec:shear}

The formulae for the deflection angle can be used to compute the two components $\gamma_1$, $\gamma_2$ of the shear $\gamma$. These are given in terms of the lens-plane-position derivatives of the deflection angle as follows:
\begin{equation}
\label{eq:gamma-1-def}
\gamma_1\equiv\gamma\cos{2\phi}=\frac{1}{2}\left(\frac{\partial\,\alpha_{01}}{\partial\,x_1}- \frac{\partial\,\alpha_{02}}{\partial\,x_2}\right)
\end{equation}
and
\begin{equation}
\label{eq:gamma-2-def}
\gamma_2\equiv\gamma\sin{2\phi}=\frac{\partial\,\alpha_{01}}{\partial\,x_2}= \frac{\partial\,\alpha_{02}}{\partial\,x_1}\,,
\end{equation}
where $\phi$ is the lens phase. Using the deflection angle components from Eqs.~(\ref{eq:deflection-1}) and (\ref{eq:deflection-2}) yields the expressions
\begin{multline}
\label{eq:gamma-1}
\gamma_1(\boldsymbol x)=\frac{4\,\kappa_\text{s}\,\sqrt{1-e^2}} {\left[(x_1-e)^2+x_2^2\right]^2\left[(x_1+e)^2+x_2^2\right]^2}\,\left\{ \left[(x_1-e)^2+x_2^2\right]\left[(x_1+e)^2+x_2^2\right](x_1^2-x_2^2-e^2)\,f_0(x_1,x_2)
\right. \\[9pt]
\left.+\left[2\,e^2(x_1^2-1)\left[(x_1^2-x_2^2-e^2)^2-4\,x_1^2 x_2^2\right]-(x_1^2-x_2^2-e^2)^3(3+e^2)/2+ 6\,x_1^2x_2^2(e^2-1)(x_1^2- x_2^2-e^2)\right]\,f_1(x_1,x_2)\,\right. \\[9pt]
\left. -\left[(x_1^2-x_2^2-e^2)\left([x_1^2+x_2^2]^2-e^4\right)-8\,e^2 x_1^2 x_2^2\right]\,f_2(x_1,x_2) +2\,x_1x_2\left[(x_1^2+x_2^2+e^2)^2-4\,e^2(x_2^2+e^2)\right]\,f_3(x_1,x_2) \right\}
\end{multline}
and
\begin{multline}
\label{eq:gamma-2}
\gamma_2(\boldsymbol x)=\frac{4\,\kappa_\text{s}\,\sqrt{1-e^2}} {\left[(x_1-e)^2+x_2^2\right]^2\left[(x_1+e)^2+x_2^2\right]^2}\,\left\{ 2\,x_1x_2\,\left[(x_1-e)^2+x_2^2\right]\left[(x_1+e)^2+x_2^2\right]\,f_0(x_1,x_2)
\right. \\[9pt]
\left.+x_1x_2\,\left([x_1^2+x_2^2+e^2]\left[(5\,e^2-3)\,x_1^2-3\,(1+e^2)\,x_2^2 +(5-3\,e^2)\,e^2\right]-4\,e^2x_1^2[1+e^2]\right)\,f_1(x_1,x_2)\right. \\[9pt]
\left. -2\,x_1x_2\,\left[(x_1^2+x_2^2+e^2)^2-4\,e^2(x_2^2+e^2)\right]\,f_2(x_1,x_2) -\left[(x_1^2-x_2^2-e^2)\left([x_1^2+x_2^2]^2-e^4\right)-8\,e^2x_1^2x_2^2\right]\,f_3(x_1,x_2) \right\}\,,
\end{multline}
where the auxiliary functions $f_1, f_2, f_3$ are defined in Eq.~(\ref{eq:deflection-functions}) and we newly introduce
\begin{equation}
\label{eq:shear-f0}
f_0(x_1,x_2)=1+\cfrac{1}{2\sqrt{1-e^2}}\, \cfrac{x_1^2+x_2^2+e^2-2+(1-e^2x_1^2)\,\mathcal{F}\left(\sqrt{x_1^2+x_2^2\,/(1-e^2)}\;\right)} {1-x_1^2-x_2^2/(1-e^2)}\,.
\end{equation}
The expressions in Eqs.~(\ref{eq:gamma-1}) and (\ref{eq:gamma-2}) can be combined to yield the shear
\begin{equation}
\label{eq:shear}
\gamma(\boldsymbol x)=\sqrt{\gamma_1^2(\boldsymbol x)+\gamma_2^2(\boldsymbol x)}
\end{equation}
and the phase $\phi(\boldsymbol x)$ can then be computed from Eqs.~(\ref{eq:gamma-1-def}) and (\ref{eq:gamma-2-def}).

\begin{figure*}
\centering
\includegraphics[width=17cm]{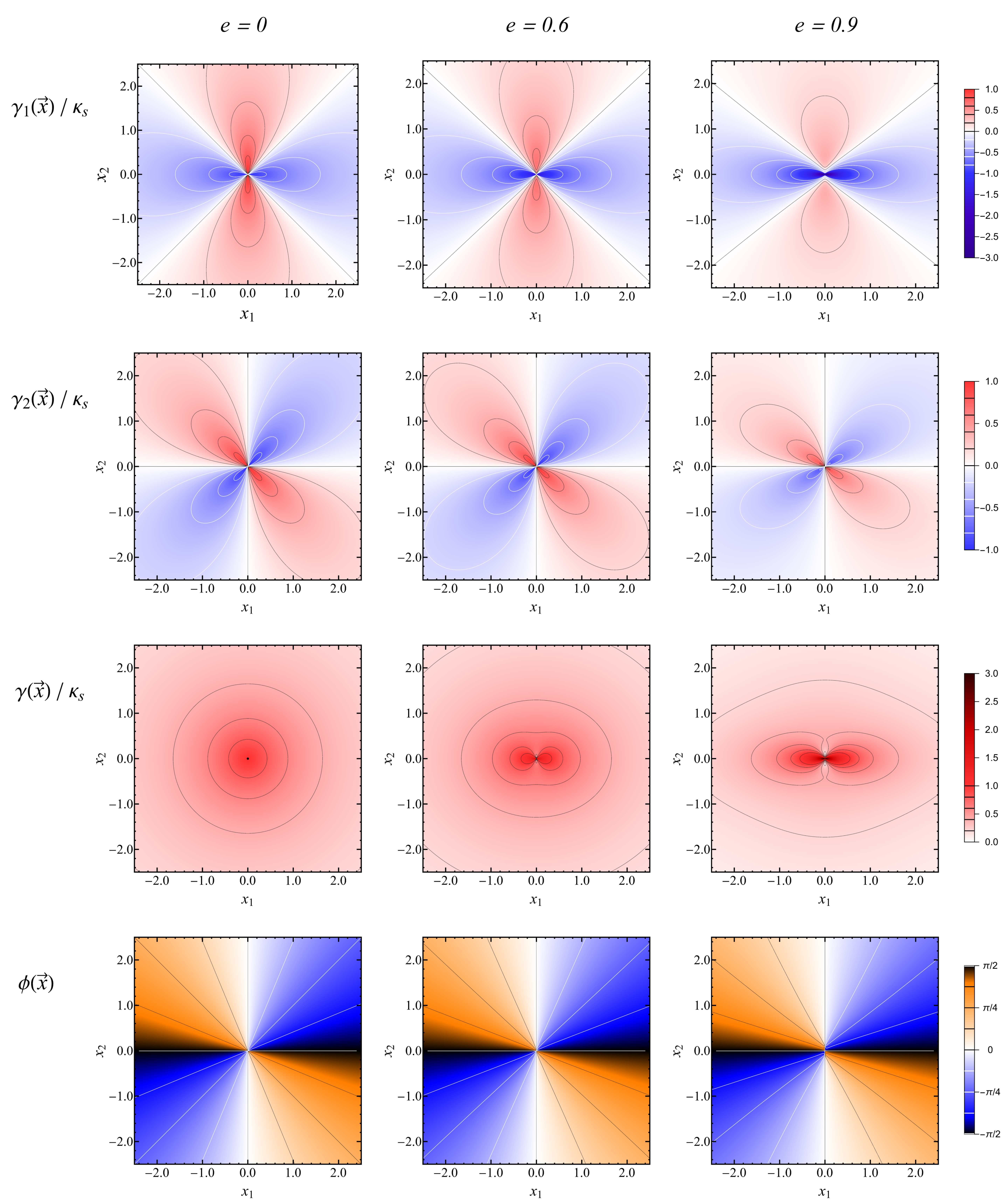}
\caption{Large-scale lens-plane maps of the shear component $\gamma_1$ (top row), shear component $\gamma_2$ (second row), shear $\gamma$ (third row), and phase $\phi$ (bottom row) of the ellipsoidal NFW lens for eccentricities $e\in\{0, 0.6, 0.9\}$, as marked above the three columns. Colors and contour values for each row are marked in the color bars on the right side; shear and its components are plotted in units of $\kappa_\text{s}$.\label{fig:shear}}
\end{figure*}

\subsubsection{Large-scale properties}
\label{sec:shear-large-scale}

Lens-plane maps of the shear components, the shear, and the phase are presented in Fig.~\ref{fig:shear} for projected eccentricities $e=0$, 0.6, and 0.9. In the zero-eccentricity case shown in the left column, the shear component $\gamma_1$ (top row) is zero along both diagonals, and it reaches its azimuthal positive maxima and negative minima along the vertical and horizontal axes, respectively. The component $\gamma_2$ (second row) has the same pattern rotated counter-clockwise by $\pi/4$: it is zero along both axes, it reaches its azimuthal extremes along the diagonals, and it alternates in the plot quadrants between negative (first and third) and positive values (second and fourth). The shear (third row) has a purely radial dependence, reaching its maximum value $\gamma=\kappa_\text{s}$ at the halo center and decreasing outward asymptotically to zero \citep[e.g.,][]{karamazov_heyrovsky22}. The phase $\phi$ (bottom row) has a purely azimuthal dependence: it is equal to the azimuthal angle measured from the horizontal axis shifted by $-\pi/2$ and limited here to the interval $(-\pi/2, \pi/2]$. The phase is thus zero along the vertical axis, $\pi/2$ along the horizontal axis, and it alternates in the plot quadrants between negative and positive values in the same manner as $\gamma_2$.

\begin{figure*}
\centering
\includegraphics[width=17cm]{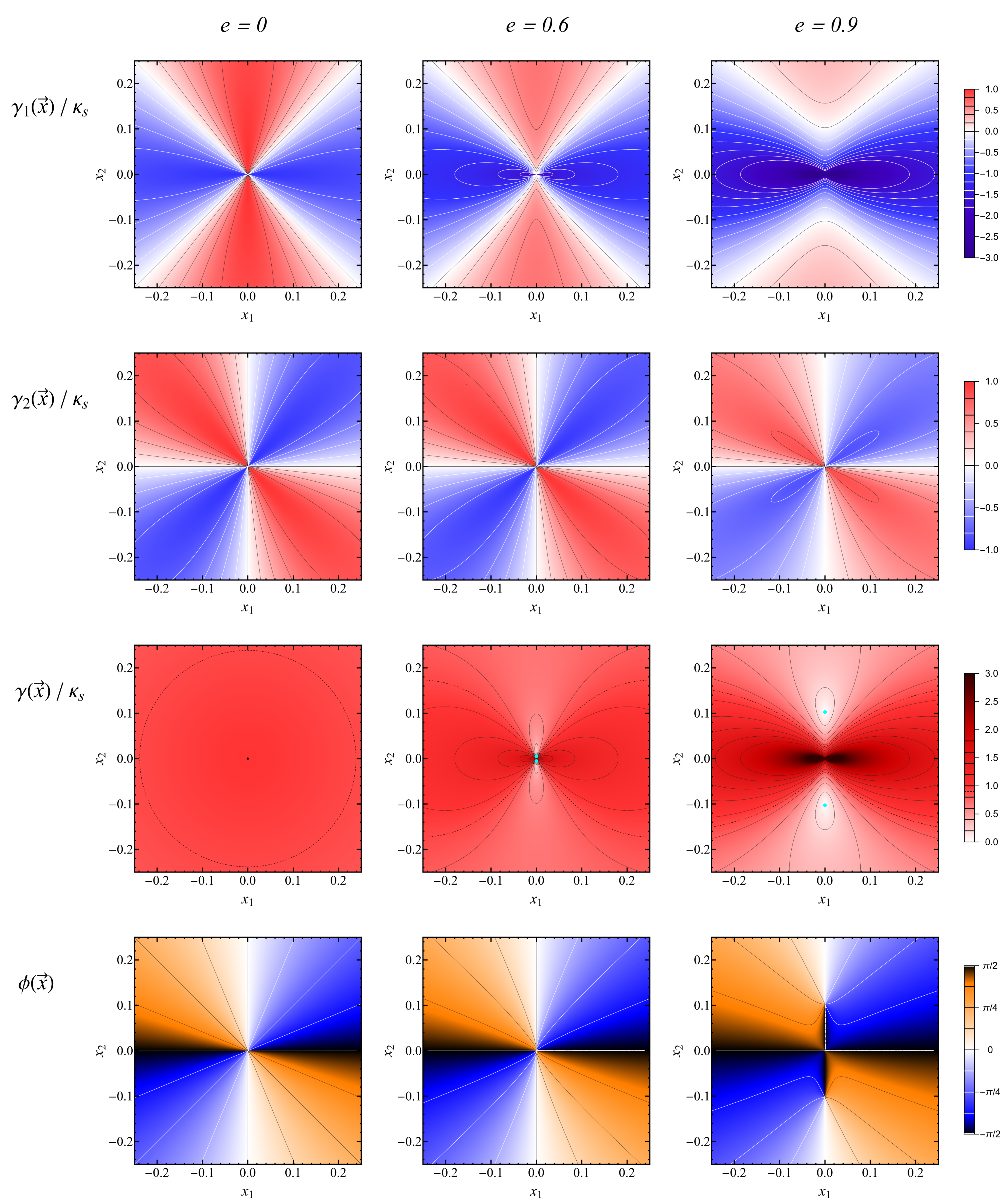}
\caption{Halo-center details of the lens-plane maps from Fig.~\ref{fig:shear}, with the same color bars and additional shear contours in the first and third rows. In the maps of the shear on the third row: the central dot in the first panel corresponds to $\gamma=\kappa_\text{s}$; due to the very low shear variation in the first panel the dotted $\gamma=0.9\,\kappa_\text{s}$ contour is added in all three panels; the cyan dots in the second and third panels mark the positions of the zero-shear points. Remaining notation as in Fig.~\ref{fig:shear}.\label{fig:shear-zoom}}
\end{figure*}

For non-zero eccentricities shown in the two right columns, the character of the plots changes significantly. Least so in the case of $\gamma_2$ (second row): the azimuthal extremes occur along lines closer to the major axis, and the absolute values in units of $\kappa_\text{s}$ decrease, as indicated by the shrinking contours. In the case of $\gamma_1$ (top row), the absolute values decrease in the positive areas along the minor axis, but they increase in the negative areas along the major axis. However, while the inner negative contours of $\gamma_1$ expand, the outer ones shrink. We note also that while the zero contours of $\gamma_2$ lie always along the coordinate axes irrespective of the eccentricity, the zero-contours of $\gamma_1$ do not reach the halo center when $e>0$. The pattern of the shear $\gamma$ (third row) changes dramatically: the outer oval contours give way to the central dipole-like pattern dominated by the character of $\gamma_1$. Along the minor axis just above and below the center there are localized low-shear regions. The phase (bottom row) preserves the large-scale pattern in the quadrants, but the contours are tilted toward the horizontal axis. More specifically, the $\gamma_2=0$ contours correspond to $\phi=0$ and $\phi=\pi/2$ contours in all the phase plots in Fig.~\ref{fig:shear} (as well as in Fig.~\ref{fig:shear-zoom}), while the $\gamma_1=0$ contours correspond to $|\phi|=\pi/4$ contours. It can be seen in the bottom right panel of Fig.~\ref{fig:shear} that non-zero eccentricities lead to deviations from the large-scale pattern in the vicinity of the halo center, which are explored in the following Sect.~\ref{sec:shear-halo-center}.

\subsubsection{Halo-center features}
\label{sec:shear-halo-center}

In order to understand the central patterns, we present in Fig.~\ref{fig:shear-zoom} the same set of plots zoomed in ten times, and interpret them using expansions of the plotted quantities in the vicinity of the halo center. In the zero-eccentricity case, the expansions of the shear and its components to leading order for $x\ll 1$ yield
\begin{equation}
\label{eq:gamma-center-sym}
\gamma(\boldsymbol x)=\kappa_\text{s}+ \mathcal{O}(x^2\,\ln{x})\,,
\end{equation}
\begin{equation}
\label{eq:gamma-1-center-sym}
\gamma_1(\boldsymbol x)=\kappa_\text{s}\,\frac{x_2^2-x_1^2}{x_1^2+x_2^2}+ \mathcal{O}(x^2\,\ln{x})\,,
\end{equation}
and
\begin{equation}
\label{eq:gamma-2-center-sym}
\gamma_2(\boldsymbol x)=-\,\kappa_\text{s}\,\frac{2\,x_1\,x_2}{x_1^2+x_2^2}+ \mathcal{O}(x^2\,\ln{x})\,,
\end{equation}
while the phase at any distance $x$ from the center fulfills
\begin{equation}
\label{eq:phase-center-sym}
\left(\,\cos{2\phi(\boldsymbol x)},\,\sin{2\phi(\boldsymbol x)}\,\right) = \left(\frac{x_2^2-x_1^2}{x_1^2+x_2^2},\,-\,\frac{2\,x_1\,x_2}{x_1^2+x_2^2}\right)\,.
\end{equation}
The expression for the shear explains the featureless bottom left plot showing the slow radial decrease of the shear. The expressions for the shear components explain their alternating sign around the center. In addition, they demonstrate that both components are undefined directly at the halo center: approaching the center from different directions leads to all values in the interval $[-\kappa_\text{s},\kappa_\text{s}]$. Nevertheless, their combination given by Eq.~(\ref{eq:shear}) that yields the shear is well defined and analytic at the halo center. The phase given by Eq.~(\ref{eq:phase-center-sym}) is always oriented perpendicular to the position vector $\boldsymbol x$ \citep[see][]{karamazov_heyrovsky22}; at the center it is undefined.

In the case of non-zero eccentricity, the central expansions require the analogous condition $a\ll 1$ plus $|x_1|,|x_2|\ll e$. Hence, the following results are valid well within the unit ellipse and at the same time well within the circle passing through its foci. In this region the shear and its components can be approximated by
\begin{multline}
\label{eq:gamma-center}
\gamma(\boldsymbol x)=\frac{2\,\kappa_\text{s}}{e^2}\left\{\left[ \left(1-\sqrt{1-e^2}\right)^2\, \left(\ln{\frac{\sqrt{x_1^2+x_2^2/(1-e^2)}}{2}}+1\right)+2\,\sqrt{1-e^2}\, \ln{\left(\frac{1+\sqrt{1-e^2}}{2}\,\sqrt{1+\frac{e^2}{1-e^2}\,\frac{x_2^2}{x_1^2+x_2^2}}\right)} \;\right]^2 \right. \\ \left. +4\,(1-e^2)\,\arctan^2{\frac{x_1 x_2(1-\sqrt{1-e^2})}{x_1^2\sqrt{1-e^2}+x_2^2}}\right\}^{1/2} +\mathcal{O}(x^2\,\ln{x})\,,
\end{multline}
\begin{equation}
\label{eq:gamma-1-center}
\gamma_1(\boldsymbol x)=\frac{2\,\kappa_\text{s}}{e^2}\left[ \left(1-\sqrt{1-e^2}\right)^2\,\left(\ln{\frac{\sqrt{x_1^2+x_2^2/(1-e^2)}}{2}}+1\right) +2\,\sqrt{1-e^2}\, \ln{\left(\frac{1+\sqrt{1-e^2}}{2}\,\sqrt{1+\frac{e^2}{1-e^2}\,\frac{x_2^2}{x_1^2+x_2^2}}\right)} \;\right]+\mathcal{O}(x^2\,\ln{x})\,,
\end{equation}
and
\begin{equation}
\label{eq:gamma-2-center}
\gamma_2(\boldsymbol x)=-4\,\kappa_\text{s}\,\frac{\sqrt{1-e^2}}{e^2}\, \arctan{\frac{x_1 x_2(1-\sqrt{1-e^2})}{x_1^2\sqrt{1-e^2}+x_2^2}}+\mathcal{O}(x^2\,\ln{x})\,,
\end{equation}
where the expression ``$x^2\,\ln{x}$" in the next higher order stands for both $x_i^2\,\ln{a}$ and $x_i^2\,\ln{x}$ terms. The phase has to be computed from these approximations using Eqs.~(\ref{eq:gamma-1-def}) and (\ref{eq:gamma-2-def}).

The central pattern of $\gamma_2$ (second row in Fig.~\ref{fig:shear-zoom}) reveals the same character as in the case of zero eccentricity, illustrating the two systematic changes with increasing $e$ noted above. First, the axes of maximum negative and positive values tilt from the diagonals toward the major axis. The position angle of the negative axis in the first quadrant is given by $\varphi_\text{neg}=\arctan{(1-e^2)^{1/4}}$, which can be shown from Eq.~(\ref{eq:gamma-2-center}). Second, the amplitude of $\gamma_2$ at the center when approached from different directions decreases from $\kappa_\text{s}$: at first slowly, then dropping to zero for $e\to 1$. The component $\gamma_2$ is thus undefined at the halo center, just as in the case of zero eccentricity.

The central pattern of $\gamma_1$ (top row) reveals that for non-zero eccentricity the zero contours do not follow the diagonals to the center, as indicated above. Instead, they cross the minor axis at points that shift away from the center as the eccentricity increases. Directly at the center, $\gamma_1\to -\infty$ from all directions. Equation~(\ref{eq:gamma-1-center}) shows that $\gamma_1\propto \ln{a}$ as $a\to 0$, so that the extreme negative region is more extended along the major axis. The divergent term vanishes for $e\to 0$, so that the next term in Eq.~(\ref{eq:gamma-1-center}) determines the overall pattern for low eccentricities. Nevertheless, despite its divergence, $\gamma_1$ is defined at the halo center in the case of non-zero eccentricity.

The central pattern of $\gamma$ (third row) reveals the $\gamma\propto -\ln{a}$ divergence driven by $\gamma_1$, extending primarily along the major axis. However, when progressing outward along the minor axis the shear drops to zero, then increases and, as seen in Fig.~\ref{fig:shear}, decreases asymptotically to zero. The immediate vicinity of the halo center thus exhibits a dramatic variation of the shear, which ranges there from $0$ to $\infty$ for any non-zero eccentricity. This behavior is in stark contrast to the near-constant shear of a halo with zero projected eccentricity (left column).

In the bottom row, we note that the phase $\phi$ switches along the minor axis from $0$ to $\pi/2$ between the zero-shear points. In the immediate vicinity of the center the phase is thus vertical along both axes and near-vertical off the axes. All contours meet at the zero-shear points, with $\phi=0$ pointing outward along the minor axis, $\phi=\pi/2$ inward along the minor axis and all other values along contours pointing sideways under different angles. Contours pointing in opposite directions have perpendicular phases ($\pm\pi/2$ difference). We conclude that in the case of non-zero eccentricity the phase is defined at the halo center, but it is undefined at the zero-shear points.

The phase pattern in combination with the position relative to the unit-convergence contour determines the orientation of image distortions. The interplay in the case of the studied ellipsoidal NFW model is demonstrated further in Sect.~\ref{sec:images}.

\begin{figure}
\centering
\resizebox{\hsize/2}{!}{\includegraphics{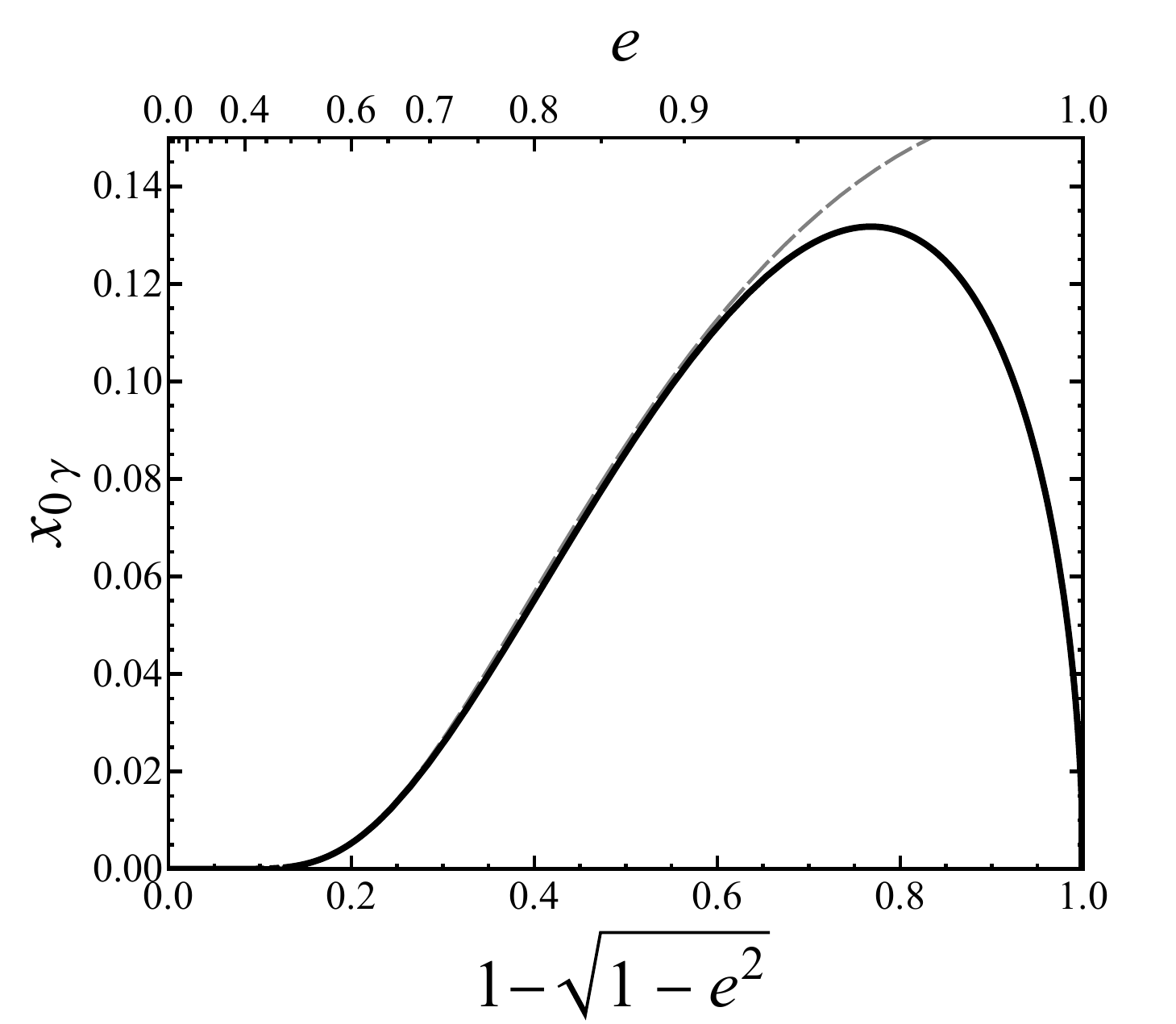}}
\caption{Position of the zero-shear point $x_{0\,\displaystyle{\gamma}}$ along the minor axis as a function of ellipticity $1-\sqrt{1-e^2}$ (linear bottom axis) and eccentricity $e$ (non-linear top axis). Solid black curve: exact value obtained by numerically solving Eq.~(\ref{eq:zero-shear-points}); dashed gray curve: low-eccentricity approximation given by Eq.~(\ref{eq:zero-shear-point-approx}).\label{fig:zero-shear-point}}
\end{figure}

\subsubsection{Zero-shear points}
\label{sec:zero-shear-points}

As shown in \cite{karamazov_heyrovsky22}, zero-shear points mark the lens-plane positions of undistorted positive-parity images, while undistorted negative-parity images occur along the unit-convergence contour. In the special case when the zero-shear points lie on the unit-convergence contour, they form umbilic points belonging to the critical curve (see Sect.~\ref{sec:curves}). In the case of our lens model the zero-shear points occur for any non-zero eccentricity at the intersections of the $\gamma_1=0$ and $\gamma_2=0$ contours. The pattern of these contours in Figs.~\ref{fig:shear} and \ref{fig:shear-zoom} shows that the zero-shear points are located at $\boldsymbol x_{0\,\displaystyle{\gamma}}=(0, \pm\,x_{0\,\displaystyle{\gamma}})$, where $x_{0\,\displaystyle{\gamma}}$ is the positive root of $\gamma_1(0, x_{0\,\displaystyle{\gamma}})=0$. Using Eq.~(\ref{eq:gamma-1}), this condition can be simplified to
\begin{equation}
\label{eq:zero-shear-points}
-2\,(x_{0\,\displaystyle{\gamma}}^2+e^2)\,f_0(0,x_{0\,\displaystyle{\gamma}}) +\left[x_{0\,\displaystyle{\gamma}}^2(3+e^2)-e^2(1-e^2)\right]\,f_1(0,x_{0\,\displaystyle{\gamma}})
+2\,(x_{0\,\displaystyle{\gamma}}^2-e^2)\,f_2(0,x_{0\,\displaystyle{\gamma}})=0\,,
\end{equation}
which can be solved numerically, taking into account the definitions of $f_1$ and $f_2$ in Eq.~(\ref{eq:deflection-functions}), and $f_0$ in Eq.~(\ref{eq:shear-f0}). Zero-shear points identified in this way are marked by the cyan dots in the $e=0.6$ and $e=0.9$ columns of the third row of Fig.~\ref{fig:shear-zoom}.

The position $x_{0\,\displaystyle{\gamma}}$ of the zero-shear points on the minor axis is shown in Fig.~\ref{fig:zero-shear-point}. For low eccentricities (marked along the top axis) the zero-shear points lie exponentially close to the halo center. They start to deviate from the center noticeably for $e\approx0.5$, reaching a peak separation of $x_{0\,\displaystyle{\gamma}}\approx0.132$ at $e\approx0.973$ and dropping back to zero for $e\to 1$. Resolving any effects associated with zero-shear points thus requires halos with projected eccentricities $e>0.5$.

For a more analytical insight into the low-eccentricity behavior, the solution of the implicit Eq.~(\ref{eq:zero-shear-points}) can be approximated by the expression
\begin{equation}
\label{eq:zero-shear-point-approx}
x_{0\,\displaystyle{\gamma}}\simeq 2\,\exp{\left[\,-2\,e^{-2}-\frac{1}{4}-\frac{7}{24}\,e^2\,\right]}\,,
\end{equation}
which is accurate to $1\%$ for $e\leq0.5$ and to $5\%$ for $e\leq 0.96$. This expression was obtained by expanding Eq.~(\ref{eq:zero-shear-points}) for $x\ll 1$, setting the combination of the two highest-order terms equal to zero, and expanding this equation in terms of $e\ll 1$. The approximation is marked by the dashed gray curve in Fig.~\ref{fig:zero-shear-point}. It traces the exact solution excellently, deviating just before the high-eccentricity peak.

\subsection{Critical curves and caustics}
\label{sec:curves}

In the present notation it is convenient to write the angular position of the source $\boldsymbol \beta$ in units of the angular projected scale semi-major axis of the halo, i.e.,
\begin{equation}
\label{eq:source-position}
\boldsymbol y = \frac{D_{\text{l}}}{\tilde{a}_\text{s}}\,\boldsymbol \beta\,.
\end{equation}
Utilizing the expression for the reduced deflection angle in the same units from Eq.~(\ref{eq:reduced-deflection-angle}), we can write the lens equation in the simple form
\begin{equation}
\label{eq:lens-equation}
\boldsymbol y = \boldsymbol x - \boldsymbol \alpha_0(\boldsymbol x)\,.
\end{equation}
In order to compute the lens-equation Jacobian we combine the expressions for the convergence from Eq.~(\ref{eq:convergence}) and for the shear components from Eqs.~(\ref{eq:gamma-1}) and (\ref{eq:gamma-2}) in the general formula
\begin{equation}
\label{eq:Jacobian}
\mathrm{det}\,J(\boldsymbol x)=\frac{\partial\,\boldsymbol y}{\partial\,\boldsymbol x}=\left[\,1-\kappa\left(a[\boldsymbol x]\right)\,\right]^2-\gamma_1^2(\boldsymbol x)-\gamma_2^2(\boldsymbol x)\,.
\end{equation}
Lens-plane points with zero Jacobian form the critical curve, which may have tangential and radial components defined by the factorization of the expression on the right-hand side of Eq.~(\ref{eq:Jacobian}). The tangential critical curve is restricted to regions with sub-critical surface density and its points $\boldsymbol x_{\text{T}}$ satisfy
\begin{equation}
\label{eq:crit-curve-tangential}
\gamma(\boldsymbol x_{\text{T}})=1-\kappa\left(a[\boldsymbol x_{\text{T}}]\right)\,.
\end{equation}
The radial critical curve is restricted to regions with super-critical surface density and its points $\boldsymbol x_{\text{R}}$ satisfy
\begin{equation}
\label{eq:crit-curve-radial}
\gamma(\boldsymbol x_{\text{R}})=\kappa\left(a[\boldsymbol x_{\text{R}}]\right)-1\,.
\end{equation}
The corresponding tangential and radial components of the caustic are obtained by mapping the critical curve to the source plane using Eq.~(\ref{eq:lens-equation}), i.e., $\boldsymbol y_{\text{T}}=\boldsymbol y(\boldsymbol x_{\text{T}})$ and $\boldsymbol y_{\text{R}}=\boldsymbol y(\boldsymbol x_{\text{R}})$, respectively.

\begin{figure*}
\centering
\includegraphics[width=18 cm]{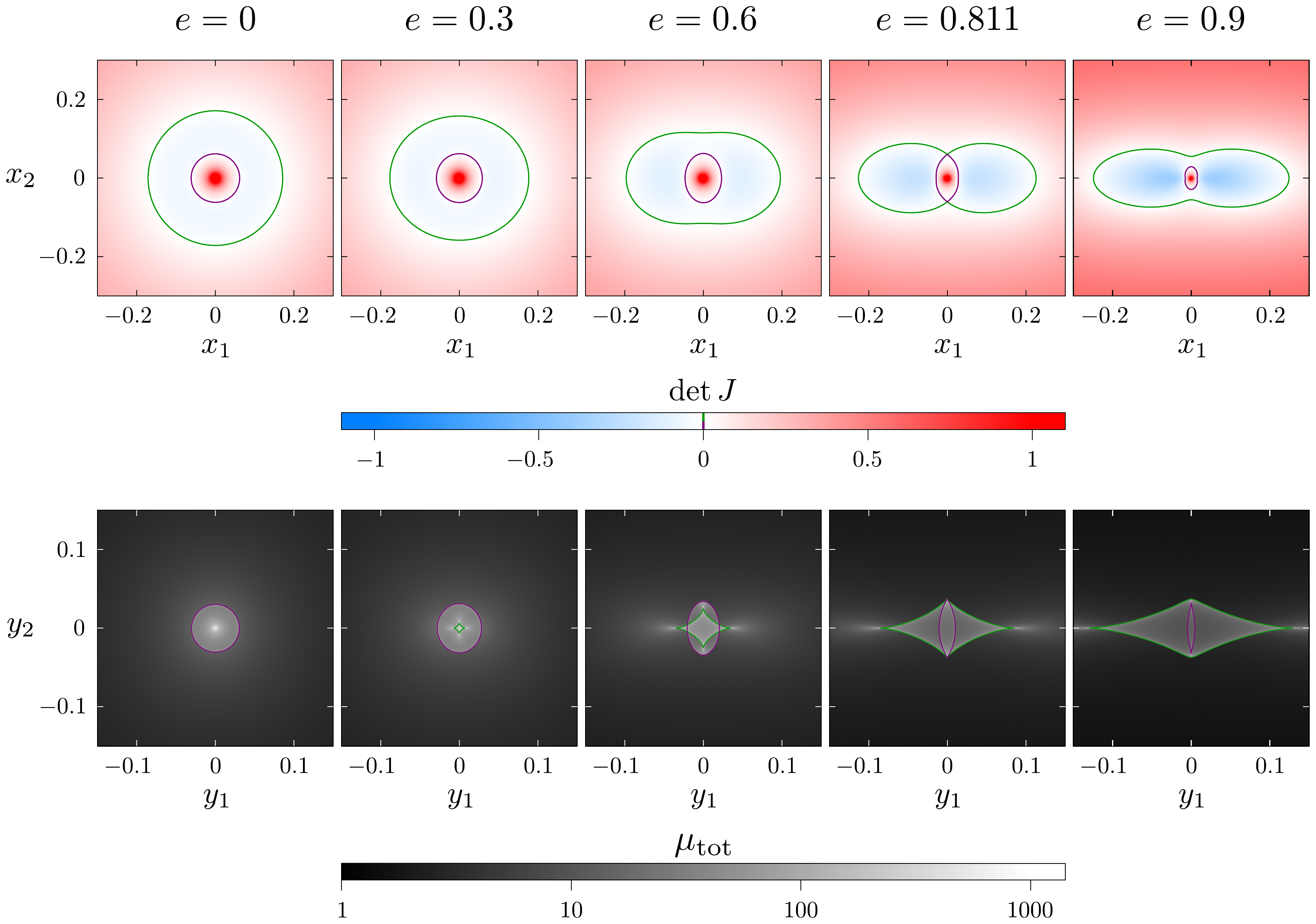}
\caption{Tangential (green) and radial (purple) critical curves (top row) and caustics (bottom row) of ellipsoidal NFW lenses with convergence parameter $\kappa_\text{s}=0.25$ and eccentricities $e\in\{0, 0.3, 0.6, 0.811, 0.9\}$, as marked above the top panels. Background: lens-plane color maps of the Jacobian $\mathrm{det}\,J(\boldsymbol x)$ in the top row; source-plane gray-scale maps of total point-source magnification $\mu_\text{tot}(\boldsymbol y)$ in the bottom row.\label{fig:gallery-basic}}
\end{figure*}

\subsubsection{Variation of geometries with eccentricity}
\label{sec:variation}

Figure~\ref{fig:gallery-basic} shows a typical sequence of critical-curve (top row) and caustic geometries (bottom row) as a function of eccentricity increasing from the left to the right column, in this example for convergence parameter $\kappa_\text{s}=0.25$. The tangential components of the curves are marked in green; the radial components in purple.

The tangential critical curve for $e=0$ is the outer circular component. At $e=0.3$ it is oval-shaped with its major axis horizontal, along the major axis of the projected mass distribution. At $e=0.6$ the oval is more compressed, locally concave in the vicinity of the vertical axis. Eccentricity $e=0.811$ corresponds to the hyperbolic-umbilic configuration for $\kappa_\text{s}=0.25$. The tangential and radial critical curves come into contact at the two umbilic points along the vertical axis. These are defined by zero shear and unit convergence, as indicated by Eqs.~(\ref{eq:crit-curve-tangential}) and (\ref{eq:crit-curve-radial}). We note that the tangential critical curve has sharp points with an undefined tangent at the umbilic points, but no inflection points. At $e=0.9$ the tangential critical curve is peanut-shaped, very elongated horizontally and locally concave close to the vertical axis.

The radial critical curve for $e=0$ is the inner circular component. At $e=0.3$ and $e=0.6$ it is oval-shaped with its major axis vertical, along the minor axis of the projected mass distribution. At the $e=0.811$ hyperbolic-umbilic configuration, the radial critical curve has sharp points at the two umbilic points along the vertical axis. At these points the radial critical curve has an undefined tangent; at the same time it is compressed and locally concave in the vicinity of the horizontal axis. At $e=0.9$ the radial critical curve is small and peanut-shaped, elongated vertically and locally concave close to the horizontal axis.

The color-map background of the critical-curve plots shows the Jacobian, with shades of red and blue indicating positive and negative values, respectively. Since image magnification is given by the absolute inverse of the Jacobian, paler regions correspond to higher magnification. We note that while these are fairly extended at low eccentricity, at high eccentricity (and the same $\kappa_\text{s}$) high magnification is limited to a narrow region along the critical curve.

In the bottom row of Fig.~\ref{fig:gallery-basic}, the tangential caustic is limited to the central point in the source plane at $e=0$. At $e=0.3$ it is a small four-cusped caustic with nearly identical extent along the horizontal and vertical axes. At $e=0.6$ the size is larger and the asymmetry is apparent, with the horizontal extent larger than the vertical and the cusps along the horizontal axis piercing the radial caustic. At the $e=0.811$ hyperbolic-umbilic configuration the tangential caustic fully encloses the radial caustic. Here the cusps along the vertical axis are replaced by hyperbolic umbilic points, shared by both caustics. At $e=0.9$ the tangential caustic is large and horizontally spindle-shaped, locally convex near the vertical axis and with only two cusps, which are located along the horizontal axis. The cusps along the vertical axis were transferred to the radial caustic in the hyperbolic-umbilic metamorphosis.

The radial caustic at $e=0$ is circular. At $e=0.3$ and $e=0.6$ it is oval-shaped with its major axis vertical, fully enclosing the tangential caustic at $e=0.3$, but pierced by its cusps along the horizontal axis at $e=0.6$. At the $e=0.811$ hyperbolic-umbilic configuration the radial caustic extends along the vertical axis to the umbilic points, at which it has an undefined tangent. At $e=0.9$ the radial caustic is small and vertically spindle-shaped, convex near the horizontal axis with two cusps along the vertical axis.

The gray-scale maps forming the background of the caustic plots show the total point-source magnification,
\begin{equation}
\label{eq:magnification}
\mu_\text{tot}(\boldsymbol y)=\sum_{i} |\mathrm{det}\,J(\boldsymbol x_i)|^{-1}\,,
\end{equation}
where $\boldsymbol x_i$ are the positions of all images formed by a point source at $\boldsymbol y$ obtained by solving the lens Eq.~(\ref{eq:lens-equation}). In the $\kappa_\text{s}=0.25$ sequence presented in Fig.~\ref{fig:gallery-basic}, sources located outside both, in between, and inside both caustics have one, three, and five images, respectively. Features to notice in the magnification maps are: the particularly high magnification inside the tangential caustic in the $e=0.3$ and $e=0.6$ maps (including within the protruding cusps); the high-magnification regions outside the cusps truncated by the radial caustic along the vertical axis in the $e=0.6$ map; or the low magnification increase from the center inside the radial caustic in the $e=0.811$ and $e=0.9$ maps.

\begin{figure*}
\centering
\includegraphics[width=18 cm]{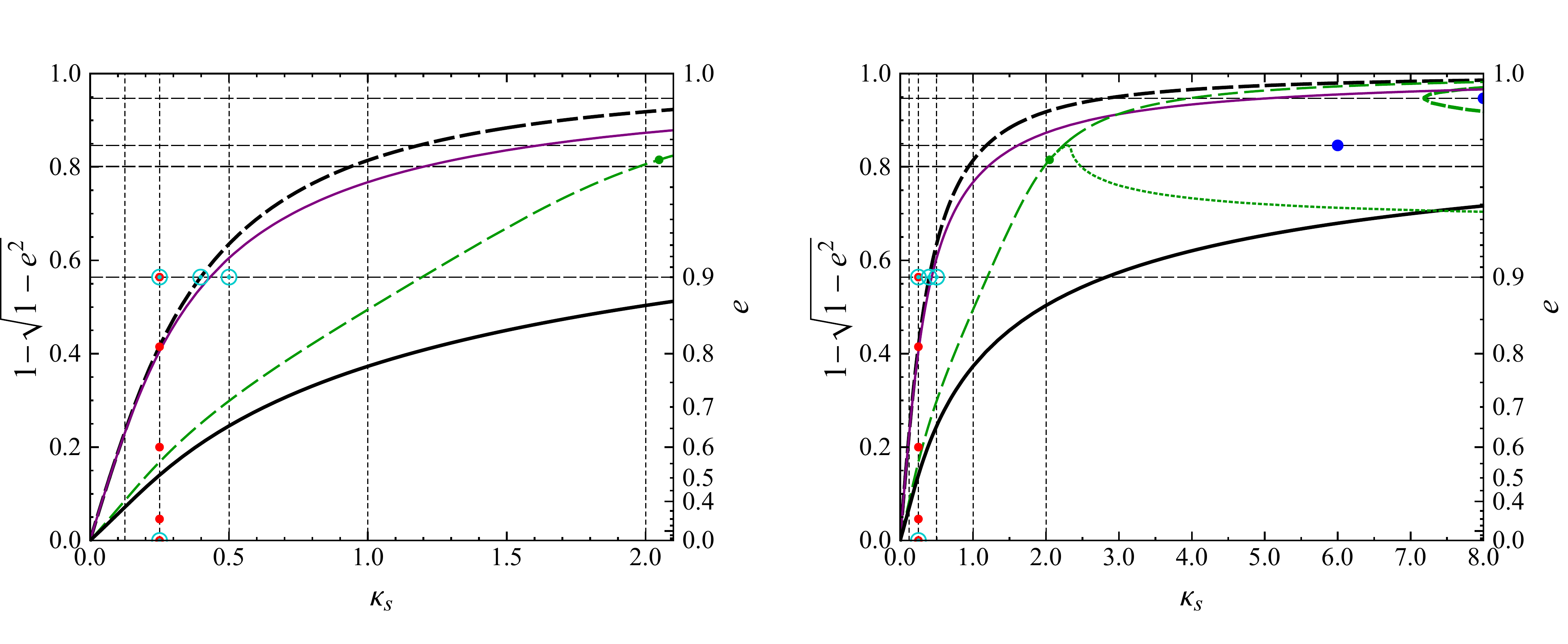}
\caption{Low- and intermediate-$\kappa_\text{s}$ lensing regimes of the ellipsoidal NFW lens mapped in its parameter space, as described in Sect.~\ref{sec:parameter-space-low} and Sect.~\ref{sec:parameter-space-intermediate}. Eccentricity $e$ is marked on the right vertical axes; ellipticity $1-\sqrt{1-e^2}$ on the left vertical axes; convergence parameter on the horizontal axis up to $\kappa_\text{s}=2.1$ in the left panel and up to $\kappa_\text{s}=8$ in the right panel. Plotted boundaries mark the following transitions: cusp piercing (bold solid black); tangential peanut (thin dashed green); radial peanut (thin solid purple); hyperbolic umbilic (dashed black); tangential bulge (thin dotted green); butterfly metamorphosis (bold dashed green). Parameter combinations of the critical-curve and caustic examples presented in Figs.~\ref{fig:gallery-basic} and \ref{fig:gallery-special} are marked by red and blue dots, respectively. Cyan dotted circles mark combinations for which image-orientation plots are presented in Fig.~\ref{fig:nfw-image-orientation}; the green dot marks the branching point of the tangential-bulge boundary from the tangential-peanut boundary. Vertical dashed lines: $\kappa_\text{s}$ values for which critical-curve and caustic dimensions are plotted in Figs.~\ref{fig:nfw-dimensions-critical-curves} and \ref{fig:nfw-dimensions-caustics}, respectively. Horizontal dashed lines: $e$ values for which the image-plane analysis is demonstrated in Figs.~\ref{fig:contours-e9} and \ref{fig:contours}.\label{fig:parameter-space}}
\end{figure*}

\subsubsection{Lensing regimes in parameter space at low $\kappa_\text{s}$}
\label{sec:parameter-space-low}

In the previous section we demonstrated the main lensing regimes of the ellipsoidal NFW lens, as defined by the corresponding critical curves and caustics, for a single value of the convergence parameter $\kappa_\text{s}=0.25$ and a sequence of eccentricities. Here we describe the parameter-space map indicating the occurrence of the different regimes for the full range of eccentricities $e\in[0,1)$ and two convergence-parameter intervals, as shown in Fig.~\ref{fig:parameter-space}. In the left panel we present a detail of the map in the low-convergence regime, shown here for $\kappa_\text{s}\in [0, 2.1]$. For convenience, we mark the ellipticity ($1-\sqrt{1-e^2}\,$) along the left axis and show the corresponding eccentricity values along the non-linear right axis.

The plotted curves mark boundaries corresponding to changes in the geometry of the critical curves and of the caustics. Green lines indicate changes on the tangential components, purple lines on the radial components, and black lines involve both. Thin lines indicate changes only on the critical curve; bold lines indicate changes on the caustic, with the top-most dashed bold line involving the critical curve as well. Solid lines indicate changes occurring along the major axis of the mass distribution, dashed lines indicate changes along the minor axis, and the dotted line indicates changes occurring off either axis. The five red dots along the $\kappa_\text{s}=0.25$ vertical line mark the configurations illustrated in Fig.~\ref{fig:gallery-basic}.

The four primary boundaries seen in the left panel all extend further beyond its range, intersecting all vertical and horizontal lines, i.e., the corresponding transitions occur for any fixed value of $\kappa_\text{s}$ or $e\in(0,1)$. We note that the boundaries do not intersect within the parameter limits of the left panel, so that the sequence of the corresponding transitions remains the same. In order of increasing eccentricity, the lowest bold solid black boundary marks the piercing of the radial caustic by the tangential caustic along the major axis. The equation identifying this condition can be derived from the lens Eq.~(\ref{eq:lens-equation}). The piercing occurs when the two caustic points along the major axis each generate one image on the tangential critical curve (on the same side from the center) and another image on the radial critical curve (on the opposite side):
\begin{equation}
\label{eq:piercing}
x_\text{T}-\alpha_0(x_\text{T},0)=-x_\text{R}-\alpha_0(-x_\text{R},0)\,,
\end{equation}
where $\alpha_0$ is the first component of the deflection angle from Eq.~(\ref{eq:deflection-major-axis}). In our choice $x_\text{T}>0$ and $x_\text{R}>0$ are the major-axis intercepts of the tangential and radial critical curves, respectively. Using Eqs.~(\ref{eq:crit-curve-tangential}) and (\ref{eq:crit-curve-radial}) and taking into account the properties of the shear components, the equations for the intercepts are
\begin{equation}
\label{eq:tangential-major}
1-\kappa(x_\text{T})+\,\gamma_1(x_\text{T},0)=0
\end{equation}
and
\begin{equation}
\label{eq:radial-major}
\kappa(x_\text{R})-1+\,\gamma_1(x_\text{R},0)=0\,.
\end{equation}
This set of three equations can be solved iteratively for a given value of $\kappa_\text{s}$ to obtain the value of $e$ for which Eq.~(\ref{eq:piercing}) is satisfied.

Crossing the thin dashed green boundary to higher eccentricities causes a ``peanut'' transition on the tangential critical curve. Two symmetric pairs of inflection points appear along the minor axis, leading to a horizontal peanut-like shape. The conditions for this transition are thus given by the common solution of the tangential-critical-curve Eq.~(\ref{eq:crit-curve-tangential}) along the minor axis
\begin{equation}
\label{eq:tangential-minor}
\kappa(\,x_2/\sqrt{1-e^2}\;)-1+\,\gamma_1(0,x_2)=0
\end{equation}
and the equation for its inflection point along the minor axis
\begin{equation}
\label{eq:inflection-tangential}
\frac{\sqrt{1-e^2}}{|x_2|}\;\frac{{\rm d}\,\kappa}{{\rm d}\,a}(\,x_2/\sqrt{1-e^2}\;)+\frac{\partial^2\gamma_1}{\partial\, x_1^2}\,(0,x_2)+\frac{1}{\gamma_1(0,x_2)} \left(\frac{\partial\,\gamma_2}{\partial\,x_1}\right)^{\!\!2}(0,x_2)=0\,,
\end{equation}
obtained by differentiating Eq.~(\ref{eq:crit-curve-tangential}). We note that while Eq.~(\ref{eq:tangential-minor}) depends on the inflection-point coordinate $x_2$, $e$, and $\kappa_\text{s}$, in Eq.~(\ref{eq:inflection-tangential}) $\kappa_\text{s}$ cancels out. It is thus advantageous to choose a value of $e$, solve this equation numerically to obtain $x_2$, and use both to directly compute the corresponding value of $\kappa_\text{s}$ from Eq.~(\ref{eq:tangential-minor}).

The thin solid purple boundary marks a similar ``peanut'' transition on the radial critical curve when crossed to higher eccentricities. Here two symmetric pairs of inflection points appear along the major axis, leading to a vertical peanut-like shape. In this case the conditions for the transition are given by the common solution of the radial-critical-curve Eq.~(\ref{eq:crit-curve-radial}) along the major axis
\begin{equation}
\label{eq:radial-major-2}
\kappa(x_1)-1+\,\gamma_1(x_1,0)=0
\end{equation}
and the equation for its inflection point along the major axis
\begin{equation}
\label{eq:inflection-radial}
\frac{1}{(1-e^2)\,|x_1|}\;\frac{{\rm d}\,\kappa}{{\rm d}\,a}(x_1)+\frac{\partial^2\gamma_1}{\partial\, x_2^2}\,(x_1,0)+\frac{1}{\gamma_1(x_1,0)} \left(\frac{\partial\,\gamma_2}{\partial\,x_2}\right)^{\!\!2}(x_1,0)=0\,,
\end{equation}
obtained by differentiating Eq.~(\ref{eq:crit-curve-radial}).  Equation~(\ref{eq:radial-major-2}) depends on the inflection-point coordinate $x_1$, $e$, and $\kappa_\text{s}$, but in Eq.~(\ref{eq:inflection-radial}) $\kappa_\text{s}$ cancels out. Here it is advantageous to choose a value of $e$, solve this equation numerically to obtain $x_1$, and use both to directly compute the corresponding value of $\kappa_\text{s}$ from Eq.~(\ref{eq:radial-major-2}).

\begin{figure}
\centering
\resizebox{0.67\hsize}{!}{\includegraphics{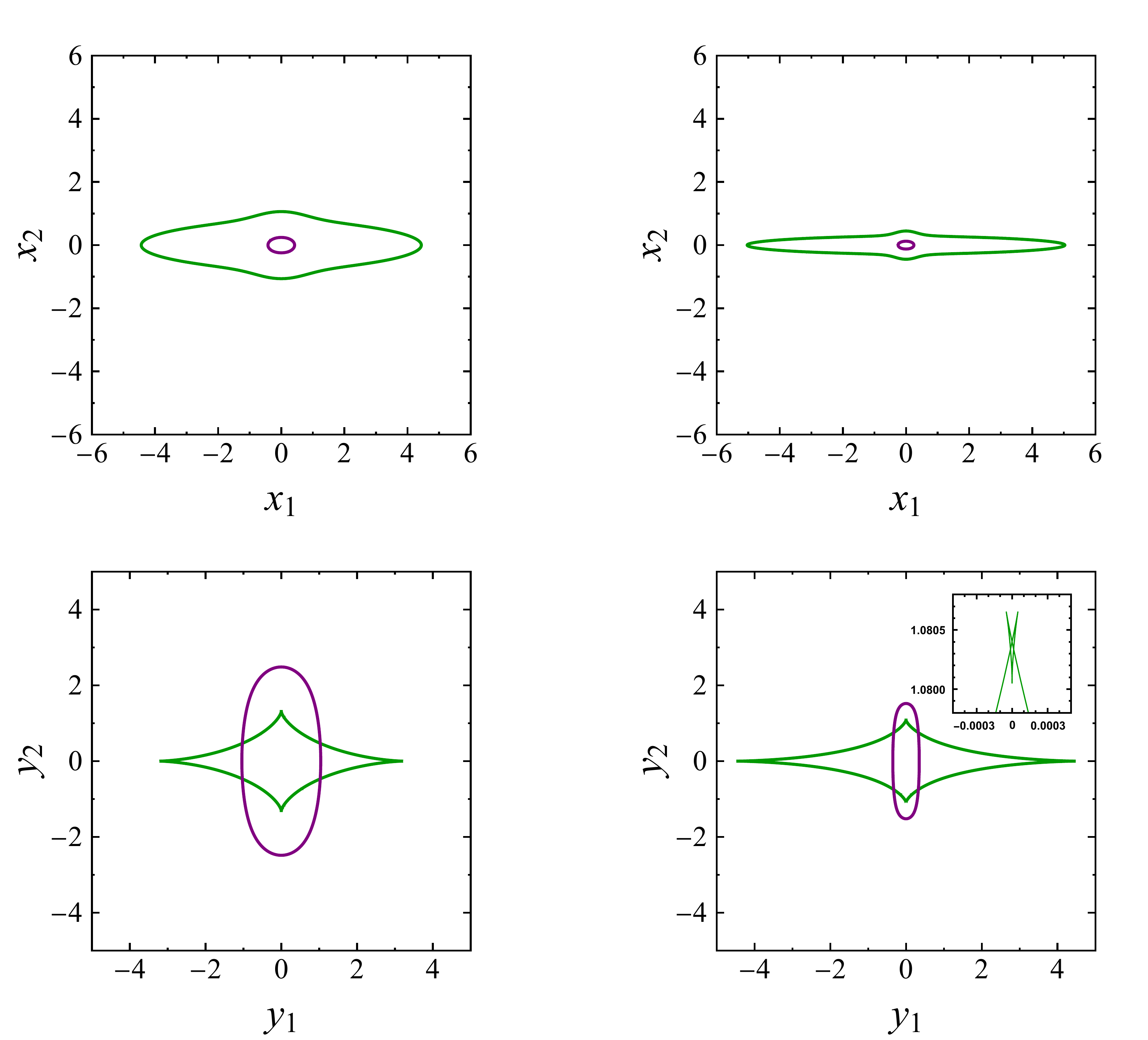}}
\caption{Critical curves (top row) and caustics (bottom row) of ellipsoidal NFW lenses with $(\kappa_\text{s},e)=(6,0.9881)$ (left column) and $(\kappa_\text{s},e)=(8,0.9986)$ (right column), with tangential and radial components marked green and purple, respectively. Both tangential critical curves are centrally bulged; both radial critical curves are flattened along the major axis. The inset in the bottom right panel shows a detail of the top part of the tangential caustic, revealing the three-cusped butterfly structure.\label{fig:gallery-special}}
\end{figure}

The topmost bold dashed black boundary marks the hyperbolic umbilic, in which the tangential and radial components come into contact on the critical curve as well as on the caustic. The conditions for this transition are defined by unit convergence and zero shear. For a given value of $\kappa_\text{s}$ we can obtain the semi-major axis of the unit-convergence ellipse by solving numerically the equation
\begin{equation}
\label{eq:unit-convergence}
\cfrac{\displaystyle 1-\mathcal{F}(a)}{\displaystyle a^2-1}=\frac{1}{2\,\kappa_\text{s}}\,.
\end{equation}
Using the obtained value of $a$, we substitute $x_{0\,\displaystyle{\gamma}}=a\,\sqrt{1-e^2}$ and solve the zero-shear-point Eq.~(\ref{eq:zero-shear-points}) numerically to obtain the corresponding eccentricity $e$.

One important result seen in the left panel of Fig.~\ref{fig:parameter-space} at low values of the convergence parameter (e.g., for $\kappa_\text{s}\lesssim 0.1$) is the dominance of the lensing regime illustrated by the last column of Fig.~\ref{fig:gallery-basic}, with an outer two-cusped tangential caustic and a small inner two-cusped radial caustic. For large values of the convergence parameter (e.g., for $\kappa_\text{s}\gtrsim 1$) the lensing regime in the second column of Fig.~\ref{fig:gallery-basic} dominates, with an outer oval radial caustic and a small inner four-cusped tangential caustic.

\subsubsection{Lensing regimes in parameter space at intermediate $\kappa_\text{s}$}
\label{sec:parameter-space-intermediate}

The right panel of Fig.~\ref{fig:parameter-space} illustrates the more complicated situation for a broader convergence-parameter interval, $\kappa_\text{s}\in [0, 8]$. The four boundaries seen in the left panel continue to rise monotonically in the right panel. However, the order of the two ``peanut'' transitions on the critical curve changes at $\kappa_\text{s}\approx 3$. For higher values of $\kappa_\text{s}$ the radial ``peanut'' appears at a lower eccentricity than the tangential ``peanut''.

More interestingly, two new boundaries can be seen in the right panel. The thin dotted green boundary branching off from the thin dashed green boundary at the green point at $(\kappa_\text{s},e)\approx(2.0487,0.9828)$ marks a ``bulge'' transition on the tangential critical curve, in which four symmetric pairs of inflection points appear off the principal axes. In the parameter-space region between the dotted and the dashed thin green boundaries the tangential critical curve thus has eight inflection points forming a central bulge. An example corresponding to the $(\kappa_\text{s},e)\approx (6,0.9881)$ parameter combination marked by the left blue dot in Fig.~\ref{fig:parameter-space} is shown in the left column of Fig.~\ref{fig:gallery-special}. The size and extent of the bulge varies substantially within its region of parameter space, from a mere bump near the minor axis to lemon-shaped critical curves. Similar bulged critical curves have been found in the singular power-law-ellipsoid lens model \citep{tessore_metcalf15} at high ellipticity.

We note that the described ``bulge'' boundary peaks at $(\kappa_\text{s},e)\approx(2.296,0.9881)$, which also identifies the highest-eccentricity parameter combination within the range of Fig.~\ref{fig:parameter-space} that has an entirely convex tangential critical curve. To the right from the green point in Fig.~\ref{fig:parameter-space} the dashed thin green boundary describes a different version of the ``peanut'' transition. When crossed from lower to higher eccentricities it corresponds to the disappearance of two symmetric pairs of inflection points along the minor axis, turning the bulged critical curve into a peanut-shaped one.

Equations describing the condition for the ``bulge'' transition can be obtained by Taylor-expanding the tangential-critical-curve Eq.~(\ref{eq:crit-curve-tangential}) around its general point $\boldsymbol x_{\text{T}0}$, and setting the quadratic and cubic terms of the expansion along its tangent equal to zero. We used an alternative approach based on the image-plane analysis of critical curves, described in Appendix~\ref{sec:Appendix-image-plane}. For a given eccentricity this method provides a tool for visualizing the existence of the transition and it indicates the corresponding value of $\kappa_\text{s}$.

The second new boundary, plotted in bold dashed green at the top right corner of the right panel of Fig.~\ref{fig:parameter-space} indicates two butterfly metamorphoses occurring at the minor-axis cusps of the tangential caustic. Inside the region to the right of the boundary, these cusps are replaced by small three-cusped self-intersecting butterfly structures. An example corresponding to the $(\kappa_\text{s},e)\approx(8,0.9986)$ parameter combination marked by the right blue dot is shown in the right column of Fig.~\ref{fig:gallery-special}, with an inset showing a blow-up of the vicinity of the top cusp revealing the butterfly. These caustic tangles lead to very localized changes in the number of images: five images in the butterfly ``wings'', and seven images in the butterfly ``proboscis'' (given that the structure lies inside the oval radial caustic). The described butterfly metamorphosis occurs for convergence-parameter values $\kappa_\text{s}\gtrapprox 7.185$. We identified and located this transition by studying the changes in the number of cusps using the image-plane analysis described in Appendix~\ref{sec:Appendix-image-plane}. We note that while the vertical extent of the butterfly features in the example in Fig.~\ref{fig:gallery-special} is merely ${\approx}0.06\%$ of the vertical extent of the tangential caustic, their prominence grows with increasing convergence parameter so that at $\kappa_\text{s}=100$ and higher eccentricities it exceeds $10\%$ of the vertical caustic extent.

The shape of the boundaries near $\kappa_\text{s}=8$ in the right panel of Fig.~\ref{fig:parameter-space} indicates that further changes in the structure of the parameter space may be expected at higher values of the convergence parameter. We describe the properties of the ellipsoidal NFW lens at high $\kappa_\text{s}$ in Appendix~\ref{sec:Appendix-parameter-space-high} and discuss the astrophysical relevance of its high-eccentricity features in Section~\ref{sec:relevance-high-eccentricity}.

\begin{figure*}
\centering
\includegraphics[width=18 cm]{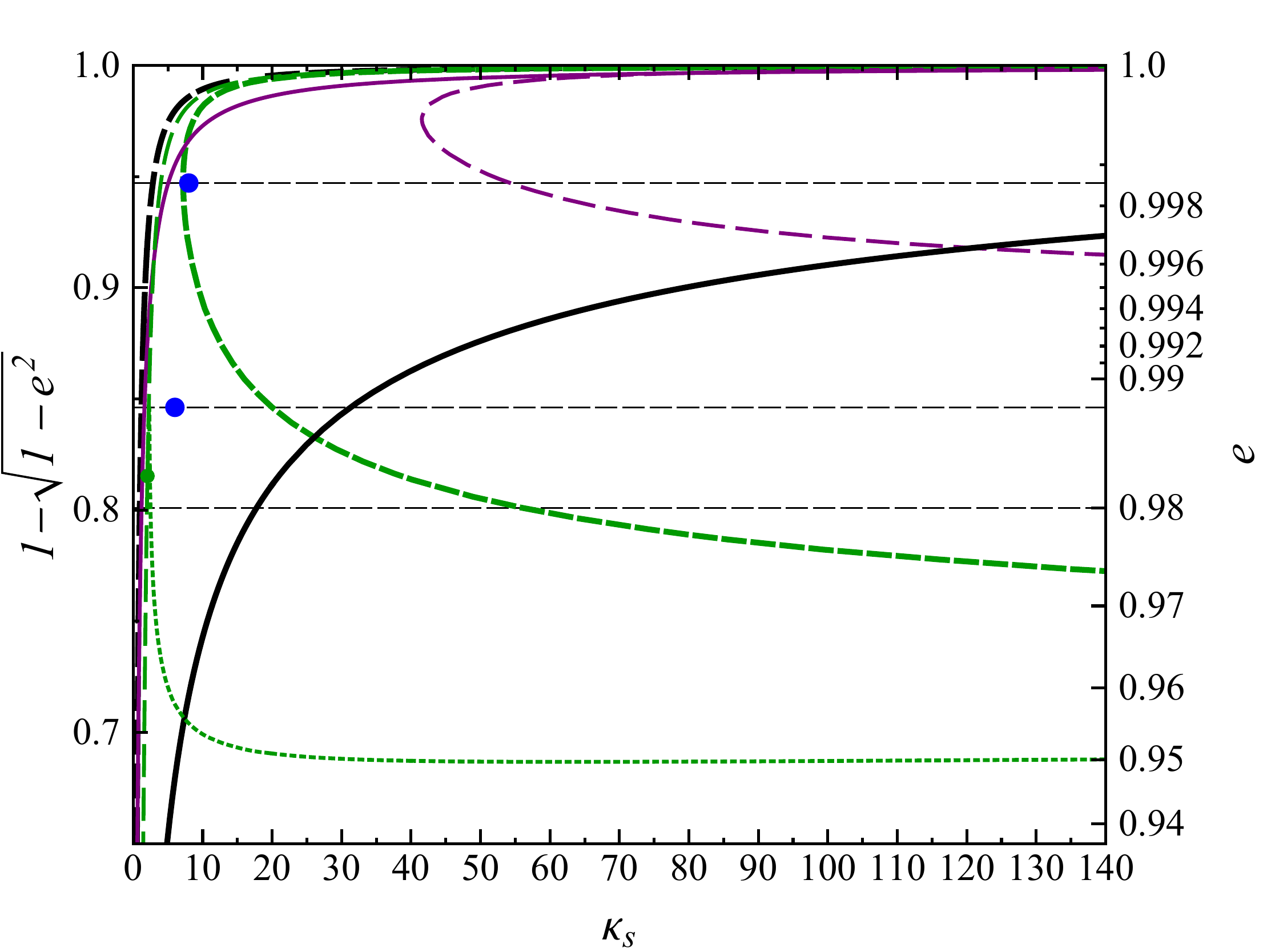}
\caption{Radii of characteristic curves of the spherical NFW lens, plotted as a function of the convergence parameter up to $\kappa_\text{s}=2.1$ (left panel) and up to $\kappa_\text{s}=8$ (right panel). Included are the radii of the tangential ($x_\text{T}$) and radial ($x_\text{R}$) critical curves, of the tangential ($y_\text{T}$) and radial ($y_\text{R}$) caustics, and of the unit-convergence circle ($x_0$). The radius of the radial limiting curve ($x_{\text{R}\infty}\approx1.32$) is marked along the right edge of the right panel. The five vertical dashed lines mark the values of $\kappa_\text{s}$ for which critical-curve and caustic dimensions are plotted for the ellipsoidal NFW lens in Figs.~\ref{fig:nfw-dimensions-critical-curves} and \ref{fig:nfw-dimensions-caustics}, respectively.
\label{fig:nfw-radii}}
\end{figure*}

\subsubsection{Critical-curve and caustic radii in the circularly symmetric case}
\label{sec:radii-circular}

In Sects.~\ref{sec:parameter-space-low}, \ref{sec:parameter-space-intermediate}, and Appendix~\ref{sec:Appendix-parameter-space-high} we described the range of possible shapes of the critical curves and caustics. In this and the following two sections we study the linear dimensions and areas of their tangential and radial components.

For initial orientation we plot in Fig.~\ref{fig:nfw-radii} the radii of the critical curves and caustics in the $e=0$ case for convergence-parameter intervals $\kappa_\text{s}\in[0,2.1]$ (left panel) and $\kappa_\text{s}\in[0,8]$ (right panel), matching the horizontal extent of the panels in Fig.~\ref{fig:parameter-space}. We note that this case includes purely spherical NFW lenses as well as ellipsoidal NFW lenses that are circular in projection, as discussed in the text following Eq.~(\ref{eq:eccentricity}). The plots are extensions of Fig.~3 from \cite{karamazov_etal21} to higher values of $\kappa_\text{s}$.

As described and illustrated in \cite{karamazov_etal21}, for convergence-parameter values $\kappa_s\lessapprox 0.2$ all the radii shrink exponentially fast to zero in the limit $\kappa_\text{s}\to 0$, and all except the tangential-caustic radius increase monotonically with increasing $\kappa_\text{s}$. The tangential and radial critical-curve radii maintain the ordering $x_\text{T}>x_0>x_\text{R}$, where $x_0$ is the radius of the unit-convergence circle defined implicitly by
\begin{equation}
\label{eq:x-0}
\kappa(x_0)=1\,.
\end{equation}

The tangential critical-curve radius is larger than the scale radius of the halo for $\kappa_\text{s}\gtrapprox0.815$. The radial critical-curve radius exceeds the scale radius for $\kappa_\text{s}\gtrapprox9.44$ (just beyond the range of the right panel) and for $\kappa_\text{s}\to\infty$ it reaches the asymptotic value $x_{\text{R}\infty}\approx1.32$ marked by the orange level mark at the right edge of the right panel of Fig.~\ref{fig:nfw-radii}. The existence of this limiting radius is a special case of the more general result that for any value of the eccentricity $e$ the radial critical curve approaches a finite curve in the limit $\kappa_\text{s}\to \infty$, as demonstrated in Appendix~\ref{sec:Appendix-image-plane}.

The tangential caustic is limited to the single point behind the lens, $y_\text{T}=0$, for any value of $\kappa_\text{s}$. The radius of the radial caustic is initially smaller than both critical-curve radii but it increases faster with $\kappa_\text{s}$, so that it exceeds the radial critical-curve radius for $\kappa_\text{s}\gtrapprox0.544$, the tangential critical-curve radius for $\kappa_\text{s}\gtrapprox4.28$, and it grows linearly with $\kappa_\text{s}$ asymptotically.

\subsubsection{Critical-curve dimensions in the general ellipsoidal case}
\label{sec:dimensions-critical-curves}

\begin{figure*}
\centering
\includegraphics[height=19 cm]{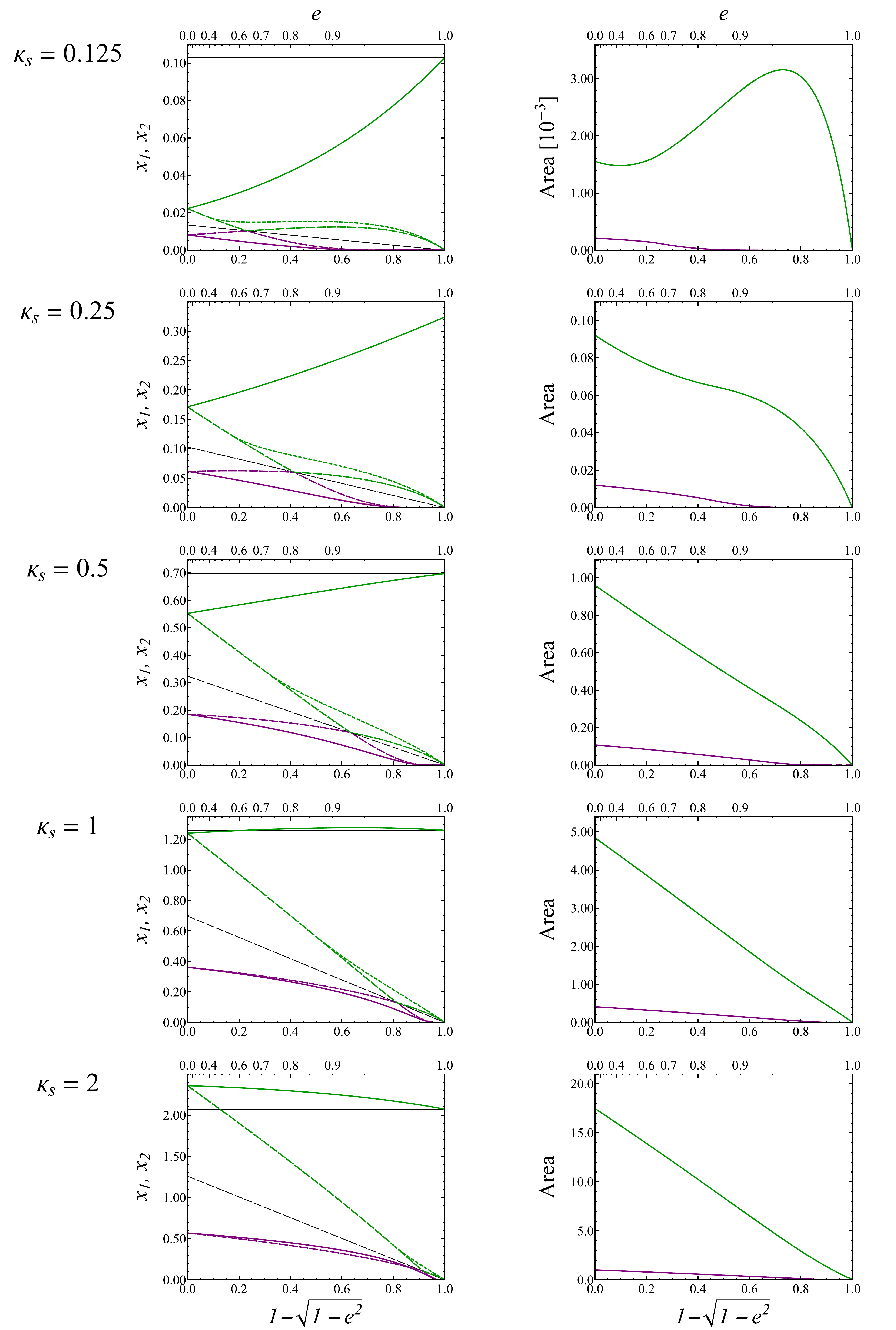}
\caption{Linear dimensions and areas of critical curves of the ellipsoidal NFW lens for five values of the convergence parameter $\kappa_\text{s}$ indicated at the left side. Plotted as a function of ellipticity (bottom axis) and eccentricity (top axis) in the left column are: tangential-critical-curve horizontal (solid green) and vertical (long-dashed green) radii plus maximum vertical extent (short-dashed green); radial-critical-curve horizontal (solid purple) and vertical (long-dashed purple) radii; unit-convergence ellipse semi-minor axis (dashed black); convergence$-1/2$ ellipse semi-major axis (solid black). The coincidence point of the purple, green, and black dashed curves marks the hyperbolic umbilic; the branching point of the short-dashed from the long-dashed green curve marks the tangential-peanut transition. The right column shows the areas enclosed by the tangential (green) and radial (purple) critical curves plotted as a function of ellipticity and eccentricity.
\label{fig:nfw-dimensions-critical-curves}}
\end{figure*}

In Fig.~\ref{fig:nfw-dimensions-critical-curves} we present the sizes of the tangential and radial critical curves for five sample values of the convergence parameter, $\kappa_\text{s}\in\{0.125,0.25,0.5,1,2\}$. Each of these values corresponds to a row in Fig.~\ref{fig:nfw-dimensions-critical-curves} with both panels showing plots as a function of ellipticity $1-\sqrt{1-e^2}\in[0,1]$, with the corresponding eccentricity values marked along the nonlinear top axis. Plotted in the left column are: the horizontal (solid) and vertical (dashed) ``radii'' (defined as the coordinates of the critical-curve intercepts with the major and minor axes of the mass distribution); the vertical maximum extent (dotted; defined as half of the maximum vertical coordinate range of the critical curve); the semi-minor axis of the unit-convergence ellipse (dashed black) and the semi-major axis of the 1/2-convergence ellipse (solid black) for orientation. Plotted in the right column are the areas enclosed by the critical curves, in units of $\tilde{a}_\text{s}^2$, squared projected scale semi-major axis of the halo.

The same five convergence-parameter values are also marked for convenience in Figs.~\ref{fig:nfw-radii} and \ref{fig:parameter-space} by the thin vertical dashed lines. The former may serve as a reference for the $e=0$ starting points along the left axes of the plots in the left column of Fig.~\ref{fig:nfw-dimensions-critical-curves}. The latter is useful for keeping track of the transitions between different lensing regimes encountered while increasing the ellipticity from zero to one.

Looking first at the green tangential critical curve at $\kappa_\text{s}=0.125$ in the left column of the first row, at the left axis its horizontal radius starts from the spherical-case radius $x_\text{T}$ (Fig.~\ref{fig:nfw-radii}) and increases monotonically with ellipticity, extending more than four times before reaching the length of the semi-major axis of the $\kappa=1/2$ contour at the $e=1$ right axis. This $e\to 1$ limit is valid for arbitrary $\kappa_\text{s}$, as can be shown analytically from Eq.~(\ref{eq:crit-curve-tangential}). The vertical radius also starts from the spherical-case radius at the left axis, decreasing monotonically until it reaches the umbilic point at ellipticity $1-\sqrt{1-e^2}\approx0.23$ (see the umbilic boundary crossing in Fig.~\ref{fig:parameter-space}). Here it is equal to the semi-minor axis of the unit-convergence ellipse and the vertical radius of the radial critical curve. For higher ellipticities, the vertical radius of the tangential critical curve rises slowly before declining to zero at $e=1$.

The dotted green curve of the maximum vertical extent branches off from the green dashed curve at ellipticity $1-\sqrt{1-e^2}\approx0.086$, corresponding to the tangential peanut boundary in Fig.~\ref{fig:parameter-space}. It remains nearly constant before dropping to zero at $e=1$. The plot of the enclosed area in the right column shows an initial mild decrease with ellipticity, followed by a prominent increase peaking at $1-\sqrt{1-e^2}\approx0.73$ before dropping to zero at $e=1$. Overall, the tangential critical curve changes with increasing ellipticity from circular to increasingly horizontally elongated, shrinking to a line segment spanning the major axis of the $\kappa=1/2$ ellipse in the unit ellipticity / unit eccentricity limit.

At $\kappa_\text{s}=0.25$ the overall change in the tangential critical curve is similar to the first row, with several differences. The increase in horizontal radius is less dramatic, the vertical radius as well as the maximum vertical extent both decrease monotonically in the full range of ellipticities, and the umbilic and peanut transitions occur at higher ellipticity as seen in Fig.~\ref{fig:parameter-space}. More notably, the area of the tangential critical curve decreases monotonically, with a mere bump around intermediately high ellipticities. Similar changes can be seen in the next row, at $\kappa_\text{s}=0.5$.

At $\kappa_\text{s}=1$ the horizontal radius stays nearly constant, increasing slightly before decreasing to the semi-major axis of the $\kappa=1/2$ contour at $e=1$. The enclosed area decreases nearly linearly with ellipticity. In the $\kappa_\text{s}=2$ case in the bottom row the horizontal radius reverses the trend seen in the top rows, decreasing monotonically with ellipticity. The tangential critical curve thus shrinks monotonically along both axes from the spherical case with an increasingly horizontally elongated shape.

The purple horizontal and vertical radii of the radial critical curve start from the spherical-case radius $x_\text{R}$ (Fig.~\ref{fig:nfw-radii}) at the left axis in the left column of Fig.~\ref{fig:nfw-dimensions-critical-curves}. The horizontal radius decreases monotonically with ellipticity to zero at $e=1$ for all values of the convergence parameter $\kappa_\text{s}$. The vertical radius initially increases with ellipticity in the first two rows until it reaches the umbilic point described above, after which it decreases to zero at $e=1$. At higher values of $\kappa_\text{s}$, from the third row downward, the vertical radius decreases monotonically even before reaching the umbilic point.

The maximum horizontal extent of the radial critical curve branches off from the horizontal radius at ellipticities larger than the radial peanut boundary in Fig.~\ref{fig:parameter-space}. However, the relative increase is so small that the difference between the two could not be seen in Fig.~\ref{fig:nfw-dimensions-critical-curves}. The area enclosed by the radial critical curve plotted in the right column decreases monotonically with ellipticity to zero at $e=1$ in all rows. The area practically vanishes at the scales of the plots already at lower ellipticities: for $1-\sqrt{1-e^2}\gtrapprox\{0.4, 0.6, 0.7, 0.8, 0.8 \}$ in the first to fifth row, respectively.

At lower values of $\kappa_\text{s}$ (first three rows), the radial critical curve changes with increasing ellipticity from circular to increasingly vertically elongated. The elongation peaks at the umbilic, after which it decreases as the two radii converge while the radial critical curve shrinks to the central point in the unit ellipticity limit. At $\kappa_\text{s}=1$, the elongation stays initially negligible and the critical curve maintains nearly circular shape until $1-\sqrt{1-e^2}\approx0.4$ corresponding to eccentricity $e\approx0.8$. At higher convergence parameters, as seen in the $\kappa_\text{s}=2$ example in the last row, the initial orientation is reversed, with the radial curve becoming increasingly horizontally elongated with increasing ellipticity. The orientation switches to vertical elongation just before the umbilic, followed by convergence of the two radii as the curve shrinks to a point in the unit ellipticity limit.

\begin{figure}
\centering
\resizebox{0.45\hsize}{!}{\includegraphics{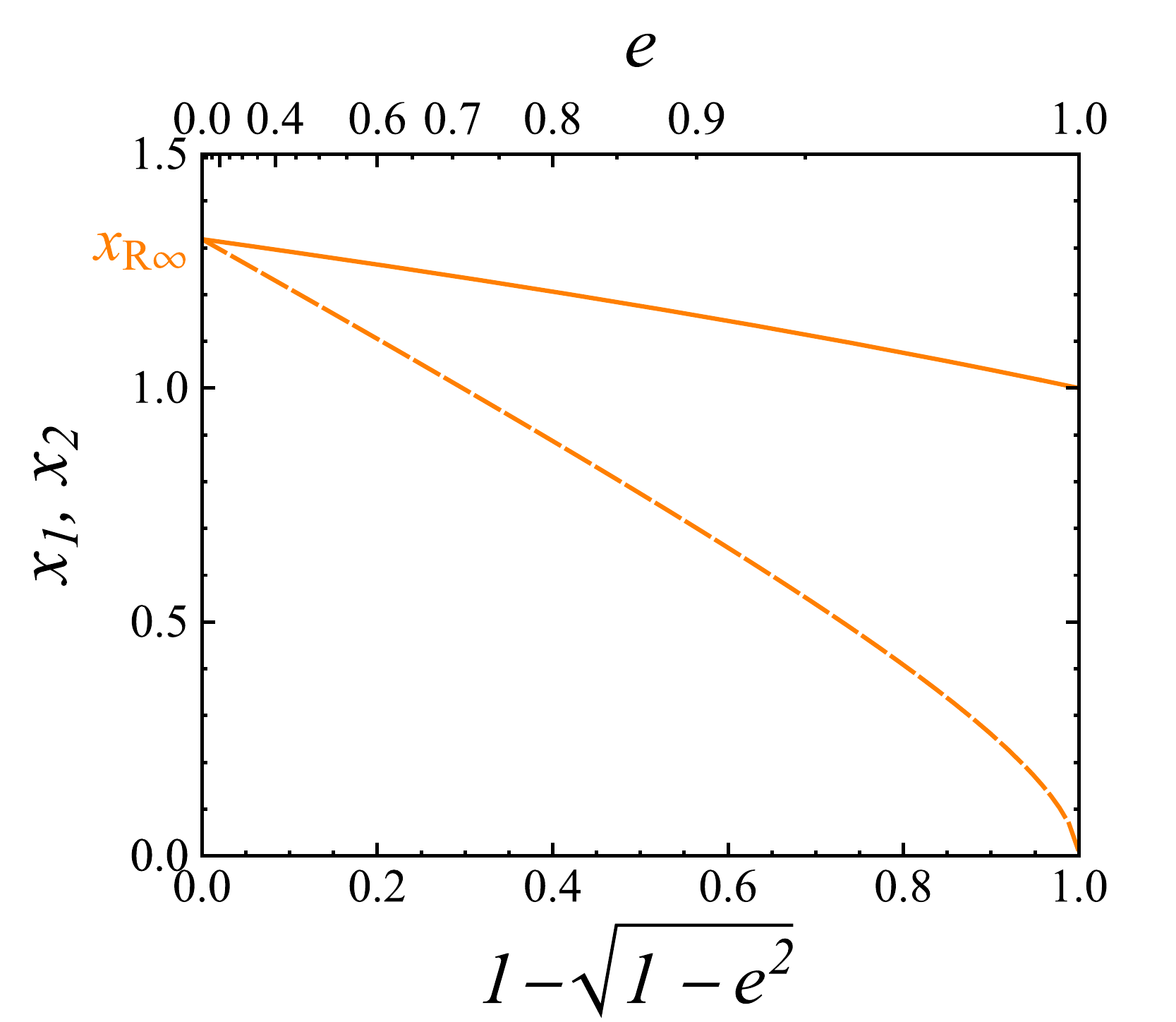}}
\caption{Linear dimensions of the radial limiting curve of the ellipsoidal NFW lens: horizontal (solid) and vertical (dashed) radius plotted as a function of ellipticity and eccentricity. The value of the radius $x_{\text{R}\infty}\approx1.32$ in the $e=0$ limit is marked along the left axis. With increasing $\kappa_\text{s}$ the dimensions of the radial critical curve approach but never exceed these values.
\label{fig:nfw-radial-limiting-curve}}
\end{figure}

With increasing $\kappa_\text{s}$ the growth of the radial critical curve slows down as it converges to a radial limiting curve, as mentioned above and described in Appendix~\ref{sec:Appendix-image-plane}. The horizontal and vertical ``radii'' of this curve are plotted in Fig.~\ref{fig:nfw-radial-limiting-curve} as a function of ellipticity. The radii start at zero ellipticity from the value $x_{\text{R}\infty}\approx1.32$ marked in the right panel of Fig.~\ref{fig:nfw-radii}. Both decrease monotonically with ellipticity: the horizontal reaches $x_1=1$; the vertical $x_2=0$ at $e=1$. The radial limiting curve shrinks from a circle at $e=0$ to an increasingly horizontally elongated shape, converging at $e=1$ to a horizontal line segment extending one scale radius to either side of the halo center. We note that although the horizontal and vertical radii of radial critical curves converge to the lines plotted in Fig.~\ref{fig:nfw-radial-limiting-curve} as $\kappa_\text{s}\to\infty$, their horizontal radius always shrinks to zero in the limit $e\to 1$, as mentioned above.

Notice also the remarkable similarity of Fig.~\ref{fig:nfw-radial-limiting-curve} and the left panel of Fig.~\ref{fig:max-sad-deflection}. The plotted curves are in fact identical, bringing us back to the properties of the deflection angle described in Sect.~\ref{sec:alpha}: the points of maximum deflection $(0,\pm x_{2 \text{max}})$ and deflection-saddle points $(\pm x_{1 \text{sad}},0)$ lie on the radial limiting curve and define its vertical and horizontal radii, respectively. Radial critical curves thus approach these points along the axes as $\kappa_\text{s}\to\infty$, but never extend beyond them.

\begin{figure*}
\centering
\includegraphics[height=19 cm]{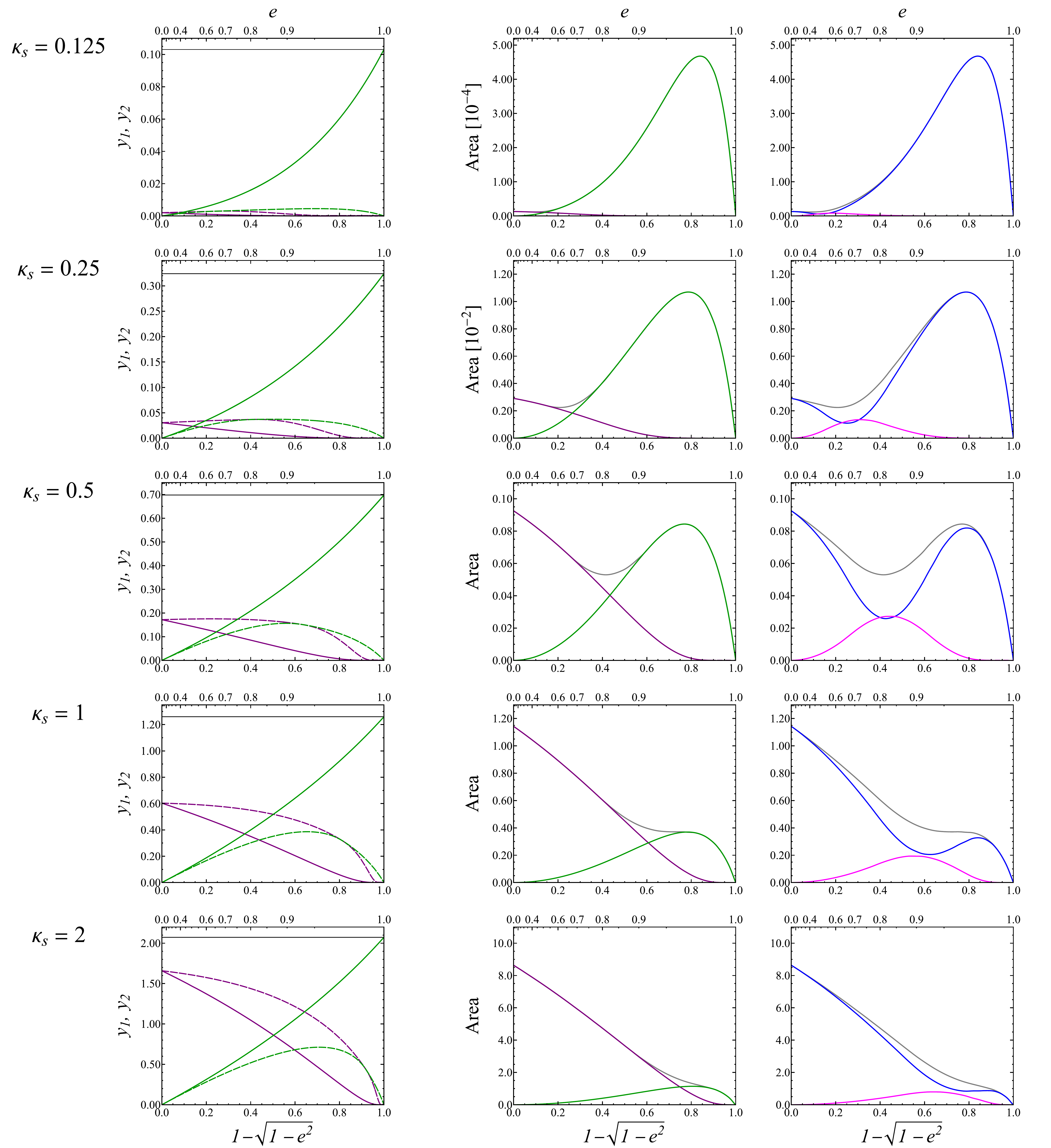}
\caption{Linear dimensions and areas of caustics of the ellipsoidal NFW lens for five values of the convergence parameter $\kappa_\text{s}$ indicated at the left side. Plotted as a function of ellipticity (bottom axis) and eccentricity (top axis) in the left column are: tangential-caustic horizontal (solid green) and vertical (dashed green) radii; radial-caustic horizontal (solid purple) and vertical (dashed purple) radii; convergence$-1/2$ ellipse semi-major axis (solid black). The intersection of the purple and green dashed curves marks the hyperbolic umbilic; the intersection of the purple and green solid curves marks the cusp-piercing transition. The middle column shows the areas enclosed by the tangential (green) and radial (purple) caustics and the total area inside the caustics (gray), which is equal to the radial-caustic area at eccentricities lower than the cusp piercing and equal to the tangential-caustic area at eccentricities higher than the hyperbolic umbilic. The right column includes similar plots showing the caustic areas yielding three (blue), five (magenta), and at least three (gray) point-source images.
\label{fig:nfw-dimensions-caustics}}
\end{figure*}

\subsubsection{Caustic dimensions in the general ellipsoidal case}
\label{sec:dimensions-caustics}

Figure~\ref{fig:nfw-dimensions-caustics} is the source-plane counterpart of Fig.~\ref{fig:nfw-dimensions-critical-curves}, including similarly designed plots describing the sizes of the tangential and radial caustics for the same set of values of the convergence parameter. Plotted in the left column are: the horizontal (solid) and vertical (dashed) ``radii'' (defined as the coordinates of the caustic intercepts with the major and minor axes of the mass distribution); the semi-major axis of the $\kappa=1/2$ ellipse (solid black) for orientation. Plotted in the central column are the areas enclosed by the caustics (including the area enclosed by the combined caustic plotted in gray), in units of $\tilde{a}_\text{s}^2$, squared projected scale semi-major axis of the halo. Plotted in the right column are the areas separated here into those enclosed by a single caustic, corresponding to point-source positions yielding three images (blue); those enclosed by both caustics, yielding five images (magenta); and the same total area as in the central column, yielding multiple images (gray).

Looking first at the green tangential caustic in any row of the left column, both radii start at zero at the left axis corresponding to the point-like nature of the caustic in the spherical case (Fig.~\ref{fig:nfw-radii}). Its horizontal radius increases monotonically with ellipticity, crosses the solid purple horizontal radius of the radial caustic at the piercing point, and reaches the length of the semi-major axis of the $\kappa=1/2$ contour at the $e=1$ right axis, the same limit reached by the horizontal radius of the tangential critical curve, as seen in Fig.~\ref{fig:nfw-dimensions-critical-curves}. The vertical radius first increases with ellipticity at the same rate as the horizontal radius, gradually slows down, reaches a peak, and decreases to zero at $e=1$. En route it crosses the dashed purple vertical radius of the radial caustic at the umbilic point. In the rows from top to bottom the piercing and umbilic transitions gradually occur at higher ellipticities, as seen in Fig.~\ref{fig:parameter-space}.

The plot of the enclosed area in the central column shows a strong increase from zero, a peak at high ellipticity and a final drop to zero at $e=1$. Overall, the tangential caustic changes with increasing ellipticity from point-like to initially symmetrically astroidal, becoming increasingly horizontally elongated, piercing the radial caustic along the major axis, losing its two cusps along the minor axis at the umbilic, then shrinking vertically to a line segment spanning the major axis of the $\kappa=1/2$ ellipse in the unit ellipticity / unit eccentricity limit.

The purple horizontal and vertical radii of the radial caustic start from the spherical-case radius $y_\text{R}$ (Fig.~\ref{fig:nfw-radii}) at the left axis in the left column of Fig.~\ref{fig:nfw-dimensions-caustics}. The horizontal radius decreases monotonically with ellipticity, crossing the solid green horizontal radius of the tangential caustic at the piercing point, and decreasing to zero at $e=1$ for all values of the convergence parameter $\kappa_\text{s}$. The vertical radius initially increases weakly with ellipticity in the first three rows, then decreases to zero at $e=1$, crossing the dashed green vertical radius of the tangential caustic at the umbilic point en route. At higher values of $\kappa_\text{s}$, in the two lowest rows, the vertical radius decreases monotonically in the full range of ellipticities.

The area enclosed by the radial caustic plotted in the central column decreases monotonically with ellipticity to zero at $e=1$ in all rows. The area practically vanishes at the scales of the plots already at lower ellipticities: for $1-\sqrt{1-e^2}\gtrapprox\{0.4, 0.65, 0.8, 0.85, 0.9 \}$ in the first to fifth row, respectively. Overall, the radial caustic changes with increasing ellipticity from circular to increasingly vertically elongated, gaining two cusps along the minor axis at the umbilic. Both radii shrink to zero, with the horizontal radius shrinking faster, so that the caustic first resembles a short vertical line segment that shrinks to zero length in the unit ellipticity limit.

The gray total area in the central column is equal to the radial caustic area at low ellipticities. It branches off at the piercing point and merges with the tangential caustic area at the umbilic point. The central column shows the dramatic change in dominance of the tangential and radial components by area. At any value of $\kappa_\text{s}$ the radial caustic dominates at low ellipticities, while the tangential caustic dominates at high ellipticities. The ellipticity at which the tangential caustic overtakes the radial caustic shifts dramatically with the convergence parameter from $1-\sqrt{1-e^2}\approx 0.14$ in the top row to $1-\sqrt{1-e^2}\approx0.75$ in the bottom row. At low convergence parameters, the total area of the caustic is absolutely dominated by the tangential caustic at all but the lowest ellipticities, while at high convergence parameters, the total area is absolutely dominated by the radial caustic at all but the highest ellipticities. This result adds an explicit quantification to the related comment at the end of the parameter-space analysis at the end of Sect.~\ref{sec:parameter-space-low}.

In the right column of Fig.~\ref{fig:nfw-dimensions-caustics}, the magenta curve representing the source-plane area yielding five point-source images is equal to the tangential caustic area at ellipticities lower than the cusp-piercing transition, and it is equal to the radial caustic area at ellipticities higher than the umbilic transition. Thus, it drops to zero at low as well as at high ellipticities. Between these two regimes the five-image area peaks, with the position of the peak shifting from low to high ellipticity with increasing convergence parameter $\kappa_\text{s}$. In the first three rows there even is a narrow interval of ellipticities around the peak at which the five-image area narrowly exceeds the three-image area marked by the blue curve. At higher $\kappa_\text{s}$ in the bottom two rows the five-image area is smaller than the three-image area at all ellipticities.

At a more basic level, the gray curve in the central and right columns vividly illustrates the change in the overall probability of multiple imaging. At low values of $\kappa_\text{s}$ the probability is negligible at low ellipticity and high at high ellipticity. At high $\kappa_\text{s}$ the opposite is true: the probability is high at low ellipticity and negligible at high ellipticity. At the same time we point out the varying scale on the vertical axes marked in the central column, illustrating the dramatic increase of the probability with the convergence parameter.

\subsection{Image properties}
\label{sec:images}

The analysis of the critical curves and caustics in Sect.~\ref{sec:curves} shows that the ellipsoidal NFW lens may produce 1, 3, 5, or 7 images of a point source lying off the caustic. Here we describe briefly the geometrical properties of the individual images. In Sect.~\ref{sec:images-magnification-flattening} we utilize convergence--shear diagrams to study the range of image-magnification and flattening values achievable by a lens with a given combination of $\kappa_\text{s}$ and $e$. In Sect.~\ref{sec:images-orientation} we illustrate the patterns of image orientation, as determined by the combination of phase and convergence. In Sect.~\ref{sec:images-examples} we present a few examples of interesting image configurations.

\subsubsection{Magnification and flattening}
\label{sec:images-magnification-flattening}

\begin{figure}
\centering
\resizebox{\hsize/2}{!}{\includegraphics{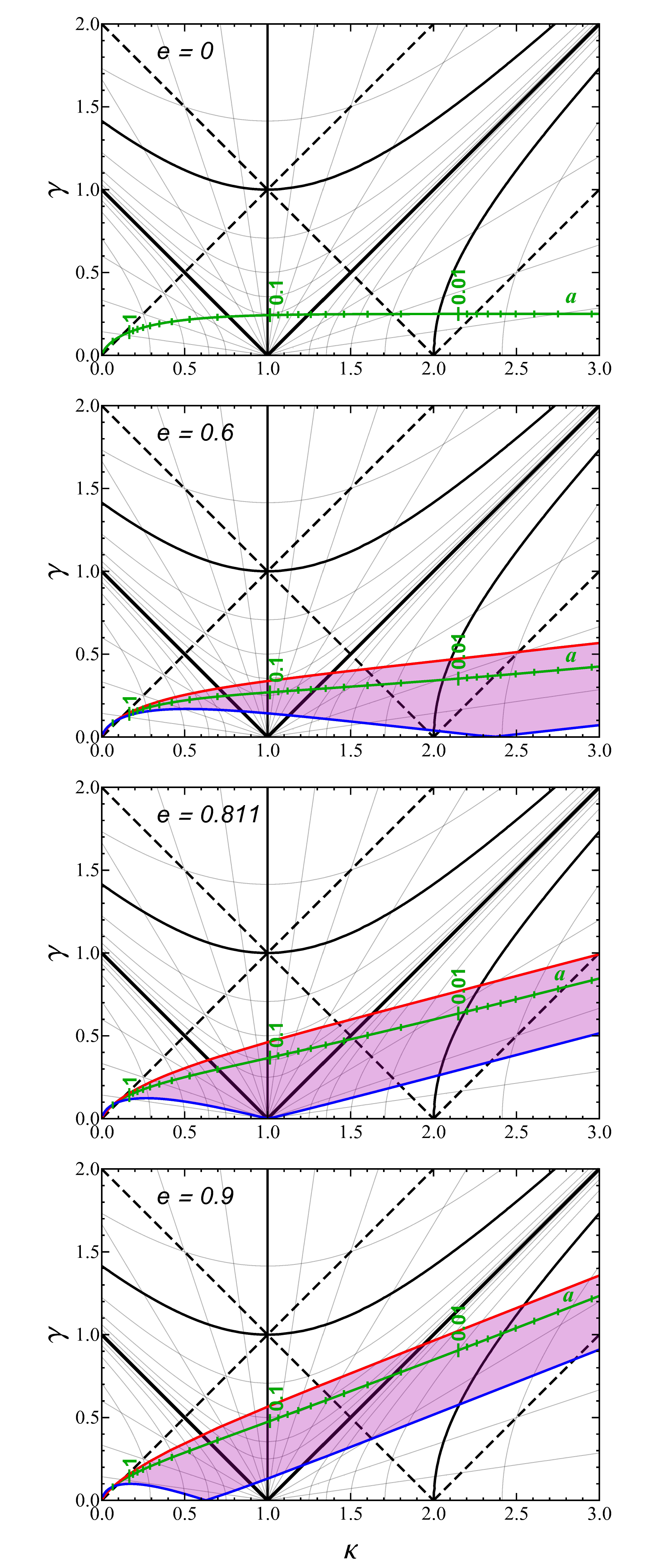}}
\caption{Convergence--shear diagrams for the ellipsoidal NFW lens with $\kappa_\text{s}=0.25$ and eccentricities $e\in\{0, 0.6, 0.811, 0.9\}$ as marked in the panels. In the top row with $e=0$ the green curve marks the $(\kappa,\gamma)$ combinations occurring in the lens plane as a function of radius $a$ marked by the green tick marks. In each of the three lower rows the magenta band marks the $(\kappa,\gamma)$ combinations occurring in the lens plane, with its vertical cross-sections indicating the shear range along the ellipse with semi-major axis $a$. The red and blue curves mark the shear along the major and minor axis, respectively. For reading the image properties from the diagrams, see Sect.~\ref{sec:images-magnification-flattening}.
\label{fig:nfw-convergence-shear-e}}
\end{figure}

The convergence--shear diagram \citep[hereafter CS diagram;][]{karamazov_heyrovsky22} is based on a simple plot with convergence $\kappa$ along the horizontal axis and shear $\gamma$ along the vertical axis, as seen in the examples in Fig.~\ref{fig:nfw-convergence-shear-e}. In such a plot, lines of constant magnification $\mu=1 / \mathrm{det}\,J$ are hyperbolae
\begin{equation}
(1-\kappa)^2-\gamma^2=1/\mu
\label{eq:cs-hyperbolae}
\end{equation}
centered around the point with the umbilic combination $(\kappa,\gamma)=(1,0)$. Their common diagonal asymptotes (marked solid bold black in Fig.~\ref{fig:nfw-convergence-shear-e}) correspond to $|\mu|=\infty$, and the absolute magnification declines outward from the diagonals. The bold black hyperbolae and the point at the origin correspond to $|\mu|=1$; the parts of the diagram above and to the right of the region between these hyperbolae correspond to demagnified images. The $(\kappa,\gamma)$ combinations in the regions under the bold diagonals yield positive magnification (i.e., positive-parity images); in the complementary region above the bold diagonals the magnification is negative (i.e., negative-parity images).

The flattening, which is defined as the ellipticity of the image of a small circular source,
\begin{equation}
f=1-\min\left({\left|\frac{1-\kappa-\gamma}{1-\kappa+\gamma}\right|, \left|\frac{1-\kappa+\gamma}{1-\kappa-\gamma}\right|}\right)\,,
\label{eq:cs-flattening}
\end{equation}
is constant along the four straight lines
\begin{equation}
\begin{matrix}
  f\,\kappa+(2-f)\,\gamma-f \\
  (2-f)\,\kappa+f\,\gamma+f-2 \\
  (2-f)\,\kappa-f\,\gamma+f-2 \\
  f\,\kappa-(2-f)\,\gamma-f
\end{matrix}
\;
\begin{matrix}
  = 0 \\
  = 0 \\
  = 0 \\
  = 0
\end{matrix}
\label{eq:cs-lines}
\end{equation}
extending radially from the point $(\kappa,\gamma)=(1,0)$. The sets of these lines are shown for $f\in\{0.25,0.5,0.75\}$ in Fig.~\ref{fig:nfw-convergence-shear-e}, where they are marked gray. The first two lines appear in the diagram in the $\kappa<1$ region, where the image is extended parallel to the phase. The third and fourth lines appear in the $\kappa>1$ region, where the image is extended perpendicular to the phase. In the case of undistorted images ($f=0$) the four lines reduce to the horizontal axis, $\gamma=0$, and the vertical line at $\kappa=1$. In the case of extremely distorted images ($f=1$) the four lines reduce to the two solid black diagonal lines, $\gamma=1-\kappa$ and $\gamma=\kappa-1$, with the former corresponding to the tangential critical curve and the latter to the radial critical curve.

Lines of constant image-scaling factor parallel to the phase $\lambda_\parallel$ correspond to the set of lines
\begin{equation}
1-\kappa-\gamma=1/\lambda_\parallel
\label{eq:cs-lambda-parallel}
\end{equation}
parallel to the tangential-critical-curve diagonal, along which $|\lambda_\parallel|=\infty$. The absolute value declines on both sides, with positive values under and negative values above the diagonal. The point at the origin corresponds to $\lambda_\parallel=1$; the dashed black $\gamma=2-\kappa$ line from the top left of the diagram corresponds to $\lambda_\parallel=-1$. Lines of constant image-scaling factor perpendicular to the phase $\lambda_\perp$ correspond to the set of lines
\begin{equation}
1-\kappa+\gamma=1/\lambda_\perp\,
\label{eq:cs-lambda-perpendicular}
\end{equation}
parallel to the radial-critical-curve diagonal, along which $|\lambda_\perp|=\infty$. The absolute value declines on both sides, with positive values above and negative values under the diagonal. The dashed black $\gamma=\kappa$ line from the origin corresponds to $\lambda_\perp=1$; the dashed black $\gamma=\kappa-2$ line at the right side of the diagram corresponds to $\lambda_\perp=-1$.

For any given point $(\kappa,\gamma)$ in the CS diagram one may thus directly read off the corresponding magnification $\mu$ (including parity as the sign of $\mu$), flattening $f$, orientation with respect to the phase, and the two scaling factors $\lambda_\parallel$ and $\lambda_\perp$ of a small image appearing in the lens plane at such a convergence--shear combination. For a CS diagram of a particular gravitational lens we utilize the plot described above, and mark in it the range of $(\kappa,\gamma)$ combinations occurring in the full image plane of the lens. In this way we may easily identify and visualize the above-mentioned properties for the full range of images possibly formed by the lens.

In Fig.~\ref{fig:nfw-convergence-shear-e} we present four CS diagrams for an ellipsoidal NFW lens with convergence parameter $\kappa_\text{s}=0.25$ and eccentricities $e\in\{0, 0.6, 0.811, 0.9\}$. The same parameter combinations plus the additional $(\kappa_\text{s},e)=(0.25,0.3)$ were used for the critical-curve and caustic examples in Fig.~\ref{fig:gallery-basic}; all five are marked by the red dots in the parameter-space maps in Fig.~\ref{fig:parameter-space}.

In the zero-eccentricity case (top panel of Fig.~\ref{fig:nfw-convergence-shear-e}) the semi-major axis is equal to the radial distance, $a=x$. The convergence and shear are purely radial functions, $\kappa(a)$ and $\gamma(a)$, so that all their combinations trace the green curve parameterized by the semi-major axis / radial distance, marked along the curve by logarithmically spaced tick marks. Starting from the asymptotic region $a\to \infty$ at the $(\kappa,\gamma)=(0,0)$ origin and progressing toward the halo center, the magnification $\mu$ initially increases from 1; the flattening $f$ increases from 0; $\lambda_\parallel$ increases from 1; $\lambda_\perp$ first decreases from 1 before reversing the trend and increasing, crossing the dashed line $\lambda_\perp=1$ at $a=x_{\text{R}\infty}\approx1.32$. Progressing further inward (to the right along the green curve), we reach the tangential critical curve at $a\approx0.17$. As we cross it and enter the negative-parity region of the diagram, the magnification diverges $\mu\to\infty$ and changes sign to $\mu=-\infty$; the flattening reaches its maximum $f=1$ and starts to decrease; $\lambda_\parallel$ exhibits the same behavior as $\mu$, diverging $\lambda_\parallel\to\infty$ and changing sign to $\lambda_\parallel=-\infty$; $\lambda_\perp\approx2.14$ at the tangential critical curve and keeps increasing. At $a\approx0.103$ the convergence exceeds $\kappa=1$ and the image distortion switches orientation from parallel to the phase to perpendicular to the phase; the flattening correspondingly reaches its minimum $f=0$ and starts to increase.

At $a\approx0.102$ the absolute magnification reaches $|\mu|\approx17$, its minimum value for negative-parity images. As we cross the radial critical curve at $a\approx0.062$ and enter the inner positive-parity region, the magnification diverges $\mu\to-\infty$ and changes sign to $\mu=\infty$; the flattening reaches its maximum $f=1$ and starts to decrease; $\lambda_\parallel\approx-2.03$ at the radial critical curve and keeps increasing; $\lambda_\perp$ exhibits similar behavior as $\mu$ with a flipped sign, diverging $\lambda_\perp\to\infty$ and changing sign to $\lambda_\perp=-\infty$. At $a\approx0.022$ the dashed line $\lambda_\parallel=-1$ is crossed, indicating that images closer to the halo center are shrunk in the direction parallel to the phase. At $a\approx0.013$ we cross the magnification $\mu=1$ hyperbola, indicating that images closer to the halo center are demagnified. At $a\approx0.0082$ the dashed line $\lambda_\perp=-1$ is crossed, indicating that images closer to the halo center are shrunk even in the direction perpendicular to the phase. Beyond the right edge of the plot, closer to the halo center, the convergence keeps increasing $\kappa\to\infty$ and the shear reaches its central value $\gamma=\kappa_\text{s}=0.25$. All the followed quantities tend to zero: $\mu\searrow0$, $f\searrow0$, $\lambda_\parallel\nearrow0$, $\lambda_\perp\nearrow0$.

\begin{figure}
\centering
\resizebox{0.67\hsize}{!}{\includegraphics{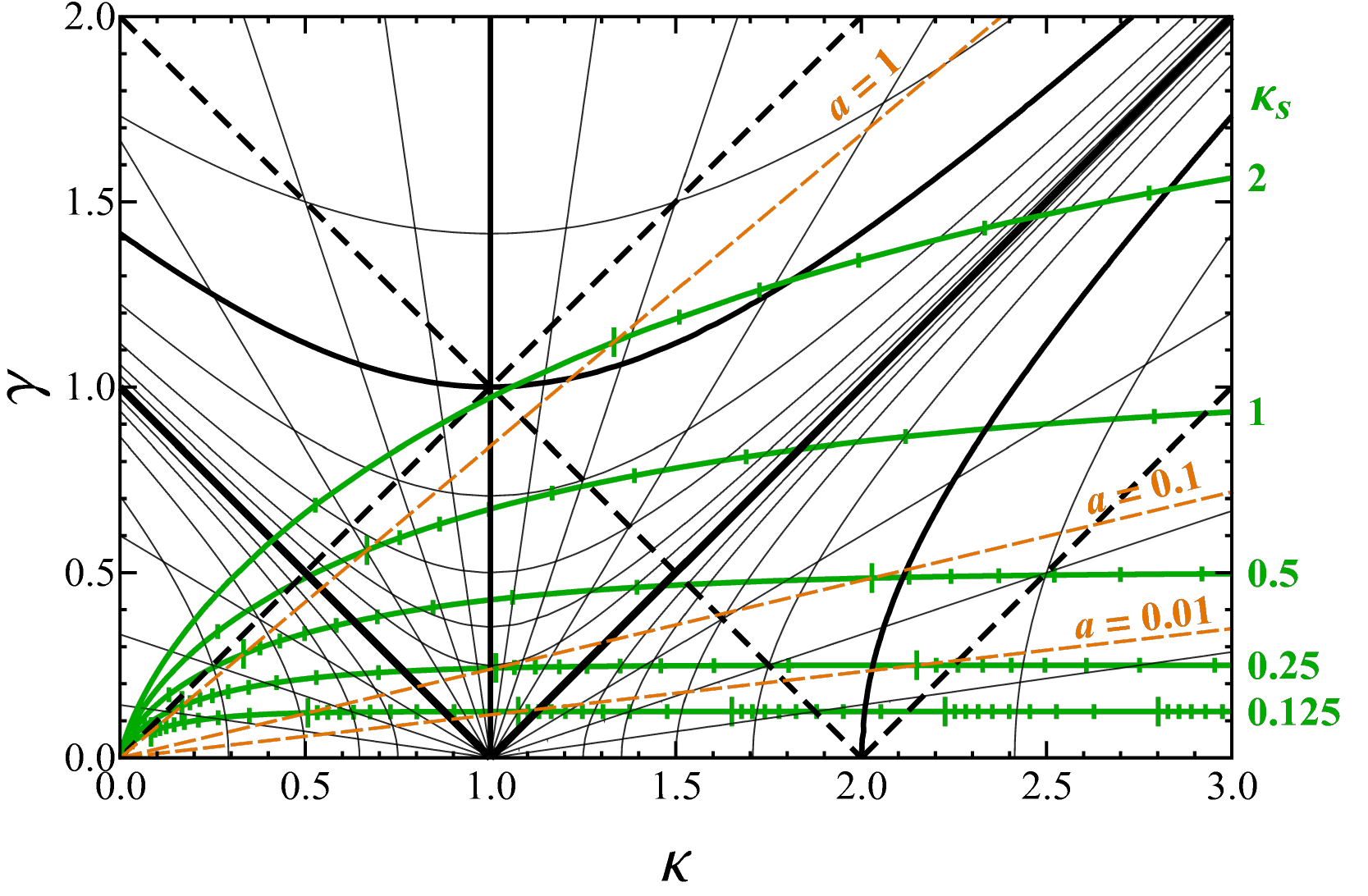}}
\caption{Convergence--shear diagrams for $e=0$ NFW lenses with five values of $\kappa_\text{s}$ as marked along the right axis. The conversion between different values of $\kappa_\text{s}$ is obtained simply by scaling the plot from the origin by the target/original $\kappa_\text{s}$ value ratio. The three dashed orange lines connect points on the different curves corresponding to radius $a=1$, to radius $a=0.1$, and to radius $a=0.01$.
\label{fig:nfw-convergence-shear-ks}}
\end{figure}

In the case of non-zero eccentricity the shear does not have the elliptical symmetry of the convergence, as described in Sect.~\ref{sec:shear}. As a result, the region in the CS diagram corresponding to such a lens is two-dimensional, as indicated by the magenta-colored areas in the lower three panels of Fig.~\ref{fig:nfw-convergence-shear-e}. Nevertheless, the connection between the semi-major axis and the convergence $\kappa(a)$ remains unchanged and independent of eccentricity. Hence, vertical lines in the diagram correspond to ellipses with semi-major axes $a$ and the eccentricity $e$ of the model. The vertical span of the colored area is thus given by the variation of the shear along these ellipses. The interval of the shear values at semi-major axis $a$ is determined by finding the minimum and maximum values of $\gamma(a\,\cos{\theta}, a\,\sqrt{1-e^2}\,\sin{\theta})$. Due to symmetry it is sufficient to vary the parameter $\theta$ from 0 along the major axis to $\pi/2$ along the minor axis.

In the lower three panels of Fig.~\ref{fig:nfw-convergence-shear-e}, the shear along the major and minor axes is marked by the red and blue lines, respectively. The green line with the semi-major-axis ticks corresponds to $\theta=\pi/4$, a line dividing the quarter-ellipse into equal-area parts. We see that in the inner region of the halo covering nearly the full extent of the plotted CS diagrams, the shear is lowest along the minor axis and highest along the major axis. In the outer region extending asymptotically to infinity, the order is reversed: minimum shear occurs along the major axis and maximum along the minor axis. In the plotted examples, the transition between the inner and outer regions occurs between $a=1$ and $a=2$. An inspection of the shear plots in Figs.~\ref{fig:shear-zoom} and \ref{fig:shear} confirms these results.

The next feature to notice in the non-zero-eccentricity panels of Fig.~\ref{fig:nfw-convergence-shear-e} is the presence of the zero-shear point along the minor axis, as described in Sect.~\ref{sec:zero-shear-points}. For $e=0.6$ in the second panel the point lies on the $a\approx0.065$ ellipse; for $e=0.811$ in the third panel on the $a\approx0.103$ unit-convergence ellipse; for $e=0.9$ in the fourth panel on the $a\approx0.24$ ellipse. These semi-major-axis values are equal to $x_{0\,\displaystyle{\gamma}}/\sqrt{1-e^2}$, and can be thus computed using the zero-shear-point positions along the minor axis plotted in Fig.~\ref{fig:zero-shear-point}. The example in the third panel with the zero-shear point lying on the unit-convergence ellipse illustrates an umbilic combination of lens parameters, at which the tangential and radial critical curves come into contact along the minor axis of the $a\approx0.103$ ellipse.

Unlike in the zero-eccentricity case, the shear in the lower three panels of Fig.~\ref{fig:nfw-convergence-shear-e} diverges at the halo center. The rate of its divergence is slower than that of the convergence $\kappa$, but it increases with increasing eccentricity as indicated by the rising slope of the magenta band. The shear can be approximated for $a\to 0$ and $e>0$ by the expression in Eq.~(\ref{eq:gamma-center}). The convergence near the halo center can be approximated by
\begin{equation}
\kappa(a)=-2\,\kappa_{\text{s}}\,(\,\ln{\frac{a}{2}}+1+\frac{3}{2}\,a^2\,\ln{\frac{a}{2}} +\frac{5}{4}\,a^2\,) +\mathcal{O}(a^4\,\ln{a})\,,
\label{eq:kappa-center}
\end{equation}
as shown by \cite{karamazov_etal21}\footnote{The expression presented here includes a correction in the third-order term.}. The two approximations can be combined to yield the dependence of the shear on the convergence and position along the ellipse $\theta$ for $a\to0$:
\begin{equation}
\gamma\simeq\frac{\left(1-\sqrt{1-e^2}\,\right)^2}{e^2}\,\kappa(a) -4\,\kappa_\text{s}\,\frac{\sqrt{1-e^2}}{e^2}\,\ln{\frac{1+\sqrt{1-e^2}}{2\,\sqrt{1-e^2\,\sin^2{\theta}}}}\,.
\label{eq:gamma-vs-kappa-center}
\end{equation}
The factor preceding the convergence explains the slope of the band in Fig.~\ref{fig:nfw-convergence-shear-e}: 0 for $e\to 0$, and 1 for $e\to 1$. The second term confirms that maximum shear $\gamma_\text{\tiny MAX}$ occurs along the major axis ($\theta=0$) and minimum shear $\gamma_\text{\tiny MIN}$ occurs along the minor axis ($\theta=\pi/2$). The vertical width of the band (at a constant $a$) is well approximated for $a\to0$ and $e>0$ by
\begin{equation}
\gamma_\text{\tiny MAX}-\gamma_\text{\tiny MIN}\simeq -2\,\kappa_{\text{s}}\,\frac{\sqrt{1-e^2}}{e^2}\,\ln{(1-e^2)}\,,
\label{eq:gamma-width-center}
\end{equation}
a slowly varying function yielding a width of $2\,\kappa_{\text{s}}$ for low eccentricities and decreasing significantly only for eccentricities $e\gtrsim0.9$.

The CS diagrams in Fig.~\ref{fig:nfw-convergence-shear-e} are all plotted for convergence parameter $\kappa_{\text{s}}=0.25$. \cite{karamazov_heyrovsky22} originally plotted the CS diagram for a spherical NFW lens with the very similar value $\kappa_{\text{s}}=0.239$ (compare their Fig. 4 with the top panel of Fig.~\ref{fig:nfw-convergence-shear-e}). The structure and properties of the CS diagrams for any particular value of $\kappa_{\text{s}}$ can be directly inferred from the diagrams for any other value, such as those in Fig.~\ref{fig:nfw-convergence-shear-e}. The expressions for the convergence and shear are directly proportional to $\kappa_{\text{s}}$, so that $\kappa(\boldsymbol x)/\kappa_{\text{s}}$ and $\gamma(\boldsymbol x)/\kappa_{\text{s}}$ are independent of $\kappa_{\text{s}}$. In terms of the CS diagram this means that increasing or decreasing $\kappa_{\text{s}}$ amounts to expanding the plot in the diagram radially away from the origin, or shrinking the plot radially toward the origin, respectively.

We illustrate this property in Fig.~\ref{fig:nfw-convergence-shear-ks}, for simplicity on the example of the circularly symmetric case from the top panel of Fig.~\ref{fig:nfw-convergence-shear-e}. The five green curves correspond to convergence-parameter values $\kappa_{\text{s}}\in\{0.125, 0.25, 0.5, 1, 2\}$ as marked along the right side of the plot (the same values were used in Figs.~\ref{fig:nfw-dimensions-critical-curves} and \ref{fig:nfw-dimensions-caustics}). For $\kappa_{\text{s}}=0.125$ all points (including their $a$ values) are moved two times closer to the origin with respect to the $\kappa_{\text{s}}=0.25$ case, for $\kappa_{\text{s}}=0.5$ all points are moved two times further from the origin, etc. For orientation, we connected points with $a=1$ on all the curves by the top dashed orange line, and added two more dashed orange lines similarly for all $a=0.1$ and all $a=0.01$ points.

The changes of imaging properties for different $\kappa_{\text{s}}$ in the circularly symmetric case can be interpreted from Fig.~\ref{fig:nfw-convergence-shear-ks} analogously to the walk-through performed for $\kappa_{\text{s}}=0.25$ above. Here we point out just two examples of features readily seen from the figure. First, as $\kappa_{\text{s}}$ increases, the minimum absolute magnification of negative-parity images decreases, so that for $\kappa_{\text{s}}=2$ we already see a zone with demagnified images ($|\mu|<1$) between the tangential and radial critical curves. Second, the crossing of the dashed line $\gamma=\kappa$ always occurs at $a=x_{\text{R}\infty}\approx1.32$, independent of $\kappa_{\text{s}}$. As discussed above, at larger radii $\lambda_\perp<1$ so that images are compressed perpendicular to the phase, narrower than the source. Since the critical-curve radii increase with increasing $\kappa_{\text{s}}$ (see Fig.~\ref{fig:nfw-radii}), for $\kappa_s=2$ this zone reaches well under the tangential critical curve, as seen from Fig.~\ref{fig:nfw-convergence-shear-ks}.

Changing $\kappa_{\text{s}}$ in the elliptical case (lower three panels in Fig.~\ref{fig:nfw-convergence-shear-e}) analogously amounts to expanding or shrinking the entire magenta area from or to the origin of the CS diagram, respectively. For example, reducing the convergence parameter in the second panel ${\sim}2.35$ times would lead to the umbilic configuration for $e=0.6$. Equation~(\ref{eq:gamma-vs-kappa-center}) confirms that the slope of the inner-halo band in the CS diagram does not change with $\kappa_s$, while the width of the band scales with $\kappa_s$ linearly, as seen from Eq.~(\ref{eq:gamma-width-center}).

\subsubsection{Orientation}
\label{sec:images-orientation}

Here we use the term ``image orientation'' as a shorthand for the orientation of the image distortion, or, even more specifically, for the orientation of the major axis of an elliptical image of a small circular source\footnote{Based on its equality to the phase in weak lensing, \cite{karamazov_heyrovsky22} use the term ``weak phase'' for the same quantity.}. As such the orientation is primarily defined by the lens phase $\phi$, described in Sect.~\ref{sec:shear} and illustrated in Figs.~\ref{fig:shear} and \ref{fig:shear-zoom}. The secondary factor is the convergence: outside the unit-convergence ellipse where $\kappa<1$, the image orientation is equal to the phase; inside the unit-convergence ellipse where $\kappa>1$, the image orientation is perpendicular to the phase. We quantify the orientation by the polar angle in the lens plane, measured counter-clockwise from the major axis of the mass distribution modulo $\pi$, using the corresponding value from the interval $[-\pi/2,\pi/2]$.

\begin{figure*}
\centering
\includegraphics[width=18 cm]{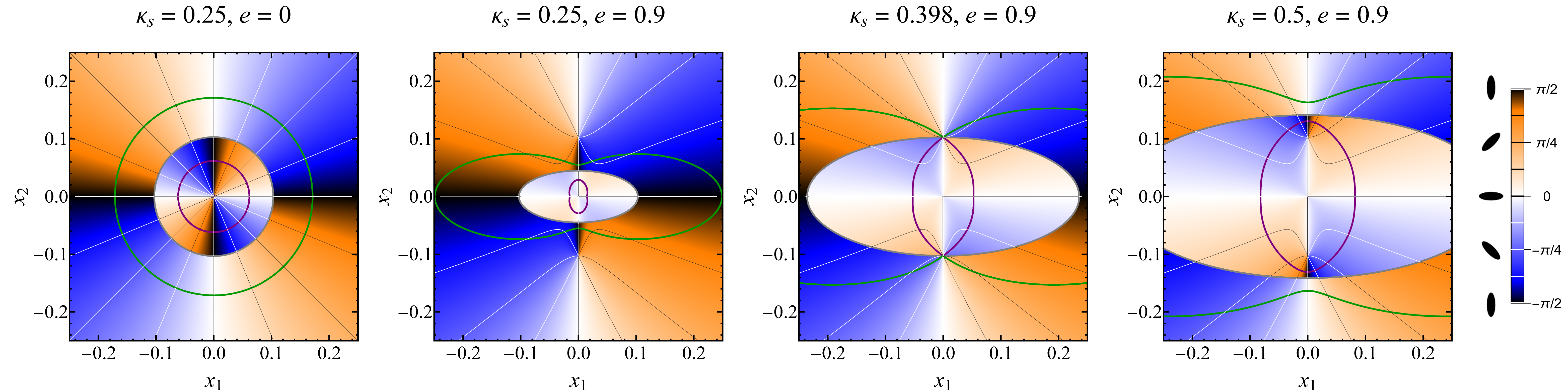}
\caption{Orientation of image distortion by the ellipsoidal NFW lens, presented in color maps with contours. The parameter combinations are indicated above the panels: the first panel shows the circularly symmetric case with convergence parameter $\kappa_\text{s}=0.25$; the three right panels show the $e=0.9$ elliptical case and a sub-umbilic, umbilic, and super-umbilic value of $\kappa_\text{s}$. The respective tangential (green) and radial (purple) critical curves are plotted in each panel. The color bar at the right side indicates the orientation angles, the contour values spaced by $\pi/8$, plus corresponding tilted-ellipse icons for visual guidance.
\label{fig:nfw-image-orientation}}
\end{figure*}

We present four examples of image-orientation maps in Fig.~\ref{fig:nfw-image-orientation}: one with circular symmetry and convergence parameter $\kappa_{\text{s}}=0.25$, and three with eccentricity $e=0.9$ and convergence parameters $\kappa_{\text{s}}\in\{0.25, 0.398, 0.5\}$. These combinations are marked in the parameter-space maps in Fig.~\ref{fig:parameter-space} by the cyan dotted circles. The circularly symmetric case in the first panel of Fig.~\ref{fig:nfw-image-orientation} illustrates the $\pi/2$ switch along the unit-convergence circle (marked in gray). Outside the circle horizontally oriented images appear along the $x_2$ axis, vertically oriented images along the $x_1$ axis, and along the green tangential critical curve images are always oriented tangentially. Inside the circle horizontally oriented images appear along the $x_1$ axis, vertically oriented images along the $x_2$ axis, and along the purple radial critical curve images are always oriented radially. The radial orientation is maintained into the halo center, where all constant-orientation contours meet.

The second panel of Fig.~\ref{fig:nfw-image-orientation} illustrates the case when the convergence parameter is lower than the umbilic value of $\kappa_{\text{s}}$ at the given eccentricity. In addition to the gray unit-convergence ellipse, the two zero-shear points outside the ellipse along the minor axis clearly play an important role here. Outside the unit-convergence ellipse horizontally oriented images appear only outside the zero-shear points along the minor axis, while vertically oriented images appear along the major axis plus along the minor axis inside the zero-shear points. Along the tangential critical curve image orientation deviates by barely more than $\pm\pi/8$ from the vertical, so that while the images are oriented tangentially near the major axis, near the minor axis they are oriented radially. In either half-plane, all constant-orientation contours meet at the zero-shear points, but the orientation changes by $\pi/2$ when crossing the point instead of staying the same as in the case of the halo center in the first panel. Inside the unit-convergence ellipse horizontally oriented images appear along both the major and the minor axis, vertically oriented images are entirely absent here and the image orientation deviates by less than $\pm\pi/8$ from the horizontal. Along the radial critical curve the images are oriented radially near the major axis, while near the minor axis they are oriented tangentially.

The third panel of Fig.~\ref{fig:nfw-image-orientation} illustrates the case when the convergence parameter is equal to the umbilic value of $\kappa_{\text{s}}$ at the given eccentricity. Here the zero-shear points lie on the unit-convergence ellipse and thus they form the umbilic points. Outside the unit-convergence ellipse horizontally oriented images appear along the minor axis, while vertically oriented images appear along the major axis. Along the tangential critical curve image orientation deviates by barely more than $\pm\pi/4$ from the vertical, so that while they are oriented tangentially near the major axis, when approaching the minor axis they are oriented nearly perpendicularly to the critical curve. Inside the unit-convergence ellipse horizontally oriented images appear along both the major and the minor axis, vertically oriented images are entirely absent here and the image orientation deviates by less than $\pm\pi/4$ from the horizontal. Along the radial critical curve the images are oriented radially near the major axis, and when approaching the minor axis they are oriented nearly perpendicularly to the critical curve. In either half-plane, constant-orientation contours with orientation in the interval $[-\pi/4,\pi/4]$ meet at the umbilic points, and there is no change in orientation when crossing the points. We note that there are no vertically oriented images in the vicinity of the umbilic points, all images deviate by no more than $\pm\pi/4$ from the horizontal.

The fourth panel of Fig.~\ref{fig:nfw-image-orientation} illustrates the case when the convergence parameter is higher than the umbilic value of $\kappa_{\text{s}}$ at the given eccentricity. In such a case the two zero-shear points appear inside the unit-convergence ellipse. Outside the ellipse horizontally oriented images appear along the minor axis, and vertically oriented images appear along the major axis. Along the tangential critical curve images are oriented tangentially near the major axis as well as near the minor axis, although between these two positions image orientation deviates by more than $\pi/4$ from the critical-curve tangent. Inside the unit-convergence ellipse horizontally oriented images appear along the major axis as well as along the minor axis inside the zero-shear points, vertically oriented images appear only along the minor-axis segments outside the zero-shear points. Only in their vicinity does the orientation deviate by more than $\pm\pi/4$ from the horizontal. Along the radial critical curve the images are oriented radially near the major axis as well as near the minor axis. In either half-plane, all constant-orientation contours meet at the zero-shear points, with the orientation changing by $\pi/2$ when crossing either point.

At low eccentricities, the zero-shear points lie very close to the halo center, as shown in Fig.~\ref{fig:zero-shear-point}. Even for very small values of $\kappa_{\text{s}}$ such a lens is in the regime of the fourth panel, but the adjacent pair of zero-shear points generates an image-orientation map resembling a distorted version of the circular case shown in the first panel.

\subsubsection{Configuration examples}
\label{sec:images-examples}

\begin{figure*}
\centering
\includegraphics[width=18 cm]{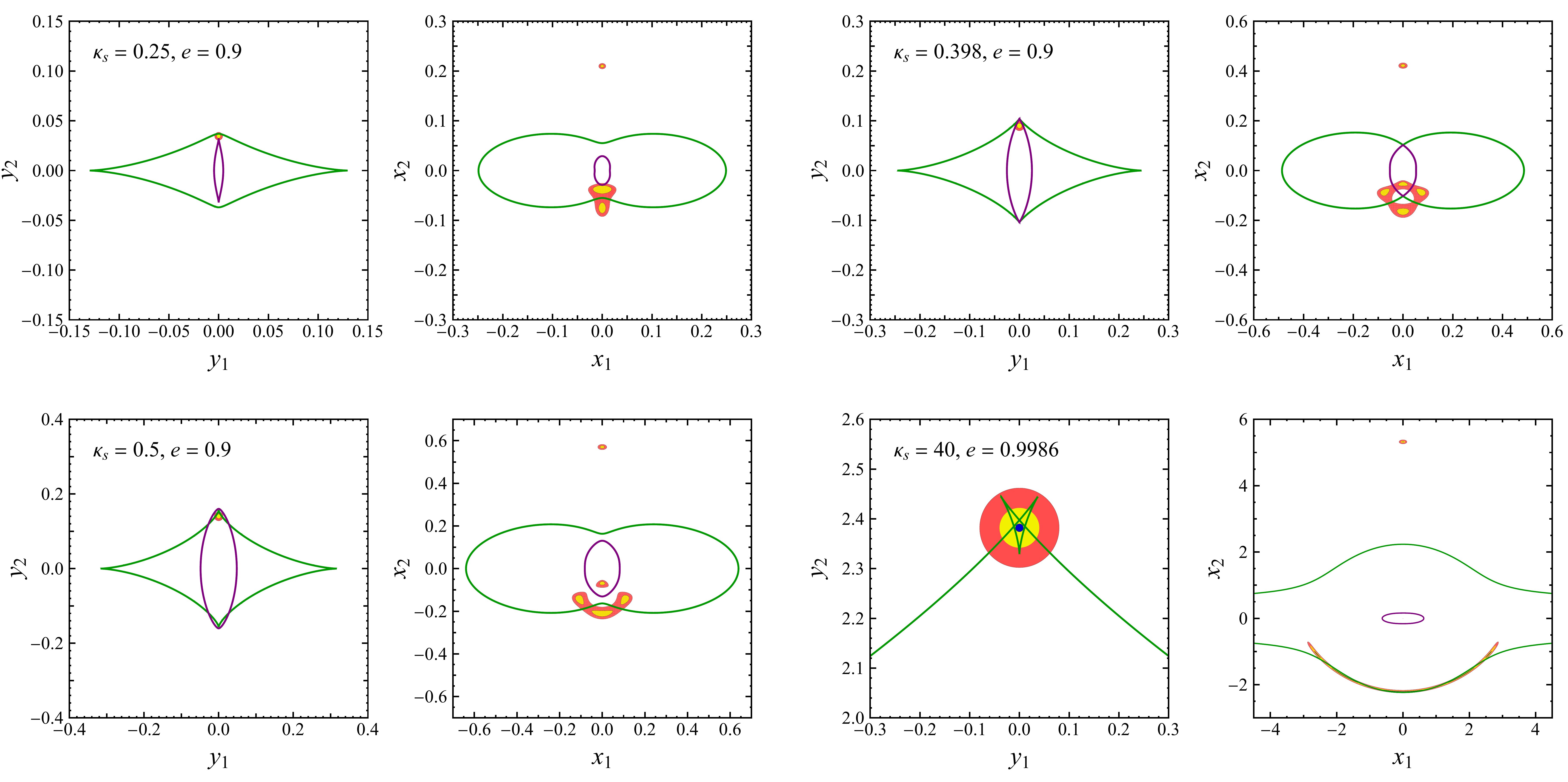}
\caption{Examples of image configurations formed by four ellipsoidal NFW lenses. Each of the four pairs of panels shows the position of the source with respect to the caustics (left), and the positions of the images with respect to the critical curves (right). Images are plotted for two source sizes (red, yellow); in the bottom right example a third size is added (blue). Lens parameters $(\kappa_\text{s},e)$ for top left example: $(0.25, 0.9)$; top right: $(0.398, 0.9)$; bottom left: $(0.5, 0.9)$; bottom right: $(40, 0.9986)$.
\label{fig:nfw-image-examples}}
\end{figure*}

In the previous parts of Sect.~\ref{sec:images} we illustrated the diversity of the imaging properties of the ellipsoidal NFW lens, their variation in the lens plane and within the parameter space of the lens. Rather than attempting to present an exhaustive overview of possible image configurations, we show in Fig.~\ref{fig:nfw-image-examples} four specific examples of images formed by lenses with four different $(\kappa_{\text{s}}, e)$ parameter combinations. In each pair of panels, the left one shows the source plane, indicating the position of the source (red and yellow circles for two source sizes) with respect to the caustics. The right panel shows the lens plane, indicating the positions of the images (colored by the corresponding source size) with respect to the critical curves.

In the top left example, the lens parameters $(\kappa_{\text{s}}, e)=(0.25,0.9)$ correspond to the fourth CS diagram in Fig.~\ref{fig:nfw-convergence-shear-e} and the second image-orientation map in Fig.~\ref{fig:nfw-image-orientation}. The circular sources with radii 0.004 (red) and 0.002 (yellow) are centered at the point $(y_1,y_2)=(0,0.0345)$, so that the yellow source lies fully between the radial and tangential caustics while the red source crosses both caustics. Due to its position, the yellow source has three full separate images. The top-most part of the red source lying outside the tangential caustic has one image, the part between the caustics has three, and the lowest part lying within the radial caustic has five images. One full image of the red source lies along the minor axis above the critical curves, while all other images are merged into one large bell-shaped macro-image under the halo center. Worth noticing is the horizontal orientation of the middle yellow image, which is tangent to the radial critical curve, and the vertical orientation of the bottom yellow image, which is perpendicular to the tangential critical curve. Both are in agreement with the second image-orientation map in Fig.~\ref{fig:nfw-image-orientation}, as discussed in Sect.~\ref{sec:images-orientation}. Similar distortions in the corresponding parts of the red macro-image lead to its peculiar shape.

In the top right example in Fig.~\ref{fig:nfw-image-examples}, the lens parameters $(\kappa_{\text{s}}, e)=(0.398,0.9)$ represent the umbilic combination seen also in the third image-orientation map of Fig.~\ref{fig:nfw-image-orientation}. The circular sources with radii 0.01 (red) and 0.005 (yellow) are centered at the point $(y_1,y_2)=(0,0.09)$, so that the yellow source lies fully inside both caustics while the red source crosses both caustics. The yellow source has five full separate images. The narrow parts of the top edge of the red source lying outside the tangential caustic have one image, the narrow parts between the caustics have three, and the rest of the red source lying inside the radial caustic has five images. One full image of the red source lies along the minor axis above the critical curves, while all other images are merged into one large bumpy ring-like macro-image around the umbilic point (rather than around the halo center). Worth noticing is the horizontal orientation of the yellow images above and below the umbilic point and the corresponding dominant horizontal distortion of the top and bottom parts of the red ring-like image. The left and right ``bumps'' on the ring correspond to the abrupt change in orientation along the unit-convergence ellipse, as seen in the third image-orientation map in Fig.~\ref{fig:nfw-image-orientation}.

In the bottom left example in Fig.~\ref{fig:nfw-image-examples}, the lens parameters $(\kappa_{\text{s}}, e)=(0.5,0.9)$ correspond to the fourth image-orientation map in Fig.~\ref{fig:nfw-image-orientation}. The circular sources with radii 0.012 (red) and 0.006 (yellow) are centered at the point $(y_1,y_2)=(0,0.14)$, so that the yellow source lies fully inside both caustics while the red source lies inside the radial caustic but crosses the tangential caustic. The yellow source has five full separate images. The narrow parts of the top edge of the red source lying outside the tangential caustic have three images, and the rest of the red source lying inside both caustics has five images. One full image of the red source lies along the minor axis above the critical curves, another full image lies inside the radial critical curve, and all other images are merged into one large boomerang-like macro-image. Worth noticing is the horizontal orientation of all yellow images along the minor axis, with the lowest one oriented tangentially to the tangential critical curve, and the near-perpendicular orientation of the two side images. The top of the red boomerang-like macro-image is distorted and flattened due to its location inside the unit-convergence ellipse, as seen in the fourth image-orientation map in Fig.~\ref{fig:nfw-image-orientation}. In comparison with the top left example, the image configuration is dramatically different even though the critical-curve geometry is superficially similar.

We include the bottom right example in Fig.~\ref{fig:nfw-image-examples} to demonstrate the special seven-image configuration arising when the tangential caustic has a pair of butterflies along the minor axis (see the bottom right panel of Fig.~\ref{fig:gallery-special} and Sect.~\ref{sec:parameter-space-intermediate}). For purposes of illustration, we use the same eccentricity and a higher convergence parameter, $(\kappa_{\text{s}}, e)=(40,0.9986)$. Instead of plotting the full extent of the caustics, we include only the part of the tangential caustic near the top butterfly noting that the entire plotted area lies inside the radial caustic. In the right panel we omit the outer parts of the tangential critical curve, which extend along the major axis out to $x_1\approx\pm12.4$. In addition to the two circular sources with radii 0.08 (red) and 0.04 (yellow) we include here a third small blue source with radius 0.008. All are centered at the point $(y_1,y_2)=(0,2.382)$. The blue source lies fully inside the lower part of the butterfly feature, the red source encloses the entire butterfly, and the intermediate yellow source encloses its main part without the three cusps. The blue source has seven full separate images: one along the minor axis above the tangential critical curve, a tiny image near the center of the radial critical curve, and five images strung out along the lower part of the tangential critical curve alternating on its sides (three inside, two outside). Moving the blue source across any of the four different segments of the tangential caustic that surround it would lead to the disappearance of one of the four pairs of neighboring images lying on opposite sides of the tangential critical curve.

The top, left, and right parts of the yellow source lying outside the tangential caustic have three images, the parts in the left and right wings and the lower left and lower right parts inside the tangential caustic but outside the butterfly have five images, and the central part adjacent to the blue source has seven images. The red source can be dissected in a similar manner. For the yellow as well as the red sources, one full image lies along the minor axis above the critical curves, another tiny full image lies inside the radial critical curve, and all other images are merged into one long arc-like macro-image with its center of curvature lying on the minor axis above the halo center.

\section{Discussion}
\label{sec:discussion}

\subsection{Relevance of high-eccentricity results}
\label{sec:relevance-high-eccentricity}

We first return to the seven-image regions described and illustrated above in Sect.~\ref{sec:images-examples}. While their existence in the source plane of the ellipsoidal NFW lens is unexpected and very interesting in the context of simple lens models, their observational significance is rather limited. First, they occur at a combination of a very high eccentricity $e$ and a very high convergence parameter $\kappa_\text{s}$. Second, even though the butterfly structures increase in size for higher values of $\kappa_\text{s}$, our tests show that at $\kappa_\text{s}=100$ their combined vertical extent reaches at most $12\%$ of the vertical width of the tangential caustic (at most $7.2\%$ for the seven-image regions of the butterflies). Due to the small extent of these regions, a lensed source would rarely fit in entirely. Two such possibilities are worth mentioning: a perfectly aligned quasar lensed by a galaxy-scale NFW halo; and an exploding supernova in a galaxy lensed by a cluster- or galaxy-scale NFW halo. For larger sources extending across the seven-image region, high-resolution imaging resolving fine structure of the source would be necessary to distinguish the individual partial images stretched along the arc.

A final point worth mentioning is the rarity of butterflies. Their occurrence requires near-perfect symmetry of the mass distribution about the butterfly axis (in our case the minor axis of the projected ellipse). Even with a highly symmetric lens, in most astrophysical settings one may expect perturbation by additional shear -- due to lens substructure, due to companion lenses, or due to larger-scale mass distributions. This influence may easily break the butterfly structure on the caustic, typically into a cusp and an adjacent swallow-tail structure \citep[e.g.,][]{schneider_etal92}. Nevertheless, the tail of the swallow tail still encloses a seven-image region in the source plane. While such perturbations may be detrimental to butterflies, they may still preserve regions with seven-image configurations.

The high-eccentricity and high-convergence-parameter region of parameter space mapped in Fig.~\ref{fig:parameter-space-high-kappa}, described in Sect.~\ref{sec:parameter-space-intermediate} and Appendix~\ref{sec:Appendix-parameter-space-high}, is the domain not only of the seven-image lens configurations, but also of bulge-shaped critical curves and other peculiarities not occurring at lower eccentricities. These properties of the ellipsoidal NFW lens could not be discovered from low-eccentricity approximations of the lens model, such as those derived from an elliptically distorted lens potential \citep[e.g.,][]{gomer_etal23}. On the other hand, the astrophysical relevance of such parameter values should be considered.

Regarding the values of the convergence parameter $\kappa_\text{s}$ in Fig.~\ref{fig:parameter-space-high-kappa}, data from \cite{merten_etal15} indicate that values substantially larger than $\kappa_\text{s}\approx1$ are too high for galaxy-cluster-scale halos. However, data from \cite{asano00} indicate that values in the range of low multiples of $10$ are relevant for galaxy-scale halos, so that at least the left part of the $\kappa_\text{s}$ range of Fig.~\ref{fig:parameter-space-high-kappa} would be relevant for them. Compact dwarf-galaxy halos (such as lens-galaxy satellites) could possibly reach up to a factor of 10 higher in $\kappa_\text{s}$, however, it would be a challenge to resolve details of imaging at their small scale.

While the eccentricity range in Fig.~\ref{fig:parameter-space-high-kappa} seems extreme ($e\gtrapprox0.937$), the conditions are less dramatic when viewed in terms of the corresponding ellipticity ($1-\sqrt{1-e^2}>0.65$) or axis ratio ($\sqrt{1-e^2}<0.35$). \cite{jing_suto02} studied the triaxial shapes of dark-matter halos arising from cosmological simulations and found that the major-to-minor axis-ratio distribution was well described by a Gaussian with mean 0.54 and standard deviation 0.113. Axis ratios lower than $0.35$ thus represent a non-negligible tail of the distribution. In principle, halos with such projected axis ratios could also represent flat mass distributions viewed (nearly) edge-on, or linear filament-like mass distributions.

The combined requirements on the convergence parameter and eccentricity are likely to deem all but the lower left region of the parameter space in Fig.~\ref{fig:parameter-space-high-kappa} astrophysically irrelevant. Nevertheless, the results presented in the figure do illustrate the complex nature of the simple underlying lens model. The techniques developed for identifying the parameter-space boundaries can be readily used in the analysis of any other lens model.

\subsection{Role of source redshift}
\label{sec:source-redshift}

The exploration of the ellipsoidal NFW lens in Sect.~\ref{sec:eNFW} was presented as a study of the properties of the model for a given combination of lens parameters. In the case of a galaxy-scale halo which typically lenses a single source (or more exactly, there is a single source plane), this approach is well substantiated. In the case of a galaxy-cluster-scale halo, there are typically many sources in many source planes -- most of them lensed weakly, some of them lensed strongly. Since the source redshift enters the lensing scenario through the critical surface density $\Sigma_{\text{cr}}$ in Eq.~(\ref{eq:critical-density}), lensing by the same halo is described by a range of parameter combinations with different values of the convergence parameter for the different source planes.

The convergence parameter $\kappa_\text{s}$ is directly proportional to the angular-diameter-distance ratio $D_\text{ls}/D_\text{s}$, which increases with source redshift at a rate that depends on the cosmological model \citep[see, e.g.,][]{umetsu20}. While some of the presented results such as those in Figs.~\ref{fig:deflection-vectorplot}--\ref{fig:zero-shear-point} or \ref{fig:nfw-radial-limiting-curve} are independent of $\kappa_\text{s}$ and thus of the source redshift, other are presented for specific values of $\kappa_\text{s}$, such as those in Figs.~\ref{fig:gallery-basic}, \ref{fig:gallery-special}, \ref{fig:nfw-convergence-shear-e}, or \ref{fig:nfw-image-orientation}--\ref{fig:nfw-image-examples}. In the parameter-space maps in Figs.~\ref{fig:parameter-space} and \ref{fig:parameter-space-high-kappa}, different source-plane distances behind a given lens correspond to points along horizontal line segments. These may intersect the plotted boundaries, leading to significant changes in the critical curves and caustics as a function of source redshift.

The dependence of the convergence--shear diagrams on the source redshift is illustrated by the dependence on $\kappa_\text{s}$ in Fig.~\ref{fig:nfw-convergence-shear-ks}. The Jacobian-factor contour plots described in Appendix~\ref{sec:Appendix-image-plane} (with examples in Figs.~\ref{fig:contours-e9} and \ref{fig:contours}) present an excellent visualization of the variation of the critical curves with source redshift. Since different values of $\kappa_\text{s}$ correspond to different nested contours in the same plot, the critical curves for a given range of source redshifts can be directly seen as the corresponding sequence of neighboring contours. For a given lens at a redshift $z_\text{l}$, the contours could just as well be labeled by the source redshift $z_\text{s}$, with $z_\text{s}=z_\text{l}$ corresponding to the point at the origin of the plots and higher $z_\text{s}$ corresponding to larger contours.

\subsection{Comparison with non-singular isothermal ellipsoid}
\label{sec:comparison-nie}

The only other fully ellipsoidal two-parameter gravitational lens model for which a detailed analysis of critical curves and caustics is available, is the non-singular isothermal ellipsoid \citep[NIE,][]{kassiola_kovner93,kormann_etal94} which includes the singular isothermal ellipsoid as its limiting case. In the \cite{kormann_etal94} parametrization, the model is described by the dimensionless core semi-minor axis $b_\text{c}$ and the axis ratio $\sqrt{1-e^2}$. The basic sequence of critical curves and caustics of the ellipsoidal NFW model with eccentricity increasing from 0 at a fixed convergence parameter $\kappa_\text{s}$, as shown in Fig.~\ref{fig:gallery-basic}, is similar to the beginning of the corresponding sequence of the NIE model with axis ratio decreasing from 1 at a fixed core parameter $b_\text{c}$ \citep[see Fig. 9 of][]{kormann_etal94}.

However, there are four notable differences between the models in this comparison. First, the NIE exhibits three additional transitions at lower values of the axis ratio (higher eccentricities): disappearance of the radial caustic and critical curve in a lips metamorphosis, disappearance of inflection points on the tangential critical curve in an ``inverse'' peanut transition, and the disappearance of the tangential caustic and critical curve in a lips metamorphosis. Second, for core parameters $b_\text{c}\geq0.5$ the NIE model has no critical curves or caustics (the Jacobian never reaches zero), while the ellipsoidal NFW model has critical curves and caustics for an arbitrary combination of its parameters. Third, the occurrence of critical-curve inflection points is different in the NIE model: the radial critical curve is always globally convex, without any inflection point, while on the tangential critical curve the inflection points disappear at low axis ratios. Fourth, the parameter-space boundaries as well as the sizes and detailed shapes of the critical curves and caustics are different due to the different functional form of the models, which also affects their lensing and multiple-imaging probabilities.

The comparison in the previous paragraph is relevant to the low-convergence-parameter regime of the ellipsoidal NFW model discussed in Sect.~\ref{sec:parameter-space-low} and mapped in the left panel of Fig.~\ref{fig:parameter-space}. The additional boundaries and lensing regimes appearing at higher $\kappa_\text{s}$, discussed in Sect.~\ref{sec:parameter-space-intermediate} and Appendix~\ref{sec:Appendix-parameter-space-high} and mapped in the right panel of Fig.~\ref{fig:parameter-space} and in Fig.~\ref{fig:parameter-space-high-kappa}, do not occur in the NIE model. A more detailed comparison of these or other ellipsoidal models is beyond the scope of the present work.

\subsection{Possible modifications of the model}
\label{sec:modifications}

One drawback of the NFW profile is the divergence of its mass at large radii. As long as the lensing occurs within a couple of scale radii from the halo center, this would not have a substantial impact on the results. However, if the model is used for describing halos of galaxies within a galaxy-cluster halo, the model should be ideally truncated at a certain distance to avoid introducing artifacts due to its increasing mass. Work on such a truncated version of the ellipsoidal NFW model is currently underway.

While the derived analytic formulae for the light deflection by the ellipsoidal NFW model present a large step toward a more astrophysically relevant model from the spherical or ellipsoidal-potential NFW models, structure-formation simulations indicate the potential further steps toward more realistic models. One would be the possible introduction of a core to eliminate the central density divergence \citep[e.g.,][]{salucci19}. Another would be to relax the homoeoidal symmetry of the model: simulations of dark matter halos such as those by \cite{jing_suto02} indicate that their nested constant-density surfaces do not preserve the same ellipsoidal shape. Rather, the axis ratios (here parameterized by the eccentricities $e_1$ and $e_2$) and axis orientations tend to vary from the center outward. However, such refinements would most probably stay in the realm of numerical modeling and their verification would likely require uniquely precise observational data.

\section{Summary}
\label{sec:summary}

Gravitational lenses with a triaxially ellipsoidal mass distribution such that the density is constant in concentric homoeoidal shells can be described by an elliptically symmetric surface density profile. We derived its eccentricity, orientation, and integral representation in Sect.~\ref{sec:triaxial} and Appendix~\ref{sec:Appendix-transformation} for an arbitrary triaxial lens with a general three-dimensional orientation. The deflection of a light ray by a lens with a surface density profile obtained from Eq.~(\ref{eq:surface_density-final}) can then be computed using the \cite{bourassa_kantowski75} formalism summarized in Sect.~\ref{sec:Bourassa}.

We applied the described approach in Sect.~\ref{sec:eNFW} to study gravitational lensing by a triaxially ellipsoidal dark-matter halo with a NFW density profile \citep{navarro_etal96}. The form of the convergence profile given by Eq.~(\ref{eq:convergence}) as a function of semi-major axis is analogous to the convergence of a spherical NFW halo. However, the convergence parameter depends on the halo shape and orientation, as seen in Eq.~(\ref{eq:kappa_s}). The main results of the present work are the analytic formulae derived in Sect.~\ref{sec:alpha}. The deflection angle for the studied ellipsoidal NFW halo is given by Eq.~(\ref{eq:NFW-scattering-function}) in terms of the complex scattering function $I$. The components of the reduced deflection angle $\boldsymbol \alpha_0$ are presented in Eqs.~(\ref{eq:deflection-1})--(\ref{eq:deflection-2}) in terms of real functions. These exact closed-form expressions are valid for arbitrary eccentricity $e\in[0,1]$ of the surface-density profile.

The availability of analytic expressions for the deflection angle enables a detailed study of the lensing properties of the model, carried out in the following sections of the article. The character of the shear, its components, and the phase is explored in Sect.~\ref{sec:shear}. The properties of the components near the halo center give rise to zero-shear points positioned symmetrically above and below the center on the minor axis of the mass distribution, as illustrated in Fig.~\ref{fig:shear-zoom}. These points exist for arbitrary non-zero eccentricity: while for $e\lesssim0.5$ they lie exponentially close to the halo center, for $e\gtrsim0.5$ they deviate noticeably, as shown in Fig.~\ref{fig:zero-shear-point}, reaching as far as $\pm0.13$ from the center in units of the projected scale semi-major axis of the NFW halo. At these points, images are undistorted and the phase switches from horizontal to vertical when progressing inward along the minor axis. In the special case when the unit-convergence contour passes through them, they form umbilic points.

The critical curves and caustics, key characteristics of lens models from the observational perspective, are studied in Sect.~\ref{sec:curves}. Their typical geometries illustrated in Fig.~\ref{fig:gallery-basic} resemble those occurring in the non-singular isothermal ellipsoid lens model \citep[e.g.,][]{kassiola_kovner93,kormann_etal94}, with notable differences pointed out in Sect.~\ref{sec:comparison-nie}. The parameter-space map indicating the occurrence of different critical-curve and caustic geometries as a function of the convergence parameter $\kappa_\text{s}$ and eccentricity $e$ in the left panel of Fig.~\ref{fig:parameter-space} shows four boundaries at low $\kappa_\text{s}$. In order from lowest to highest eccentricity, these correspond to the following transitions: cusp piercing (tangential caustic piercing radial caustic), tangential peanut (tangential critical curve turning concave at minor axis), radial peanut (radial critical curve turning concave at major axis), hyperbolic umbilic (tangential and radial components touching at minor axis).

Distinct new critical-curve and caustic features appear at higher $\kappa_\text{s}$: tangential critical curves may be centrally bulged (top row of Fig.~\ref{fig:gallery-special}); tangential caustics may have butterfly tangles along the minor axis (bottom right panel of Fig.~\ref{fig:gallery-special}); radial critical curves may turn concave at the minor axis. The corresponding additional parameter-space boundaries are mapped in the right panel of Fig.~\ref{fig:parameter-space} and in Fig.~\ref{fig:parameter-space-high-kappa}. Most of these new features appear at high eccentricities; we discuss their astrophysical relevance in Sect.~\ref{sec:relevance-high-eccentricity}. We identified and tracked the new features in parameter space using the new technique of image-plane analysis of critical curves and caustics, described in Appendix~\ref{sec:Appendix-image-plane}. We showed that plotting the contours of the tangential or radial Jacobian factor is equivalent to plotting the tangential or radial critical curves, respectively, for different values of the convergence parameter $\kappa_\text{s}$. For a given eccentricity, it takes just two plots to convey the full variation of the critical curves with $\kappa_\text{s}$: one for the tangential and another for the radial critical curves. Adding special curves to these plots helps keep track of particular characteristics: we introduced the inflection curve identifying the location of inflection points, and the cusp curve identifying the locations of cusp images along the critical curves. The technique also helped us to determine the existence of a radial limiting curve (marked orange in the bottom row of Fig.~\ref{fig:contours}), a finite outer limiting curve approached by radial critical curves as $\kappa_\text{s}\to\infty$.

The variation of the linear dimensions and enclosed areas of the critical curves and caustics with eccentricity $e$ is demonstrated in Figs.~\ref{fig:nfw-dimensions-critical-curves} and \ref{fig:nfw-dimensions-caustics}, respectively. A comparison of the top and bottom rows shows there is a substantial change from a low-convergence to a higher-convergence regime. For low values of $\kappa_\text{s}$: with increasing $e$ the tangential critical curve grows prominently horizontally and its area initially declines before increasing to a high-eccentricity peak; the radial critical curve is elongated vertically and its vertical extent initially grows with $e$ before reaching the umbilic transition; the tangential caustic has a larger area than the radial caustic at all but the lowest eccentricities; there is a narrow eccentricity interval in which the probability of five-image configurations just exceeds that of three-image configurations. For higher values of $\kappa_\text{s}$: with increasing $e$ the tangential critical curve shrinks vertically and horizontally and its area drops nearly linearly with increasing ellipticity; the radial critical curve is elongated horizontally and its dimensions decrease monotonically with $e$; the radial caustic has a larger area than the tangential caustic at all but the highest eccentricities; the probability of five-image configurations never exceeds that of three-image configurations. Radial and tangential critical curves and caustics exist for any lens-parameter combination $(\kappa_\text{s},e)$. In the limit $e\to 1$ the radial critical curve and caustic both shrink to the point at the origin, while the tangential critical curve and caustic both shrink to a line segment extending from the origin along the major axis out to points with convergence $\kappa=1/2$.

In Sect.~\ref{sec:images} we investigated the properties of individual images occurring in the image plane of the ellipsoidal NFW lens for different parameter combinations. First, we used the technique of convergence--shear diagrams \citep{karamazov_heyrovsky22} in Figs.~\ref{fig:nfw-convergence-shear-e} and \ref{fig:nfw-convergence-shear-ks} to identify the magnification and flattening of images occurring along concentric ellipses parameterized by their semi-major axis $a$ in the image plane. Second, we mapped the orientation of the image distortions in Fig.~\ref{fig:nfw-image-orientation}. Along the major axis, outside the unit-convergence ellipse images are always distorted vertically (i.e., tangentially) and inside they are always distorted horizontally (i.e., radially). However, along the minor axis the situation is more complicated: images are generally distorted horizontally except in the segments between the ellipse and the zero-shear points. As seen in the second panel of Fig.~\ref{fig:nfw-image-orientation}, images along the minor axis may be distorted radially at the tangential critical curve and tangentially at the radial critical curve.

In this work we developed and employed a range of methods to illustrate the properties of the ellipsoidal NFW lens. These techniques may be applied in a similar manner to elucidate the lensing characteristics of other models of gravitational lenses.

\begin{acknowledgements}

We thank Niek Wielders for insightful discussions and helpful comments on the manuscript.

\end{acknowledgements}

\begin{appendix}

\section{Transformations between reference frames}
\label{sec:Appendix-transformation}

We start with the principal-axes frame of the triaxially ellipsoidally symmetric mass distribution described by Eq.~(\ref{eq:density}), with coordinate axes $\hat{x}_1$, $\hat{x}_2$, and $\hat{x}_3$ parallel to its major, median, and minor axes, respectively. This frame has an arbitrary spatial orientation, as shown in Fig.~\ref{fig:ellipsoid} for a sample constant-density surface. We define the observer's frame with coordinate axes $x_1'$ and $x_2'$ in the plane of the sky and the line-of-sight axis $x_3'$ pointing toward the observer.

In order to transform from the observer's to the principal-axes coordinates we use Euler angles
describing three successive rotations: by $\varphi$ about $x_3'$ so that $x_1'$ rotates to $x_1''$ that is perpendicular to $\hat{x}_3$; by $\vartheta$ about $x_1''$ so that $x_3'$ rotates to $\hat{x}_3$; by $\psi$ about $\hat{x}_3$ so that $x_1''$ rotates to $\hat{x}_1$. In this definition angle $\vartheta$ thus corresponds to the angle between the minor axis and the line of sight. The product of the three corresponding rotation matrices yields the coordinate transformation equation
\begin{equation}
\begin{pmatrix}
  \hat{x}_1 \\
  \hat{x}_2 \\
  \hat{x}_3
\end{pmatrix}\,=\,
\begin{pmatrix}
  \text{c}\varphi\,\text{c}\psi-\text{s}\varphi\,\text{s}\psi\,\text{c}\vartheta & \text{s}\varphi\,\text{c}\psi+\text{c}\varphi\,\text{s}\psi\,\text{c}\vartheta & \text{s}\psi\,\text{s}\vartheta \\
  -\text{c}\varphi\,\text{s}\psi-\text{s}\varphi\,\text{c}\psi\,\text{c}\vartheta & -\text{s}\varphi\,\text{s}\psi+\text{c}\varphi\,\text{c}\psi\,\text{c}\vartheta & \text{c}\psi\,\text{s}\vartheta \\
  \text{s}\varphi\,\text{s}\vartheta & -\text{c}\varphi\,\text{s}\vartheta & \text{c}\vartheta
\end{pmatrix}\;
\begin{pmatrix}
  x_1' \\
  x_2' \\
  x_3'
\end{pmatrix}\,,
\label{eq:transformation}
\end{equation}
where we used abbreviated notation for the trigonometric functions: $\text{s}\rightarrow\sin$; $\text{c}\rightarrow\cos$.

We may now substitute the principal-frame coordinates from Eq.~(\ref{eq:transformation}) into Eq.~(\ref{eq:semi-major-3D}) for the semi-major axis $\hat{a}$ of a constant-density ellipsoid passing through a given point. In the latter equation, $e_1$ is the eccentricity of elliptical density contours in planes perpendicular to the minor axis $\hat{x}_3$, $e_2$ is the eccentricity of elliptical density contours in planes perpendicular to the median axis $\hat{x}_2$, and $0\leq e_1 \leq e_2 < 1$. The equation can be solved in terms of the observer's coordinates to yield the line-of-sight intersections with the ellipsoid at a given plane-of-the-sky position,
\begin{equation}
x_{3\pm}'=\frac{e_1^2(1-e_2^2)\,\text{s}\psi\,\text{c}\psi\, (x_1'\,\text{c}\varphi+x_2'\,\text{s}\varphi)+\text{c}\vartheta\, (e_2^2-e_1^2\,\text{c}^2\psi-e_1^2\,e_2^2\,\text{s}^2\psi)\, (x_2'\,\text{c}\varphi-x_1'\,\text{s}\varphi)} {\lambda_1\,\lambda_2}\, \text{s}\vartheta
\pm\sqrt{\frac{(1-e_1^2)(1-e_2^2)} {\lambda_1\,\lambda_2}}\, \sqrt{\hat{a}^2-\frac{\tilde{x}_1^2}{\lambda_1}-\frac{\tilde{x}_2^2}{\lambda_2}}\,.
\label{eq:intersections}
\end{equation}
Here we introduced the parameters
\begin{equation}
\lambda_{1,2}=\frac{1}{2}\,\left\{2-e_2^2\,\text{s}^2\vartheta-(1-\text{c}^2\psi\,\text{s}^2\vartheta)\,e_1^2\pm \sqrt{\left[e_2^2\,\text{s}^2\vartheta+(1-\text{c}^2\psi\,\text{s}^2\vartheta)\,e_1^2\,\right]^2- 4\,e_1^2\,e_2^2\,\text{s}^2\psi\,\text{s}^2\vartheta}\;\right\}\,,
\label{eq:axis_factors}
\end{equation}
where the plus and minus signs of the square-root term correspond to $\lambda_1$ and $\lambda_2$, respectively. Their values are limited by $1\geq\lambda_1\geq\lambda_2>0$, and their product simplifies to
\begin{equation}
\lambda_1\,\lambda_2= (1-e_1^2)\,\text{c}^2\vartheta+(1-e_2^2)(1-e_1^2\,\text{s}^2\psi)\,\text{s}^2\vartheta\,.
\label{eq:factor_product}
\end{equation}
The coordinates $\tilde{x}_1, \tilde{x}_2$ appearing in the final term of Eq.~(\ref{eq:intersections}) are defined by
\begin{equation}
\begin{pmatrix}
  \tilde{x}_1 \\
  \tilde{x}_2
\end{pmatrix}\,=\,\frac{1}{\sqrt{(1-e_1^2\,\text{s}^2\psi-\lambda_2)^2 +e_1^4\,\text{s}^2\psi\,\text{c}^2\psi\,\text{c}^2\vartheta}}
\begin{pmatrix}
  1-e_1^2\,\text{s}^2\psi-\lambda_2 & e_1^2\,\text{s}\psi\,\text{c}\psi\,\text{c}\vartheta  \\
  -e_1^2\,\text{s}\psi\,\text{c}\psi\,\text{c}\vartheta & 1-e_1^2\,\text{s}^2\psi-\lambda_2
\end{pmatrix}\;
\begin{pmatrix}
  \text{c}\varphi & \text{s}\varphi \\
  -\text{s}\varphi & \text{c}\varphi
\end{pmatrix}\;
\begin{pmatrix}
  x_1' \\
  x_2'
\end{pmatrix}\,.
\label{eq:rotated_coordinates}
\end{equation}
In Fig.~\ref{fig:ellipsoid} their orientation is indicated by the axes $x_1$ and $x_2$. These are plane-of-the-sky coordinates rotated to the principal axes of the two-dimensional projected ellipse, which can be seen by setting $x_{3+}'=x_{3-}'$ and using Eq.~(\ref{eq:intersections}). The semi-major axis of the ellipse is $\hat{a}\sqrt{\lambda_1}$ and it is oriented along the $\tilde{x}_1$ axis; the semi-minor axis is $\hat{a}\sqrt{\lambda_2}$ and it is oriented along the $\tilde{x}_2$ axis. We note that the first matrix in Eq.~(\ref{eq:rotated_coordinates}) including the square-root factor reduces to the identity matrix for any ellipsoid with $\psi=k\,\pi/2$ or $\vartheta=\pi/2$, or for an oblate ellipsoid ($e_1=0$) in an arbitrary orientation. In these cases Eq.~(\ref{eq:rotated_coordinates}) purely corresponds to the Eulerian rotation by $\varphi$ about $x_3'$.

Equation~(\ref{eq:intersections}) can be used to convert the integration of a density distribution along the line-of-sight at a fixed plane-of-the-sky position to integration over the corresponding range of the ellipsoidal semi-major axis $\hat{a}$. Its differentiation yields
\begin{equation}
\frac{\text{d}\,x_{3\pm}'}{\text{d}\,\hat{a}}=\pm\,\sqrt{\frac{(1-e_1^2)(1-e_2^2)} {\lambda_1\,\lambda_2}}\; \frac{\hat{a}}{\sqrt{\hat{a}^2-\frac{\tilde{x}_1^2}{\lambda_1}-\frac{\tilde{x}_2^2}{\lambda_2}}}\,,
\label{eq:derivative}
\end{equation}
an expression used in Sect.~\ref{sec:triaxial} to derive the surface density of the studied mass distribution.

The formulae presented above may be particularly useful when combining multiple halos with different spatial orientations. Examples of such systems include the halos of galaxies in a group, or the subhalos of individual galaxies in a galaxy-cluster halo.

\section{Lensing quantities at the foci of the $a = 1$ ellipse}
\label{sec:Appendix-foci}

The expressions for the deflection angle derived in Sect.~\ref{sec:alpha} and for the shear derived in Sect.~\ref{sec:shear} seem to be undefined at $\boldsymbol x =(\pm e,0)$, the foci of an ellipse with semi-major axis equal to the projected scale semi-major axis of the halo ($a=1$). Nevertheless, all lensing quantities are finite and continuous at these points, which can be shown by computing the respective limits. We present here the obtained expressions which can be used in case of running into numerical problems at the foci.

Equation~(\ref{eq:deflection-major-axis}) for the deflection angle yields the result
\begin{equation}
\boldsymbol \alpha_0(\pm e, 0)= \left( \pm\,\frac{2\,\kappa_\text{s}}{e} \left[ e^2\,\mathcal{F}(e)+\sqrt{1-e^2}-1\right],\,0\right)\,.
\label{eq:deflection-foci}
\end{equation}

The shear components from Eqs.~(\ref{eq:gamma-1}) and (\ref{eq:gamma-2}) attain the values
\begin{equation}
\gamma_1(\pm e, 0)= -\,\kappa_\text{s}\,\frac{1-e^2\,\mathcal{F}(e)}{1-e^2}
\label{eq:gamma-1-foci}
\end{equation}
and
\begin{equation}
\gamma_2(\pm e, 0)= 0\,,
\label{eq:gamma-2-foci}
\end{equation}
so that the shear and phase are
\begin{equation}
\gamma(\pm e, 0)= \,\kappa_\text{s}\,\frac{1-e^2\,\mathcal{F}(e)}{1-e^2}
\label{eq:gamma-foci}
\end{equation}
and
\begin{equation}
\phi(\pm e, 0)= \frac{\pi}{2}\,.
\label{eq:phase-foci}
\end{equation}

The expression for the convergence in Eq.~(\ref{eq:convergence}) is well behaved at these foci. We include the corresponding value here for completeness:
\begin{equation}
\kappa(e)=2\,\kappa_\text{s}\,\frac{\mathcal{F}(e)-1}{1-e^2}\,.
\label{eq:convergence-foci}
\end{equation}

\section{Image-plane analysis of critical curves and caustics}
\label{sec:Appendix-image-plane}

We present here a new technique for analyzing and visualizing the properties of critical curves and caustics of gravitational lens models. The method is related to the image-plane analysis of lenses consisting purely of a set of point masses introduced by \cite{danek_heyrovsky15a}, and applied to the analysis of triple lenses by \cite{danek_heyrovsky15b,danek_heyrovsky19}. In this appendix we thus use the term ``image plane'' instead of the synonymous ``lens plane'' used in the rest of the paper. We illustrate the technique on the example of the ellipsoidal NFW lens and demonstrate its usefulness for identifying the parameter-space boundaries in Figs.~\ref{fig:parameter-space} and \ref{fig:parameter-space-high-kappa}.

In Sect.~\ref{sec:Appendix-image-plane-contours} we demonstrate the equivalence between Jacobian-factor contours in the image plane and critical curves with all possible values of the convergence parameter $\kappa_\text{s}$. Various properties of these critical curves and their corresponding caustics can be studied by plotting additional curves over the contours. In Sect.~\ref{sec:Appendix-image-plane-inflection} we introduce the inflection curve which connects all inflection points occurring along these critical curves and helps to identify the ``peanut'' transitions discussed in Sect.~\ref{sec:curves}. In Sect.~\ref{sec:Appendix-image-plane-cusps} we define the cusp curve which connects all images of caustic cusps along these critical curves, permitting us to keep track of the numbers of cusps on different components of the caustic. It also led us to identify the butterfly metamorphoses discussed in Sect.~\ref{sec:parameter-space-intermediate}.

\subsection{Critical curves as Jacobian-factor contours}
\label{sec:Appendix-image-plane-contours}

The Jacobian from Eq.~(\ref{eq:Jacobian}) can be written as the product of two factors,
\begin{equation}
\label{eq:Jacobian-factorized}
\mathrm{det}\,J(\boldsymbol x)=j_{\text{T}}(\boldsymbol x)\,j_{\text{R}}(\boldsymbol x)\,,
\end{equation}
where the tangential factor
\begin{equation}
\label{eq:Jacobian-tangential-factor}
j_{\text{T}}(\boldsymbol x)=1/\lambda_\parallel(\boldsymbol x)=1-\kappa(\boldsymbol x)-\gamma(\boldsymbol x)
\end{equation}
is the inverse of $\lambda_\parallel$, the image-scaling factor in the direction parallel to the phase. The tangential factor is equal to zero along the tangential critical curve, and its values span the range $1-\max(\kappa+\gamma) \leq j_{\text{T}} \leq 1-\min(\kappa+\gamma)$. In the case of the ellipsoidal NFW lens, $j_{\text{T}}\in(-\infty,1)$ for any combination of convergence parameter $\kappa_\text{s}$ and eccentricity $e$. The minimum is achieved at the halo center; the tangential critical curve exists for all lens-parameter combinations; and the maximum value of 1 is achieved asymptotically far from the halo center.

A constant-tangential-factor contour $j_{\text{T}}(\boldsymbol x)=j_{\text{T}}$ is defined by the equation
\begin{equation}
\label{eq:contour-tangential-factor-0}
1-\kappa(\kappa_\text{s},e,\boldsymbol x)-\gamma(\kappa_\text{s},e,\boldsymbol x)=j_{\text{T}}\,,
\end{equation}
where we explicitly indicate the parameter dependence of the convergence and the shear. Subtracting $j_{\text{T}}$ and dividing the resulting form by $1-j_{\text{T}}$ leads to
\begin{equation}
\label{eq:contour-tangential-factor-1}
1-\frac{1}{1-j_{\text{T}}}\,[\,\kappa(\kappa_\text{s},e,\boldsymbol x)+\gamma(\kappa_\text{s},e,\boldsymbol x)\,]=0\,.
\end{equation}
In the case of the ellipsoidal NFW model, the equation can be simplified to
\begin{equation}
\label{eq:contour-tangential-factor}
1-\kappa\left(\frac{\kappa_\text{s}}{1-j_{\text{T}}},e,\boldsymbol x\right)-\gamma\left(\frac{\kappa_\text{s}}{1-j_{\text{T}}},e,\boldsymbol x\right)=0\,,
\end{equation}
because $\kappa_\text{s}$ appears purely as a multiplicative factor in the convergence and in the shear, as seen from Eqs.~(\ref{eq:convergence}) and (\ref{eq:shear}), respectively.

The interpretation of Eq.~(\ref{eq:contour-tangential-factor}) is the following: the Jacobian tangential-factor contour $j_{\text{T}}(\boldsymbol x)=j_{\text{T}}$ is identically equal to the tangential critical curve of a lens with the same eccentricity $e$ and a re-scaled convergence parameter $\kappa_\text{s}/(1-j_{\text{T}})$. By plotting the contours of $j_{\text{T}}(\boldsymbol x)$ for a given eccentricity $e$ we thus obtain the tangential critical curves for the full range of convergence-parameter values, with $j_{\text{T}}\to -\infty$ corresponding to the low-mass limit $\kappa_\text{s}\to 0$, and $j_{\text{T}}\nearrow 1$ corresponding to the high-mass limit $\kappa_\text{s}\to \infty$.

\begin{figure}
\centering
\resizebox{\hsize/2}{!}{\includegraphics{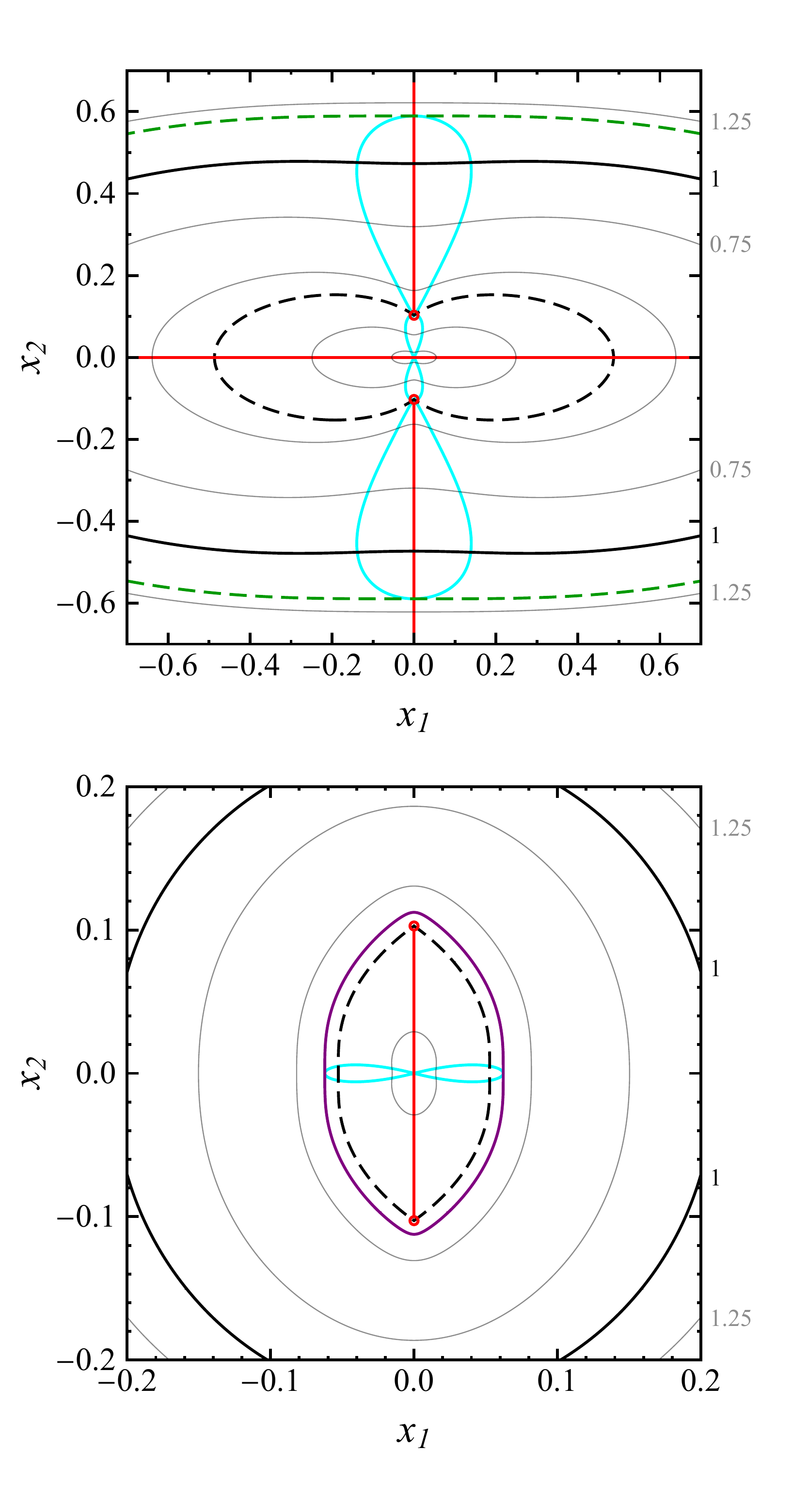}}
\caption{Inflection curves (cyan) and cusp curves (red) plotted in the top panel over contours of the Jacobian tangential factor $j_{\text{T}}(\boldsymbol x)$ showing the tangential critical curves, and in the bottom panel over contours of the Jacobian radial factor $j_{\text{R}}(\boldsymbol x)$ showing the radial critical curves of the ellipsoidal NFW lens with eccentricity $e=0.9$. Contours are plotted in both panels for $\kappa_\text{s}$ from 0.25 to 1.25 in steps of 0.25 with $\kappa_\text{s}=1$ marked black, and with the additional dashed-black umbilic $\kappa_\text{s}\approx0.398$ contours. The top panel includes the smallest $\kappa_\text{s}=0.125$ contour; its radial counterpart is too small to be visible in the bottom panel. The dashed green $\kappa_\text{s}\approx1.195$ contour in the top panel marks the tangential-peanut transition parameter combination. The solid purple $\kappa_\text{s}\approx0.432$ contour in the bottom panel marks the radial-peanut transition parameter combination.
\label{fig:contours-e9}}
\end{figure}

An example for eccentricity $e=0.9$ is shown in the top panel of Fig.~\ref{fig:contours-e9}. The smallest gray central contour corresponds to the $\kappa_\text{s}=0.125$ tangential critical curve, the other gray contours are plotted from $\kappa_\text{s}=0.25$ in steps of $0.25$, with $\kappa_\text{s}=1$ plotted in black. The dashed black contour corresponds to the $\kappa_\text{s}\approx0.398$ hyperbolic-umbilic configuration. The properties of all contours / tangential critical curves can be read off the $e=0.9$ dashed horizontal line in the parameter-space maps in Fig.~\ref{fig:parameter-space}, and their dimensions can be compared with the $e=0.9$ points along the green curves in Fig.~\ref{fig:nfw-dimensions-critical-curves}.

The radial factor
\begin{equation}
\label{eq:Jacobian-radial-factor}
j_{\text{R}}(\boldsymbol x)=1/\lambda_\perp(\boldsymbol x)=1-\kappa(\boldsymbol x)+\gamma(\boldsymbol x)
\end{equation}
is the inverse of $\lambda_\perp$, the image-scaling factor in the direction perpendicular to the phase. The radial factor is equal to zero along the radial critical curve, and its values span the range $1-\max(\kappa-\gamma) \leq j_{\text{R}} \leq 1-\min(\kappa-\gamma)$. In the case of the ellipsoidal NFW lens, $j_{\text{R}}\in(-\infty,1+\max(\gamma-\kappa)]$, where the upper limit depends on the combination of $\kappa_\text{s}$ and $e$. The minimum is achieved at the halo center; the radial critical curve exists for all lens-parameter combinations; the maximum value is greater than 1 and it is achieved at a distance from the halo center where the shear exceeds the convergence.

A constant-radial-factor contour $j_{\text{R}}(\boldsymbol x)=j_{\text{R}}$ is defined by the equation
\begin{equation}
\label{eq:contour-radial-factor-0}
1-\kappa(\kappa_\text{s},e,\boldsymbol x)+\gamma(\kappa_\text{s},e,\boldsymbol x)=j_{\text{R}}\,.
\end{equation}
Subtracting $j_{\text{R}}$ and dividing the resulting form by $1-j_{\text{R}}$ leads to
\begin{equation}
\label{eq:contour-radial-factor-1}
1-\frac{1}{1-j_{\text{R}}}\,[\,\kappa(\kappa_\text{s},e,\boldsymbol x)-\gamma(\kappa_\text{s},e,\boldsymbol x)\,]=0\,.
\end{equation}
In the case of the ellipsoidal NFW model, the equation can be simplified for $j_{\text{R}}<1$ to
\begin{equation}
\label{eq:contour-radial-factor}
1-\kappa\left(\frac{\kappa_\text{s}}{1-j_{\text{R}}},e,\boldsymbol x\right)+\gamma\left(\frac{\kappa_\text{s}}{1-j_{\text{R}}},e,\boldsymbol x\right)=0\,,
\end{equation}
for the same reason as in Eq.~(\ref{eq:contour-tangential-factor}) as long as $1-j_{\text{R}}$ remains positive.

The interpretation of Eq.~(\ref{eq:contour-radial-factor}) is analogous to the tangential-factor case above: the Jacobian radial-factor contour $j_{\text{R}}(\boldsymbol x)=j_{\text{R}}$ with $j_{\text{R}}<1$ is identically equal to the radial critical curve of a lens with the same eccentricity $e$ and a re-scaled convergence parameter $\kappa_\text{s}/(1-j_{\text{R}})$. By plotting the contours of $j_{\text{R}}(\boldsymbol x)$ for a given eccentricity $e$ we thus obtain the radial critical curves for the full range of convergence-parameter values, with $j_{\text{R}}\to -\infty$ corresponding to the low-mass limit $\kappa_\text{s}\to 0$, and $j_{\text{R}}\nearrow 1$ corresponding to the high-mass limit $\kappa_\text{s}\to \infty$.

An example for eccentricity $e=0.9$ is shown in the bottom panel of Fig.~\ref{fig:contours-e9}. The smallest gray central contour corresponds to the $\kappa_\text{s}=0.25$ radial critical curve; we omitted $\kappa_\text{s}=0.125$ which would appear as a mere dot here. The other gray contours are plotted in steps of $0.25$, with $\kappa_\text{s}=1$ plotted in black. The dashed black contour corresponds to the $\kappa_\text{s}\approx0.398$ hyperbolic-umbilic configuration. The properties of all contours / radial critical curves can be read off the $e=0.9$ dashed horizontal line in the parameter-space maps in Fig.~\ref{fig:parameter-space}, and their dimensions can be compared with the $e=0.9$ points along the purple curves in Fig.~\ref{fig:nfw-dimensions-critical-curves}.

While in the case of $j_{\text{T}}(\boldsymbol x)$ all contours correspond to tangential critical curves, in the case of $j_{\text{R}}(\boldsymbol x)$ only the contours with $j_{\text{R}}< 1$ correspond to radial critical curves while those with $j_{\text{R}}\geq 1$ do not. Among these dropouts, the contour $j_{\text{R}}=1$ has a special significance. As seen from Eq.~(\ref{eq:contour-radial-factor-0}), along this contour the convergence is equal to the shear. This curve depends only on the eccentricity $e$, because $\kappa_\text{s}$ cancels out as a multiplicative factor of both sides of the equality. We can thus define the radial limiting curve by the equation
\begin{equation}
\label{eq:radial-limiting-curve}
\kappa(1,e,\boldsymbol x)=\gamma(1,e,\boldsymbol x)\,.
\end{equation}
Radial critical curves approach this curve in the limit $\kappa_\text{s}\to\infty$, which can be seen by rewriting the radial-critical-curve equation in the form
\begin{equation}
\label{eq:crit-curve-radial-alt}
\kappa(1,e,\boldsymbol x)=\gamma(1,e,\boldsymbol x)+\frac{1}{\kappa_\text{s}}\,.
\end{equation}
In the case of the ellipsoidal NFW lens, the radial limiting curve for a given eccentricity $e$ thus plays the role of an outer envelope of all radial critical curves with the same eccentricity. In the circularly symmetric case, the radial limiting curve is a circle of radius $x_{\text{R}\infty}\approx1.32$, marked in the right panel of Fig.~\ref{fig:nfw-radii} by the orange dash along the right edge. In the general ellipsoidal case, the horizontal and vertical dimensions of the curve are plotted in Fig.~\ref{fig:nfw-radial-limiting-curve} and discussed in the last two paragraphs of Sect.~\ref{sec:dimensions-critical-curves}.

\subsection{Inflection points of critical curves}
\label{sec:Appendix-image-plane-inflection}

The concept of visualizing an entire one-parameter class of critical curves in a single tangential or radial contour plot can be developed further by tracking specific properties of the critical curves to identify parameter-space boundaries such as those plotted in Figs.~\ref{fig:parameter-space} and \ref{fig:parameter-space-high-kappa}. Here we demonstrate this approach on the occurrence of inflection points along the critical curves.

For a curve given by the general implicit expression
\begin{equation}
\label{eq:function}
g(\boldsymbol x)=0\,,
\end{equation}
inflection points can be found by Taylor-expanding $g(\boldsymbol x)$ and setting the quadratic term equal to zero in the tangential direction. This necessary condition can be expressed by the equation
\begin{equation}
\label{eq:inflection-curve}
\frac{\partial^2 g}{\partial x_1^2}\,\left(\frac{\partial g}{\partial x_2}\right)^{\!2} -2\,\frac{\partial^2 g}{\partial x_1\,\partial x_2}\,\frac{\partial g}{\partial x_1}\,\frac{\partial g}{\partial x_2} +\frac{\partial^2 g}{\partial x_2^2}\,\left(\frac{\partial g}{\partial x_1}\right)^{\!2}=0\,.
\end{equation}
For simplicity, we call the curve defined by Eq.~(\ref{eq:inflection-curve}) the inflection curve.

In the case of the tangential critical curve we set $g(\boldsymbol x)=j_{\text{T}}(\boldsymbol x)$ and compute the first derivatives using Eq.~(\ref{eq:Jacobian-tangential-factor}), obtaining
\begin{equation}
\label{eq:tangential-factor-derivative-1}
\frac{\partial j_{\text{T}}}{\partial x_1}=-\,\frac{x_1}{a}\,\frac{{\rm d}\,\kappa}{{\rm d}\,a}-\frac{1}{\gamma}\,\left[\gamma_1\,\frac{\partial \gamma_1}{\partial x_1}+\gamma_2\,\frac{\partial \gamma_2}{\partial x_1}\right]
\end{equation}
and
\begin{equation}
\label{eq:tangential-factor-derivative-2}
\frac{\partial j_{\text{T}}}{\partial x_2}=-\,\frac{x_2}{a(1-e^2)}\,\frac{{\rm d}\,\kappa}{{\rm d}\,a}-\frac{1}{\gamma}\,\left[\gamma_1\,\frac{\partial \gamma_1}{\partial x_2}+\gamma_2\,\frac{\partial \gamma_2}{\partial x_2}\right]\,.
\end{equation}
The derivatives of the convergence and of the shear components can be computed from Eqs.~(\ref{eq:convergence}), (\ref{eq:gamma-1}), and (\ref{eq:gamma-2}); the second derivatives can be computed analogously. All derivatives of $j_{\text{T}}(\boldsymbol x)$ are directly proportional to the convergence parameter $\kappa_\text{s}$. The left-hand side of Eq.~(\ref{eq:inflection-curve}) is thus clearly a homogeneous function of $\kappa_\text{s}$. Division by $\kappa_\text{s}^3$ shows that the inflection curve is independent of $\kappa_\text{s}$. We can therefore plot it over the contours of $j_{\text{T}}$ for a given eccentricity $e$; its intersections with the contours then indicate the locations of inflection points along tangential critical curves for all values of $\kappa_\text{s}$.

In the $e=0.9$ tangential contour plot in the top panel of Fig.~\ref{fig:contours-e9} the inflection curve is marked by the cyan double-hourglass-shaped line. The smallest, low-$\kappa_\text{s}$ critical curves have four intersections with the inflection curve, corresponding to two pairs of inflection points positioned symmetrically above and below the major axis. At the dashed black $\kappa_\text{s}\approx0.398$ umbilic critical curve the pairs merge and vanish along the minor axis at the umbilic points marked by the small red circles. This value corresponds to the $e=0.9$ point along the dashed black boundary in Fig.~\ref{fig:parameter-space}. As $\kappa_\text{s}$ increases, the pairs of inflection points reappear, and finally vanish along the minor axis at $\kappa_\text{s}\approx1.195$ for the dashed green critical curve. This value corresponds to the tangential ``peanut'' transition discussed in Sect.~\ref{sec:parameter-space-low}, as well as the $e=0.9$ point along the dashed green boundary in Fig.~\ref{fig:parameter-space}. For higher values of $\kappa_\text{s}$, the $e=0.9$ tangential critical curves are globally convex, with no inflection points.

In the case of the radial critical curve we set $g(\boldsymbol x)=j_{\text{R}}(\boldsymbol x)$ and compute the first derivatives using Eq.~(\ref{eq:Jacobian-radial-factor}), obtaining
\begin{equation}
\label{eq:radial-factor-derivative-1}
\frac{\partial j_{\text{R}}}{\partial x_1}=-\,\frac{x_1}{a}\,\frac{{\rm d}\,\kappa}{{\rm d}\,a}+\frac{1}{\gamma}\,\left[\gamma_1\,\frac{\partial \gamma_1}{\partial x_1}+\gamma_2\,\frac{\partial \gamma_2}{\partial x_1}\right]
\end{equation}
and
\begin{equation}
\label{eq:radial-factor-derivative-2}
\frac{\partial j_{\text{R}}}{\partial x_2}=-\,\frac{x_2}{a(1-e^2)}\,\frac{{\rm d}\,\kappa}{{\rm d}\,a}+\frac{1}{\gamma}\,\left[\gamma_1\,\frac{\partial \gamma_1}{\partial x_2}+\gamma_2\,\frac{\partial \gamma_2}{\partial x_2}\right]\,.
\end{equation}
The derivatives of the convergence and of the shear components as well as the second derivatives can be computed analogously as in the tangential case. The inflection curve given by Eq.~(\ref{eq:inflection-curve}) is independent of $\kappa_\text{s}$, for the same reasons presented in the tangential case above. When plotted over the contours of $j_{\text{R}}$ for a given eccentricity $e$, its intersections with the contours indicate the locations of inflection points along radial critical curves for all values of $\kappa_\text{s}$.

In the $e=0.9$ radial contour plot in the bottom panel of Fig.~\ref{fig:contours-e9} the inflection curve is marked by the cyan bow-tie-shaped line. The smallest, low-$\kappa_\text{s}$ critical curves have four intersections with the inflection curve, corresponding to two pairs of inflection points positioned symmetrically on either side of the minor axis. The pairs merge and vanish along the major axis at $\kappa_\text{s}\approx0.432$ for the purple critical curve. This value corresponds to the radial ``peanut'' transition discussed in Sect.~\ref{sec:parameter-space-low}, as well as the $e=0.9$ point along the purple boundary in Fig.~\ref{fig:parameter-space}. For higher values of $\kappa_\text{s}$, the $e=0.9$ radial critical curves are globally convex, with no inflection points.

\subsection{Cusp number}
\label{sec:Appendix-image-plane-cusps}

Another property that can be directly studied from the contour plots is the number of cusps along the caustics. The cusp number on a caustic is equal to the number of cusp images on the corresponding critical curve, and thus it can be determined even in the image plane.

Cusps occur on the caustic at points where the tangent vector to the caustic $\boldsymbol T_{\boldsymbol y}$ vanishes. We can express $\boldsymbol T_{\boldsymbol y}$ as a function of the image-plane position using the lens-equation Jacobi matrix and the tangent vector to the critical curve $\boldsymbol T_{\boldsymbol x}$,
\begin{equation}
\label{eq:tangent-vector-caustic}
\boldsymbol T_{\boldsymbol y} (\boldsymbol x)=J(\boldsymbol x) \cdot \boldsymbol T_{\boldsymbol x}(\boldsymbol x)=
\begin{pmatrix}
  1-\kappa-\gamma_1 & -\gamma_2 \\
  -\gamma_2 & 1-\kappa+\gamma_1
\end{pmatrix}\;
\begin{pmatrix}
  -\,\frac{\partial}{\partial x_2}\mathrm{det}\,J \\
  \frac{\partial}{\partial x_1}\mathrm{det}\,J
\end{pmatrix}\,,
\end{equation}
where we expressed $\boldsymbol T_{\boldsymbol x}(\boldsymbol x)$ as a vector perpendicular to the gradient of the Jacobian, since the critical curve is defined as a line of constant (zero) Jacobian. Taking into account the factorization of the Jacobian from Eq.~(\ref{eq:Jacobian-factorized}), we may express the general cusp condition in the form
\begin{equation}
\label{eq:cusp-condition}
\boldsymbol T_{\boldsymbol y} (\boldsymbol x)=
\begin{pmatrix}
  -(1-\kappa-\gamma_1)\,\frac{\partial(j_{\text{T}}\,j_{\text{R}})}{\partial x_2} -\gamma_2\,\frac{\partial(j_{\text{T}}\,j_{\text{R}})}{\partial x_1} \\
  \gamma_2\,\frac{\partial(j_{\text{T}}\,j_{\text{R}})}{\partial x_2}+ (1-\kappa+\gamma_1)\,\frac{\partial(j_{\text{T}}\,j_{\text{R}})}{\partial x_1}
\end{pmatrix}=
\begin{pmatrix}
 0 \\
 0
\end{pmatrix}\,.
\end{equation}
We note that this zero-tangent-vector condition is automatically satisfied also by umbilic points, for which the entire Jacobi matrix is zero. However, the umbilic points on the caustic are not cusps, so we have to sort them out from the solutions of Eq.~(\ref{eq:cusp-condition}).

For non-umbilic points along the tangential critical curve, $j_{\text{T}}=0$ and $j_{\text{R}}\neq 0$. Equation~(\ref{eq:cusp-condition}) can then be simplified to the equation set
\begin{eqnarray}
\label{eq:cusp-condition-tangential-1}
   (\gamma-\gamma_1)\,\frac{\partial j_{\text{T}}}{\partial x_2} + \gamma_2\,\frac{\partial j_{\text{T}}}{\partial x_1}  &=& 0 \\
\label{eq:cusp-condition-tangential-2}
  \gamma_2\,\frac{\partial j_{\text{T}}}{\partial x_2} +  (\gamma+\gamma_1)\,\frac{\partial j_{\text{T}}}{\partial x_1} &=& 0\;.
\end{eqnarray}
We note that the two equations are equivalent off the major and minor axis, which can be seen for example by multiplying the first equation by $(\gamma+\gamma_1)/\gamma_2$. Along the axes, $\gamma_2=0$ and one of the equations reduces to the trivial $0=0$ while the other requires one partial derivative of $j_{\text{T}}$ to be zero. Hence, for finding all cusps both equations should be preserved.

Squaring the two equations and adding them leads to the equivalent single equation
\begin{equation}
\label{eq:cusp-condition-tangential-single}
(\gamma+\gamma_1)\left(\frac{\partial j_{\text{T}}}{\partial x_1}\right)^{\!2} +2\,\gamma_2\,\frac{\partial j_{\text{T}}}{\partial x_1}\,\frac{\partial j_{\text{T}}}{\partial x_2} + (\gamma-\gamma_1)\left(\frac{\partial j_{\text{T}}}{\partial x_2}\right)^{\!2}=0\,.
\end{equation}
However, the set of Eqs.~(\ref{eq:cusp-condition-tangential-1}) and (\ref{eq:cusp-condition-tangential-2}) is better suited for numerical solution, since solving Eq.~(\ref{eq:cusp-condition-tangential-single}) amounts to the more challenging task of finding a curve minimizing the non-negative function on the left-hand side. Finally, we note that expressing the shear components in terms of shear and phase using Eqs.~(\ref{eq:gamma-1-def}) and (\ref{eq:gamma-2-def}) leads to a more geometric formulation of the tangential-caustic cusp condition,
\begin{equation}
\label{eq:cusp-condition-tangential-geometric}
\left(\frac{\partial j_{\text{T}}}{\partial x_1},\frac{\partial j_{\text{T}}}{\partial x_2}\right)\cdot\left(\cos{\phi},\sin{\phi}\right)=0\,,
\end{equation}
i.e., cusp images occur along the tangential critical curve at points where the gradient of the Jacobian tangential factor is perpendicular to the lens phase. Since the gradient is also perpendicular to the critical curve, the condition can be restated more simply: cusp images occur along the tangential critical curve at points where the lens phase is parallel to the critical curve.

We call the curve defined by the common solution of Eqs.~(\ref{eq:cusp-condition-tangential-1}) and (\ref{eq:cusp-condition-tangential-2}) the tangential cusp curve. Division of both equations by $\kappa_\text{s}^2$ shows that the tangential cusp curve is independent of $\kappa_\text{s}$. In a similar manner as we did in the case of the inflection curve in Sect.~\ref{sec:Appendix-image-plane-inflection}, we can thus plot it over the contours of $j_{\text{T}}$ for a given eccentricity $e$ and its intersections with the contours will indicate the locations of cusp images along tangential critical curves for all values of $\kappa_\text{s}$.

In the $e=0.9$ tangential contour plot in the top panel of Fig.~\ref{fig:contours-e9} the tangential cusp curve is marked by the red line. Here it consists of the entire major axis and the minor axis except the segment between the umbilic points. The smallest, low-$\kappa_\text{s}$ critical curves have two intersections with the cusp curve on the major axis, corresponding to two cusps on the corresponding caustic (e.g., the green caustic in the bottom right panel of Fig.~\ref{fig:gallery-basic}). For $\kappa_\text{s}\gtrapprox0.398$, above the black dashed umbilic tangential critical curve, two additional cusp images occur along the minor axis, corresponding to a total of four cusps on the tangential caustic (e.g., the green caustic in the second and third bottom panels of Fig.~\ref{fig:gallery-basic}).

For non-umbilic points along the radial critical curve, $j_{\text{R}}=0$ and $j_{\text{T}}\neq 0$. In this case Eq.~(\ref{eq:cusp-condition}) can be simplified to the equation set
\begin{eqnarray}
\label{eq:cusp-condition-radial-1}
   (\gamma+\gamma_1)\,\frac{\partial j_{\text{R}}}{\partial x_2} - \gamma_2\,\frac{\partial j_{\text{R}}}{\partial x_1}  &=& 0 \\
\label{eq:cusp-condition-radial-2}
  \gamma_2\,\frac{\partial j_{\text{R}}}{\partial x_2} - (\gamma-\gamma_1)\,\frac{\partial j_{\text{R}}}{\partial x_1} &=& 0\;.
\end{eqnarray}
The two equations are equivalent off the major and minor axis, which can be seen for example by multiplying the first equation by $(\gamma-\gamma_1)/\gamma_2$. Along the axes, $\gamma_2=0$ and one of the equations reduces to the trivial $0=0$ while the other requires one partial derivative of $j_{\text{R}}$ to be zero. Hence, for finding all cusps both equations should be preserved.

The equivalent single equation has the form
\begin{equation}
\label{eq:cusp-condition-radial-single}
(\gamma-\gamma_1)\left(\frac{\partial j_{\text{R}}}{\partial x_1}\right)^{\!2} -2\,\gamma_2\,\frac{\partial j_{\text{R}}}{\partial x_1}\,\frac{\partial j_{\text{R}}}{\partial x_2} + (\gamma+\gamma_1)\left(\frac{\partial j_{\text{R}}}{\partial x_2}\right)^{\!2}=0\,.
\end{equation}
Just as in the tangential case, the set of Eqs.~(\ref{eq:cusp-condition-radial-1}) and (\ref{eq:cusp-condition-radial-2}) is better suited for numerical solution. Expressing the shear components in terms of shear and phase using Eqs.~(\ref{eq:gamma-1-def}) and (\ref{eq:gamma-2-def}) leads to a more geometric formulation of the radial-caustic cusp condition
\begin{equation}
\label{eq:cusp-condition-radial-geometric}
\left(\frac{\partial j_{\text{R}}}{\partial x_1},\frac{\partial j_{\text{R}}}{\partial x_2}\right)\cdot\left(\sin{\phi},-\cos{\phi}\right)=0\,,
\end{equation}
i.e., cusp images occur along the radial critical curve at points where the gradient of the Jacobian radial factor is parallel to the lens phase. The condition can be restated more simply: cusp images occur along the radial critical curve at points where the lens phase is perpendicular to the critical curve.

We call the curve defined by the common solution of Eqs.~(\ref{eq:cusp-condition-radial-1}) and (\ref{eq:cusp-condition-radial-2}) the radial cusp curve. This curve is also independent of $\kappa_\text{s}$, and we can thus plot it over the contours of $j_{\text{R}}$ for a given eccentricity $e$, so that its intersections with the contours will indicate the locations of cusp images along radial critical curves for all values of $\kappa_\text{s}$.

In the $e=0.9$ radial contour plot in the bottom panel of Fig.~\ref{fig:contours-e9} the radial cusp curve is marked by the red line. Here it consists purely of the segment of the minor axis between the umbilic points. The smallest, low-$\kappa_\text{s}$ critical curves have two intersections with the cusp curve on the minor axis, corresponding to two cusps on the corresponding caustic (e.g., the purple caustic in the bottom right panel of Fig.~\ref{fig:gallery-basic}). For $\kappa_\text{s}\gtrapprox0.398$, above the black dashed umbilic radial critical curve, the cusp images along the minor axis vanish and the radial caustic thus has no cusps (e.g., the purple caustic in the first three bottom panels of Fig.~\ref{fig:gallery-basic}).

\begin{figure*}
\centering
\includegraphics[width=18 cm]{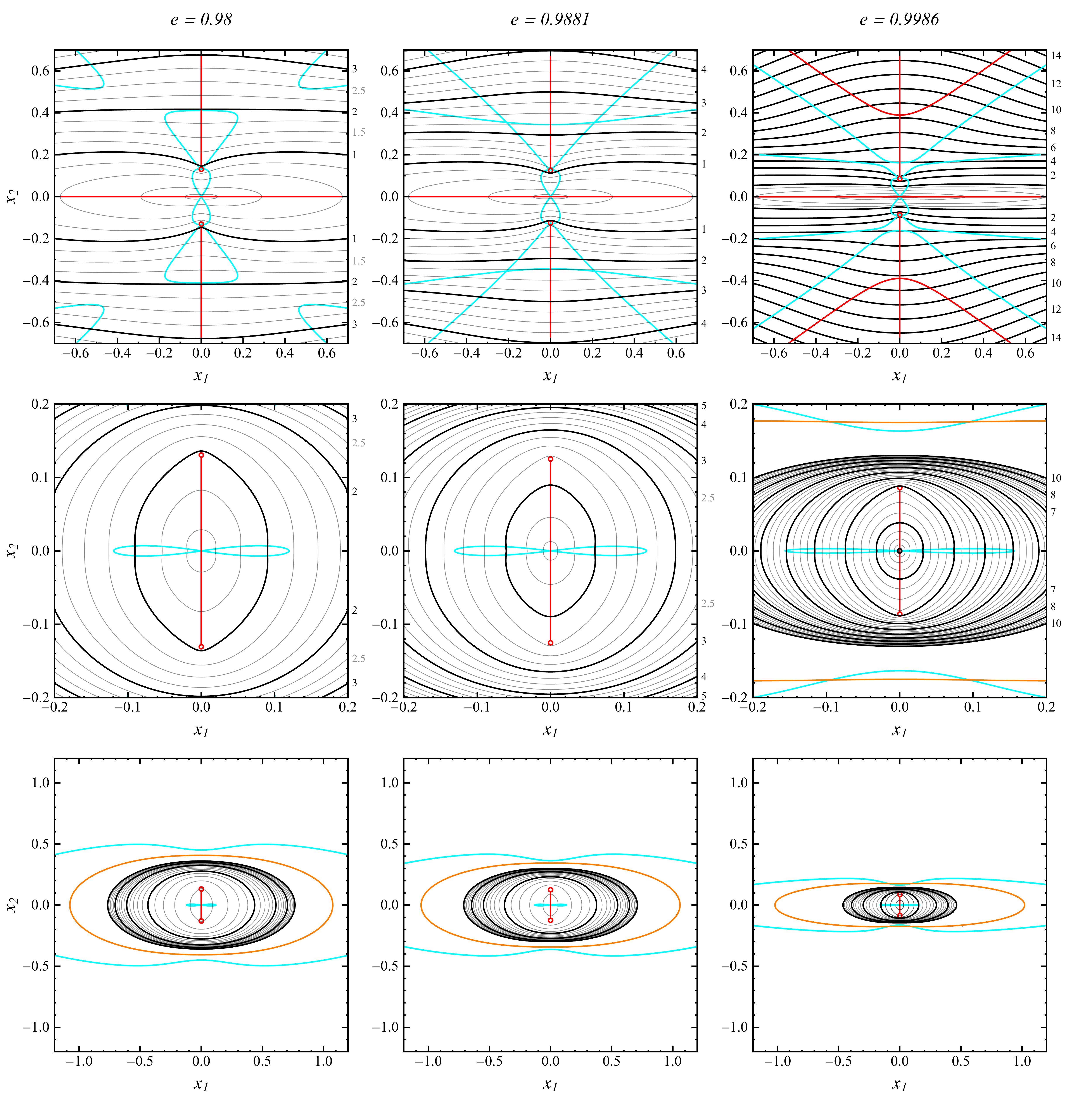}
\caption{Contour plots showing the tangential (top row) and radial (middle row, six times larger scale in bottom row) critical curves of the ellipsoidal NFW lens with eccentricity $e=0.98$ (left column), $e=0.9881$ (central column), and $e=0.9986$ (right column). Additionally plotted are: inflection curves (cyan), cusp curves (red), radial limiting curves (orange). Contours are plotted in the top row for $\kappa_\text{s}=0.125$ and from 0.25 up in steps of 0.25 (integer-valued contours in black); in the top right panel only integer-valued contours are plotted for $\kappa_\text{s}\geq1$. The middle row includes contours from $\kappa_\text{s}=0.5$ up in steps of 0.25 (integer-valued contours in black); in the middle right panel only contours in the interval $1\leq\kappa_\text{s}\leq10$ are plotted. The bottom row includes integer-valued contours from $\kappa_\text{s}=1$ to 20 (multiples of 5 in black).
\label{fig:contours}}
\end{figure*}

\subsection{Sample contour plots}
\label{sec:Appendix-image-plane-samples}

In the case of the ellipsoidal NFW lens, the tangential and radial contour plots with the inflection and cusp curves have the same general structure seen in the plots in Fig.~\ref{fig:contours-e9} for eccentricities up to $e\approx0.9496$. The plots can be used to determine the dashed black, solid purple, and dashed green boundaries seen in the left panel of Fig.~\ref{fig:parameter-space}. The only exception is the solid black cusp-piercing boundary, which cannot be directly identified from the contour plots. The reason can be seen from Eq.~(\ref{eq:piercing}) identifying the boundary: it is not homogeneous in $\kappa_\text{s}$, and thus it cannot be used to define a characteristic curve in the contour plots valid for all $\kappa_\text{s}$.

For eccentricities $e\gtrapprox0.9497$ (or, equivalently, for ellipticities $\gtrapprox0.687$), the changing structure of the contour plots can be used to identify all the additional parameter-space boundaries seen in the right panel of Fig.~\ref{fig:parameter-space} and in Fig.~\ref{fig:parameter-space-high-kappa}. In the three columns of Fig.~\ref{fig:contours} we present sample contour plots for eccentricities $e\in\{0.98,0.9881,0.9986\}$, which are also marked for orientation by the top three horizontal dashed lines in the parameter-space maps in Figs.~\ref{fig:parameter-space} and \ref{fig:parameter-space-high-kappa}. The top row of Fig.~\ref{fig:contours} shows the tangential contour plots, the middle row shows the central part of the radial contour plots, and the bottom row shows the large-scale radial contour plots out to the respective radial limiting curves. The contours in the top row correspond to the same $\kappa_\text{s}$ values and spacings as in the top panel of Fig.~\ref{fig:contours-e9}, except in the right column, in which we included only integer-valued contours for $\kappa_\text{s}>1$ to avoid crowding. In the middle row, contours starting from $\kappa_\text{s}=0.5$ are plotted for the same values and spacings as in the bottom panel of Fig.~\ref{fig:contours-e9}, except in the right column, in which we included only contours from $\kappa_\text{s}=1$ to $\kappa_\text{s}=10$. In the bottom row, contours are plotted from $\kappa_\text{s}=1$ to $\kappa_\text{s}=20$, with bold black contours marking multiples of 5.

The main changes occur on the tangential critical curves, illustrated in the top row of Fig.~\ref{fig:contours}. For eccentricities $e\gtrapprox0.9497$, four symmetric pairs of off-axis inflection points appear first at $\kappa_\text{s}\approx 62.6$, progressing with increasing eccentricity rapidly to high $\kappa_\text{s}$ and, more slowly, to lower $\kappa_\text{s}$. In the $e=0.98$ top left panel of Fig.~\ref{fig:contours}, the corresponding lobes of the inflection curve can be seen in the corners of the plot. The critical curve tangent to these lobes (here $\kappa_\text{s}\approx 2.48$ just inside the plotted $\kappa_\text{s}=2.5$ contour) identifies the dotted green ``bulge'' boundary seen in the right panel of Fig.~\ref{fig:parameter-space} and in Fig.~\ref{fig:parameter-space-high-kappa}. These inflection points lead to the centrally bulged critical curves described in Sect.~\ref{sec:parameter-space-intermediate}. The contour plot also illustrates the narrow interval of convergence parameters $\kappa_\text{s}\in(1.98,2.48)$, for which the tangential critical curves are purely convex.

At eccentricity $e\approx0.9881$, shown in the central column of Fig.~\ref{fig:contours}, the outer lobes of the inflection curve extend inward and connect with its central component. The $\kappa_\text{s}\approx2.30$ critical curve passing through the connection points is the highest-eccentricity ellipsoidal NFW tangential critical curve that is purely convex. The parameter-space map in the right panel of Fig.~\ref{fig:parameter-space} shows that this eccentricity corresponds to the peak of the dotted green ``bulge'' boundary. The $\kappa_\text{s}=6$ tangential critical curve shown in the top left panel of Fig.~\ref{fig:gallery-special}, with parameters marked by the left blue point in Fig.~\ref{fig:parameter-space}, would appear in a zoomed-out version of the top central contour plot, which extends only to $\kappa_\text{s}\approx 4.5$.

In the $e=0.9986$ example in the right column of Fig.~\ref{fig:contours}, the inflection curve is reconnected, with a separate part close to the major axis and the top and bottom parts offset along the minor axis. Their intersections with the minor axis indicate that the dashed green boundary in the right panel of Fig.~\ref{fig:parameter-space} here plays the combined role of a ``peanut''/``bulge'' boundary: critical curves with lower $\kappa_\text{s}$ are peanut-shaped; those with higher $\kappa_\text{s}$ are bulge-shaped. This contour plot also shows an interesting feature on the cusp curve: the $\kappa_\text{s}\approx7.185$ branching points that can be seen on the minor axis at $x_2\approx\pm0.39$. This feature marks the butterfly transition, indicated by the bold dashed green boundary in Fig.~\ref{fig:parameter-space-high-kappa}: while for lower $\kappa_\text{s}$ the tangential caustic has four cusps, for higher $\kappa_\text{s}$ it has eight cusps. This can be seen on the $\kappa_\text{s}=8$ tangential caustic in the bottom right panel of Fig.~\ref{fig:gallery-special}, with parameters marked by the right blue point in Fig.~\ref{fig:parameter-space}. The corresponding tangential critical curve from the top right panel of Fig.~\ref{fig:gallery-special} can be partly seen in the top right contour plot as the first contour outward of the butterfly branching point. As seen from Fig.~\ref{fig:parameter-space-high-kappa}, the butterfly transition occurs even for the eccentricities of the left two columns of Fig.~\ref{fig:contours}, where it can be found in larger-scale versions of the contour plots.

The radial critical curves seen in the bottom two rows of Fig.~\ref{fig:contours} show less variability, with the same character of the central inflection and cusp curves as in the bottom panel of Fig.~\ref{fig:contours-e9}. What can be seen newly in the bottom and middle right plots are the orange radial limiting curves, defined by Eq.~(\ref{eq:radial-limiting-curve}) and with dimensions plotted in Fig.~\ref{fig:nfw-radial-limiting-curve}. We note that the inflection curve has an outer component lying beyond the radial limiting curve in the left two columns, and thus of no relevance to the corresponding radial critical curves. However, for $e=0.9986$ in the right column this component of the inflection curve reaches inside the radial limiting curve, so that radial critical curves with $\kappa_\text{s}\gtrapprox54.3$ are pinched in near the minor axis. The corresponding second radial ``peanut'' transition is indicated by the dashed purple boundary in Fig.~\ref{fig:parameter-space-high-kappa}. This highest-eccentricity feature of the ellipsoidal NFW lens occurs only for eccentricities $e\gtrapprox0.9946$. At lower eccentricities this component of the inflection curve does not cross the radial limiting curve and radial critical curves remain fully convex for any $\kappa_\text{s}$ above the solid purple boundary in Fig.~\ref{fig:parameter-space}.

\subsection{Final comments}
\label{sec:Appendix-image-plane-comments}

The properties of the convergence parameter $\kappa_\text{s}$ permit other useful interpretations of the image-plane contour sequences. For example, $\kappa_\text{s}$ is the only model parameter that depends on the source redshift, as discussed in Sect.~\ref{sec:source-redshift}. The different contours may then be viewed as the critical curves of the same lensing object for sources lying at different distances behind the lens, which is particularly relevant for galaxy-cluster-scale halos. Similarly, and closer to its definition, $\kappa_\text{s}$ is directly proportional to the halo density parameter and the virial halo mass. The different contours may then be viewed as the critical curves for the same source but for lenses with different densities or masses.

While we demonstrated the method for the ellipsoidal NFW model, it can be directly applied as described to any other two-parameter lens model in which one of the parameters appears as a purely multiplicative factor of the convergence and the shear. This requirement is hardly restrictive, since most lens models involve such a density-scale parameter. The method can be used just as well for similar models with $n>2$ parameters: each contour plot would then show the variation of critical curves with the multiplicative parameter for a fixed combination of the remaining $n-1$ parameters.

\section{Lensing regimes in parameter space at high $\kappa_\text{s}$}
\label{sec:Appendix-parameter-space-high}

\begin{figure}
\centering
\resizebox{0.67\hsize}{!}{\includegraphics{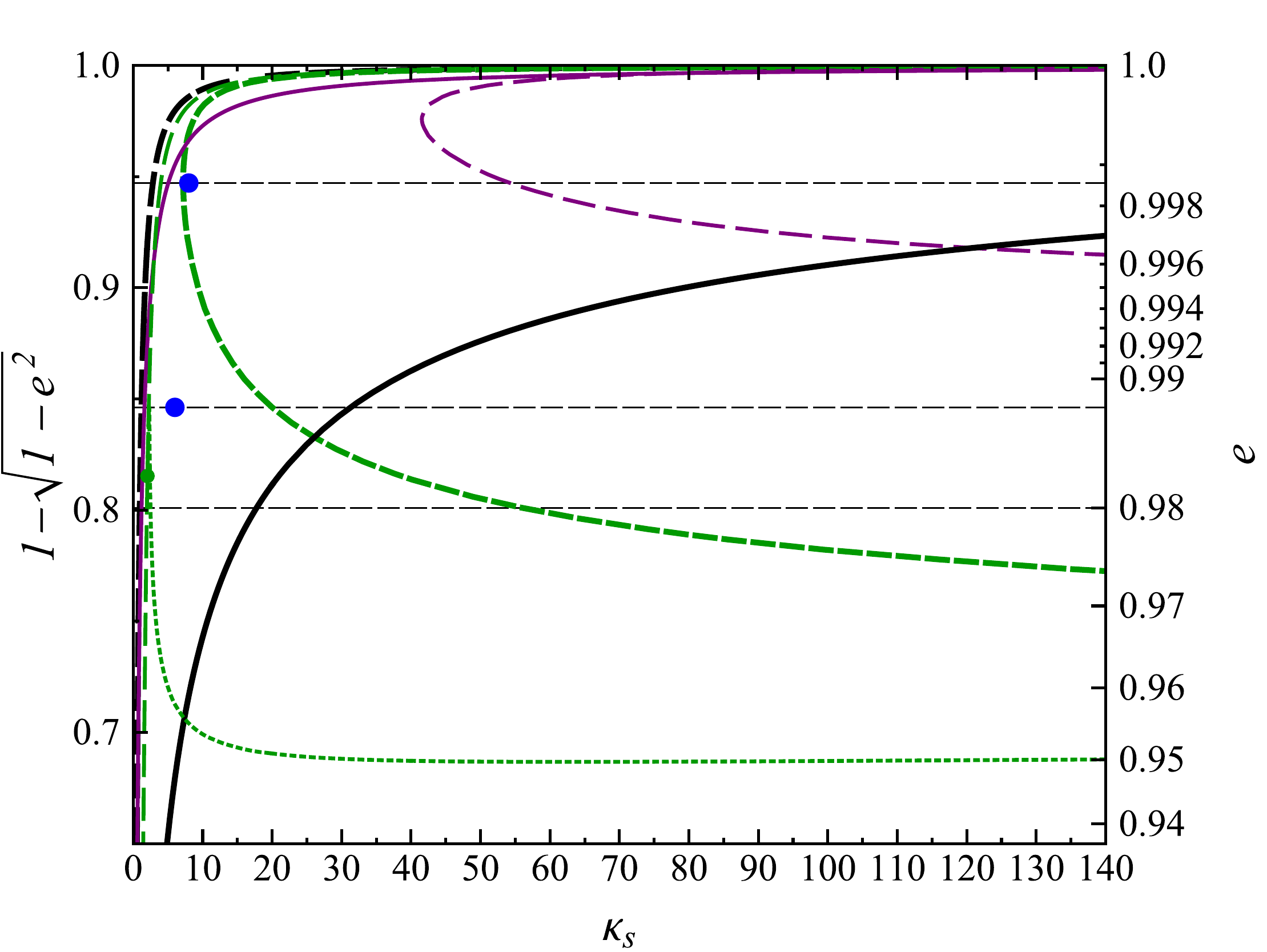}}
\caption{High-$\kappa_\text{s}$ lensing-regime boundaries of the ellipsoidal NFW lens mapped in its parameter space, as described in Appendix~\ref{sec:Appendix-parameter-space-high}, plotted here for ellipticities $1-\sqrt{1-e^2}>0.65$ and for convergence parameters up to $\kappa_\text{s}=140$. Notation same as in Fig.~\ref{fig:parameter-space}, with the additional thin dashed purple boundary marking the minor-axis radial-peanut transition.\label{fig:parameter-space-high-kappa}}
\end{figure}

The structure of the parameter-space map of the ellipsoidal NFW lens at higher convergence-parameter values, further to the right of the range of Fig.~\ref{fig:parameter-space} is shown in Fig.~\ref{fig:parameter-space-high-kappa} for $\kappa_\text{s}\in[0,140]$. We show here only the region of higher ellipticity, $1-\sqrt{1-e^2}\geq 0.65$, corresponding to eccentricities $e\gtrsim0.937$. No boundaries occur at lower eccentricity for $\kappa_\text{s}>5$ within the range of Fig.~\ref{fig:parameter-space-high-kappa}. The plot illustrates the changes of the boundaries from Fig.~\ref{fig:parameter-space} as well as the appearance of another new boundary. The thin dashed purple boundary extending to the right from $\kappa_\text{s}\approx41.6$ corresponds to another ``peanut'' transition on the radial critical curve. In this one two symmetric pairs of inflection points appear at the minor axis of the mass distribution, so that to the right of the boundary the radial critical curve is pinched in vertically.

Regarding the sequence of the boundaries, a closer inspection reveals there are only two boundaries that gradually intersect others within the range of Fig.~\ref{fig:parameter-space-high-kappa}; the remaining boundaries do not cross each other. The solid black cusp-piercing boundary intersects the green dotted ``bulge'' boundary at $\kappa_\text{s}\approx7.37$, the bold green dashed butterfly boundary at $\kappa_\text{s}\approx26.1$, and the purple dashed radial ``peanut'' boundary at $\kappa_\text{s}\approx121$. The solid purple radial ``peanut'' boundary crosses the thin green dashed ``peanut'' boundary at $\kappa_\text{s}\approx2.98$, the bold green dashed butterfly boundary at $\kappa_\text{s}\approx7.62$, and the dashed purple radial ``peanut'' boundary $\kappa_\text{s}\approx78.3$. Over most of the range of Fig.~\ref{fig:parameter-space-high-kappa} the dotted green ``bulge'' boundary represents the first transition when increasing the eccentricity from zero. The boundary drops to a minimum at $(\kappa_\text{s},e)\approx(62.6,0.950)$, then rises slowly with increasing $\kappa_\text{s}$. In general, we note the entirely different sequence of nine transitions with increasing eccentricity at the right edge of Fig.~\ref{fig:parameter-space-high-kappa} in comparison with the generic low-convergence sequence of four transitions as seen in the left panel of Fig.~\ref{fig:parameter-space}.

We have not explored the parameter space for higher values of $\kappa_\text{s}$ beyond the range of Fig.~\ref{fig:parameter-space-high-kappa} comprehensively. Overall, the boundaries tend to level out, their rate of change declines with increasing $\kappa_\text{s}$. The lower part of the new purple dashed peanut boundary from Fig.~\ref{fig:parameter-space-high-kappa} declines monotonically asymptotically to $e\approx0.9945$. The declining lower part of the bold green dashed butterfly boundary eventually crosses the slowly rising dotted green bulge boundary at $(\kappa_\text{s},e)\approx(8000,0.956)$. For higher values of $\kappa_\text{s}$ the butterfly boundary marks the first transition to occur when increasing the eccentricity from zero. However, at such high values of the convergence parameter these results purely illustrate the changing properties of the simple lens model.

\end{appendix}


\begin{thebibliography}{}

\bibitem[Asano(2000)]{asano00}
Asano, K.\ 2000, \pasj, 52, 99

\bibitem[Bartelmann(1996)]{bartelmann96}
Bartelmann, M.\ 1996, \aap, 313, 697

\bibitem[Bourassa \& Kantowski(1975)]{bourassa_kantowski75}
Bourassa, R.~R. \& Kantowski, R.\ 1975, \apj, 195, 13

\bibitem[Brada{\v{c}} et al.(2005)]{bradac_etal05}
Brada{\v{c}}, M., Schneider, P., Lombardi, M., et al.\ 2005, \aap, 437, 39

\bibitem[Bray(1984)]{bray84}
Bray, I.\ 1984, \mnras, 208, 511

\bibitem[Dan{\v{e}}k \& Heyrovsk{\'y}(2015a)]{danek_heyrovsky15a}
Dan{\v{e}}k, K. \& Heyrovsk{\'y}, D.\ 2015a, \apj, 806, 63

\bibitem[Dan{\v{e}}k \& Heyrovsk{\'y}(2015b)]{danek_heyrovsky15b}
Dan{\v{e}}k, K. \& Heyrovsk{\'y}, D.\ 2015b, \apj, 806, 99

\bibitem[Dan{\v{e}}k \& Heyrovsk{\'y}(2019)]{danek_heyrovsky19}
Dan{\v{e}}k, K. \& Heyrovsk{\'y}, D.\ 2019, \apj, 880, 72

\bibitem[Du et al.(2020)]{du_etal20}
Du, W., Zhao, G.-B., Fan, Z., et al.\ 2020, \apj, 892, 62

\bibitem[Einasto(1965)]{einasto65}
Einasto, J.\ 1965, Trudy Astrofizicheskogo Instituta Alma-Ata, 5, 87

\bibitem[El{\'\i}asd{\'o}ttir et al.(2007)]{eliasdottir07}
El{\'\i}asd{\'o}ttir, {\'A}., Limousin, M., Richard, J., et al.\ 2007, ArXiv e-prints [arXiv:0710.5636]

\bibitem[Etherington et al.(2024)]{etherington_etal24}
Etherington, A., Nightingale, J.~W., Massey, R., et al.\ 2024, \mnras, 531, 3684

\bibitem[Ettori et al.(2013)]{ettori_etal13}
Ettori, S., Donnarumma, A., Pointecouteau, E., et al.\ 2013, \ssr, 177, 119

\bibitem[Finney et al.(2018)]{finney_etal18}
Finney, E.~Q., Brada{\v{c}}, M., Huang, K.-H., et al.\ 2018, \apj, 859, 58

\bibitem[Golse \& Kneib(2002)]{golse_kneib02}
Golse, G. \& Kneib, J.-P.\ 2002, \aap, 390, 821

\bibitem[Gomer et al.(2023)]{gomer_etal23}
Gomer, M.~R., Sluse, D., Van de Vyvere, L., et al.\ 2023, \aap, 679, A128

\bibitem[Grogin \& Narayan(1996)]{grogin_narayan96}
Grogin, N.~A. \& Narayan, R.\ 1996, \apj, 464, 92

\bibitem[Jauzac et al.(2018)]{jauzac_etal18}
Jauzac, M., Eckert, D., Schaller, M., et al.\ 2018, \mnras, 481, 2901

\bibitem[Jing \& Suto(2002)]{jing_suto02}
Jing, Y.~P. \& Suto, Y.\ 2002, \apj, 574, 538

\bibitem[Karamazov \& Heyrovsk{\'y}(2022)]{karamazov_heyrovsky22}
Karamazov, M. \& Heyrovsk{\'y}, D.\ 2022, \apj, 927, 101

\bibitem[Karamazov et al.(2021)]{karamazov_etal21}
Karamazov, M., Timko, L., \& Heyrovsk\'{y}, D.\ 2021, \apj, 922, 72

\bibitem[Kassiola \& Kovner(1993)]{kassiola_kovner93}
Kassiola, A. \& Kovner, I.\ 1993, \apj, 417, 450

\bibitem[Keeton(2001)]{keeton01}
Keeton, C.~R.\ 2001, ArXiv e-prints [arXiv:astro-ph/0102341]

\bibitem[Kormann et al.(1994)]{kormann_etal94}
Kormann, R., Schneider, P., \& Bartelmann, M.\ 1994, \aap, 284, 285

\bibitem[Limousin et al.(2007)]{limousin_etal07}
Limousin, M., Richard, J., Jullo, E., et al.\ 2007, \apj, 668, 643

\bibitem[Ludlow et al.(2013)]{ludlow_etal13}
Ludlow, A.~D., Navarro, J.~F., Boylan-Kolchin, M., et al.\ 2013, \mnras, 432, 1103

\bibitem[Meena \& Bagla(2023)]{meena_bagla23}
Meena, A.~K. \& Bagla, J.~S.\ 2023, \mnras, 526, 3902

\bibitem[Meneghetti et al.(2003)]{meneghetti_etal03}
Meneghetti, M., Bartelmann, M., \& Moscardini, L.\ 2003, \mnras, 340, 105

\bibitem[Merten et al.(2015)]{merten_etal15}
Merten, J., Meneghetti, M., Postman, M., et al.\ 2015, \apj, 806, 4

\bibitem[Natarajan \& Kneib(1997)]{natarajan_kneib97}
Natarajan, P. \& Kneib, J.-P.\ 1997, \mnras, 287, 833

\bibitem[Natarajan et al.(2024)]{natarajan_etal24}
Natarajan, P., Williams, L.~L.~R., Brada{\v{c}}, M., et al.\ 2024, \ssr, 220, 19

\bibitem[Navarro et al.(1996)]{navarro_etal96}
Navarro, J.~F., Frenk, C.~S., \& White, S.~D.~M.\ 1996, \apj, 462, 563

\bibitem[Newman et al.(2013)]{newman_etal13}
Newman, A.~B., Treu, T., Ellis, R.~S., et al.\ 2013, \apj, 765, 24

\bibitem[Oguri(2021)]{oguri21}
Oguri, M.\ 2021, \pasp, 133, 074504

\bibitem[Oguri et al.(2003)]{oguri_etal03}
Oguri, M., Lee, J., \& Suto, Y.\ 2003, \apj, 599, 7

\bibitem[Okabe et al.(2013)]{okabe_etal13}
Okabe, N., Smith, G.~P., Umetsu, K., et al.\ 2013, \apjl, 769, L35

\bibitem[O'Riordan et al.(2020)]{oriordan_etal20}
O'Riordan, C.~M., Warren, S.~J., \& Mortlock, D.~J.\ 2020, \mnras, 496, 3424

\bibitem[O'Riordan et al.(2021)]{oriordan_etal21}
O'Riordan, C.~M., Warren, S.~J., \& Mortlock, D.~J.\ 2021, \mnras, 501, 3687

\bibitem[Rodrigues et al.(2017)]{rodrigues_etal17}
Rodrigues, D.~C., del Popolo, A., Marra, V., et al.\ 2017, \mnras, 470, 2410

\bibitem[Salucci(2019)]{salucci19}
Salucci, P.\ 2019, \aapr, 27, 2

\bibitem[Schneider et al.(1992)]{schneider_etal92}
Schneider, P., Ehlers, J., \& Falco, E.~E.\ 1992, Gravitational Lenses (Berlin: Springer-Verlag)

\bibitem[Schramm(1990)]{schramm90}
Schramm, T.\ 1990, \aap, 231, 19

\bibitem[Shajib et al.(2021)]{shajib_etal21}
Shajib, A.~J., Treu, T., Birrer, S., et al.\ 2021, \mnras, 503, 2380

\bibitem[Shajib et al.(2022)]{shajib_etal22}
Shajib, A.~J., Vernardos, G., Collett, T.~E., et al.\ 2022, \ssr, submitted [arXiv:2210.10790]

\bibitem[Suyu et al.(2012)]{suyu_etal12}
Suyu, S.~H., Hensel, S.~W., McKean, J.~P., et al.\ 2012, \apj, 750, 10

\bibitem[Tessore \& Metcalf(2015)]{tessore_metcalf15}
Tessore, N. \& Metcalf, R.~B.\ 2015, \aap, 580, A79

\bibitem[Treu(2010)]{treu10}
Treu, T.\ 2010, \araa, 48, 87

\bibitem[Umetsu(2020)]{umetsu20}
Umetsu, K.\ 2020, \aapr, 28, 7

\bibitem[Umetsu \& Diemer(2017)]{umetsu_diemer17}
Umetsu, K. \& Diemer, B.\ 2017, \apj, 836, 231

\bibitem[van de Ven et al.(2009)]{van_de_Ven_etal09}
van de Ven, G., Mandelbaum, R., \& Keeton, C.~R.\ 2009, \mnras, 398, 607

\bibitem[Wagner(2020)]{wagner20}
Wagner, J.\ 2020, General Relativity and Gravitation, 52, 61

\bibitem[Wright \& Brainerd(2000)]{wright_brainerd00}
Wright, C.~O. \& Brainerd, T.~G.\ 2000, \apj, 534, 34

\end{thebibliography}
\end{document}